\documentclass[twocolumn]{aa}

\usepackage{lscape}
\usepackage{rotating}
\usepackage{graphicx}
\usepackage[final]{hyperref}
\usepackage{rotating}
\usepackage{float}
\usepackage{txfonts}
\usepackage{float}
\usepackage{siunitx}
\usepackage{tabularx}
\usepackage{color,hyperref}
\usepackage{geometry,fancyhdr}
\usepackage{multicol,multirow}
\usepackage{arydshln}
\usepackage[flushleft]{threeparttable}
\hypersetup{colorlinks,breaklinks,
            linkcolor=blue,urlcolor=blue,
            anchorcolor=blue,citecolor=blue}
\usepackage{caption}
\usepackage{longtable}
\bibpunct{(}{)}{;}{a}{}{,}
\usepackage{graphicx}
\usepackage{txfonts}
\bibpunct{(}{)}{;}{a}{}{,} % to follow the A&A style

\begin{document}

   \title{Gaia20bdk -- a new FUor in Sh~2-301 Star Forming Region}
   \titlerunning{Gaia20bdk -- a new FUor}
   \authorrunning{M. Siwak, Á. Kóspál}
%   \subtitle{I. Overviewing the $\kappa$-mechanism}

   \author{M. Siwak\inst{1,2},
        \'A. K\'osp\'al\inst{1,3,4},
        P. \'Abrah\'am\inst{1,3,5},
        G. Marton\inst{1},
        P. Zieli{\'n}ski\inst{6},  
        M. Gromadzki\inst{7},
        \L. Wyrzykowski\inst{7},
        Z. Nagy\inst{1},
        M. Szil\'agyi\inst{1},
        S. B. Potter\inst{8,9},
        R. Sefako\inst{8},        
        H. L. Worters\inst{8},
        D. A. H. Buckley\inst{8,10,11},
        T. Giannini\inst{12},
        E. Fiorellino\inst{13,14},
        F. Cruz-S\'aenz de Miera\inst{15,1},
        M. Kun\inst{1},
        Zs. M. Szab\'o\inst{16,17,1},
        P. W. Lucas\inst{18},        
        J. Krzesi\'nski\inst{19},
        B. Zakrzewski\inst{2},
        W. Og{\l}oza\inst{2},
        A. P\'al\inst{1},
        B. Cseh\inst{1},	
        \'A. Horti-D\'avid\inst{1},
        A. Jo\'o\inst{1},
        Cs. Kalup\inst{1},
        L. Kriskovics\inst{1},
        \'A. S\'odor\inst{1},
        R. Szak\'ats\inst{1},	
        J. Vink\'o\inst{1}
        }
   \institute{Konkoly Observatory, HUN-REN Research Centre for Astronomy and Earth Sciences, MTA Centre of Excellence, Konkoly-Thege Mikl\'os \'ut 15-17, 1121 Budapest, Hungary \email{michal.siwak@csfk.org}
   \and
   Mt. Suhora Astronomical Observatory, University of the National Education Commission, ul. Podchor\k{a}\.zych 2, 30-084 Krak{\'o}w, Poland
   \and
   Institute of Physics and Astronomy, ELTE E\"otv\"os Lor\'and University, P\'azm\'any P\'eter s\'et\'any 1/A, H-1117 Budapest, Hungary
   \and
   Max-Planck-Institut für Astronomie, K{\"o}nigstuhl 17, D-69117 Heidelberg, Germany
   \and
   Department of Astrophysics, 1180 Vienna, T{\"u}rkenschanzstraße 17, Austria
   \and
   Institute of Astronomy, Faculty of Physics, Astronomy and Informatics, Nicolaus Copernicus University in Toru{\'n}, ul. Grudzi\k{a}dzka 5, PL-87-100 Toru{\'n}, Poland
   \and
   Astronomical Observatory, University of Warsaw, Al. Ujazdowskie 4, 00-478 Warszawa, Poland
   \and
   South African Astronomical Observatory, PO Box 9, Observatory 7935, Cape Town, South Africa
   \and
   Department of Physics, University of Johannesburg, PO Box 524, Auckland Park 2006, South Africa
   \and
   Department of Physics, University of the Free State, PO Box 339, Bloemfontein 9300, South Africa
   \and
   Department of Astronomy, University of Cape Town, Private Bag X3, Rondebosch 7701, South Africa
%   \and
%   Department of Physics, Adiyaman University, 02040 Adiyaman, Turkey
   \and 
   INAF-Osservatorio Astronomico di Roma, via di Frascati 33, 00040, Monte Porzio Catone, Italy
   \and
   Instituto de Astrofísica de Canarias, IAC, Vía Láctea s/n, 38205 La Laguna (S.C.Tenerife), Spain
   \and
   Departamento de Astrofísica, Universidad de La Laguna, 38206 La Laguna (S.C.Tenerife), Spain
   \and
   Institut de Recherche en Astrophysique et Planétologie, Université de Toulouse, UT3-PS, OMP, CNRS, 9 av. du Colonel-Roche, 31028 Toulouse Cedex 4, France
   \and 
   Scottish Universities Physics Alliance (SUPA), School of Physics and Astronomy, University of St Andrews, North Haugh, St Andrews, KY16 9SS, UK
   \and
   Max-Planck-Institut für Radioastronomie, Auf dem Hügel 69, 53121 Bonn, Germany 
   \and
   Centre for Astrophysics, University of Hertfordshire, College Lane, Hatfield, AL10 9AB, UK
   \and
   Astronomical Observatory, Jagiellonian University, ul. Orla 171, PL-30-244 Krak{\'o}w, Poland
}

% \abstract{}{}{}{}{} 
% 5 {} token are mandatory
 
  \abstract
   {We analyse multi-colour photometric and spectroscopic observations of a Young Stellar Object Gaia20bdk.}
   {We aim to investigate the exact nature of the eruptive phenomenon that the star has been experiencing since 2018.}
   {We use public-domain archival photometry to characterise the quiescent phase in order to establish major physical parameters of the progenitor. Then, we use our and public-domain optical and infrared photometry and spectroscopy to study the outburst.}
   {Gaia20bdk is a member of the Sharpless~2-301 star-forming region, at a distance of 3.3~kpc. The progenitor is a rather massive $2.7\pm0.5$~M$_{\sun}$, G7-type Class~I young star having an effective temperature of $5300^{+500}_{-300}$~K and bolometric luminosity of $11\pm2$~L$_{\sun}$. The optical and infrared photometric and spectroscopic data obtained during the outburst show a variety of signatures commonly found in classical FUors. Our disc modelling results in a bolometric luminosity of $100-200$~L$_{\sun}$ and mass accretion rate of $1-2\times10^{-5}$~M$_{\sun}$~yr$^{-1}$, also confirming the object's FUor classification. Further monitoring is necessary to track the light changes, accretion rate and spectral variations, as well as to understood the mechanisms behind the disc flickering.}
   {}

   \keywords{Stars: formation, variables: T Tauri, Herbig Ae/Be, Accretion, accretion discs}
   
   \maketitle
%-------------------------------------------------------------------

\section{Introduction}
\label{sec:intro}

FU~Ori-type stars (FUors) were recognized by \citet{Herbig1977} as classical T Tauri-type stars (CTTS) undergoing enhanced disc-matter accretion onto the forming star for a few decades or longer (see also \citealt{Hartmann1996}). According to \citet{fischer2023}, in terms of the accumulated mass and luminosity rise, FUors represent the most spectacular end of the continuum of eruptive accretion in young stars. It is followed by EX~Lupi-type stars (EXors), showing less dramatic eruptive events lasting a few months to a year, but re-appearing on the time scale of a few years \citep{Herbig1989}, and ''intermediate'' cases \citep{Audard_2014prpl.conf387A} of eruptive stars showing properties of both FUors and EXors in various proportions , e.g. V1647~Ori \citep{Andrews2004, Acosta2007, Muzerolle2015}, 2MASS~22352345+7517076 \citep{Kun2019MNRAS.483.4424K}, V899~Mon \citep{Ninan2015, Park2021}, Gaia19ajj \citep{Hillenbrand2019b}, Gaia19bey \citep{Hodapp2020}, Gaia18cjb \citep{Fiorellino2024}, ASASSN-13db \citep{Holoien2014, Sicilia2017}, ASASSN-15qi \citep{Herczeg2016} and a few dozens of other objects discovered by \citealt{Guo2024MNRAS}.

According to \citet{Contreras-Pena2019, Contreras-Pena2024}, \citet{Guo2021}, and \citet{Fiorellino2023}, eruptive events are much more common for Class~I than Class~II of Young Stellar Objects (YSO), and all these stars should undergo at least a dozen eruptive accretion events \citep{Contreras-Pena2019, Contreras-Pena2024}. These events have a major impact on the inner disc temperature causing the expansion of the snow line of various volatiles \citep{Cieza2016}. Therefore, multi-wavelength studies of young eruptive stars -- especially FUor events lasting for several decades and centuries and thus having a major impact on the protoplanetary disc properties -- are crucial for understanding their chemical evolution and eventually also the composition and morphology of planetary systems, including our Solar System \citep{Abraham2009_Natur459224A, Abraham2019_ApJ887156A, Hubbard2017, Hubbard2017ApJ...840L...5H, Molyarova2018, Wiebe2019, Kospal2020, Kospal2023}. 

The recently most recognised list of 26 FUors compiled by \citet{Audard_2014prpl.conf387A} has been expanding constantly thanks to the ground- and the space-based sky surveys. Since 2014 the Gaia spacecraft observes each part of the sky in broad-band filters several dozen times a year \citep{Gaia2016}. After a few years of continuous observations, Gaia became an effective tool for catching long-term brightness changes in various astronomical objects down to 21~mag. The community is notified about significant brightness variations via the Gaia Photometric Science Alert system \citep{Hodgkin2021}. 

Gaia17bpi \citep{Hillenbrand2018_ApJ869146H} and Gaia18dvy \citep{SzegediElek2020_ApJ899130S} are the first two FUors discovered 
by Gaia. The spacecraft also played a significant role in recognizing ESO-H$\alpha$~148 (Gaia21elv) as a bona fide FUor \citep{Nagy2023}, for which in contrast to the previous stars, the alert was triggered due to brightness decrease. Gaia was instrumental in the discovery of Gaia21bty, tentatively identified as a FUor by \citet{Siwak2023}. Other FUors (or at least FUor candidates), among them Class 0 objects, were recently discovered from the ground and space-based infrared telescopes, such as PTF14jg \citep{Hillenbrand2019a}, PGIR~20dci \citep{Hillenbrand2021}, IRAS~16316-1540 \citep{Yoon2021}, WTP 10aaauow \citep{Tran2023}, SST-gbs~J21470601+4739394 \citep{Ashraf2024MNRAS.527.11651}, VVVv721 \citep{Contreras-Pena2017, Guo2020}, SPICY~99341 \& SPICY~100587 \citep{Contreras-Pena2023JKAS...56..253C}, L222-78 \citep{Guo2024MNRAS-L}, while in the accompanying paper \citet{Guo2024MNRAS} presented 15+ confirmed FUors. Finally, RNO~54 was recently recognized as a post-outburst FUor by \citet{Magakian2023A&A...675A..79M} and \citet{Hillenbrand2023}. 

The object of interest here, Gaia20bdk 
($\alpha_{\rm J2000}$ = 7$^{\rm h}$10$^{\rm m}$14$\fs$92, $\delta_{\rm J2000}$=$-$18$^{\circ}$27$'$01$\farcs$04), was identified by the Gaia Photometric Science Alert system on 2020 March 2 as a 1.2~mag brightening of a red star  2MASS~07101491--1827010 or Gaia~DR3~2934263554310209536 \citep{GaiaDR3}. The star is located in the outskirts of an optically bright nebula in Canis Major, only about 4 degrees from the Galactic mid-plane. 
The progenitor of Gaia20bdk was initially classified as Class I/II YSO candidate by \citet{Marton2016}, and Class~II by \citet{Pandey2022}. The authors listed the star among a cluster of 37 identified YSOs, north-east of H~II region Sh~2-301. % The authors found that these YSOs are clustering in the newly identified ''NE-cluster''.
During the modelling of these 37 YSO's spectral energy distributions (SEDs), they determined typical extinction at $A_V=5.8$~mag and age $2.5\pm1.6$~Myr. The mass of our target was determined at $2.67\pm1.5M_{\odot}$. They also determined the distance to the cluster at $3.54\pm0.54$~kpc. The distance is in agreement with another determination $4.3\pm0.6$~kpc to the nearby bubble [HKS2019]~E93 \citep{Hanaoka2019,Hou2014}. This is important as the Gaia DR3 catalogue does not offer well-determined parallax specifically for our target; the photogeometric distance obtained by \citet{Bailer2021} is $1646_{-247}^{+1829}$~pc.

%In spite of the above determinations, we decided to take another approach to refine the distance issue. Most recent study of open clusters observed by Gaia shows that Gaia20bdk shares common proper motions with the members of the young open cluster [DBS2003]~3 \citep{Hunt2023}, overlapping with the ''NE-cluster'' distinguished in Sh~2-301 by \citet{Pandey2022}. GaiaDR3 catalogue provides very accurate $3.32\pm0.04$~kpc distance to the members of [DBS2003]~3, which in principle are young and therefore appear to be the more evolved members of the ''NE-cluster'' from the entire H~II region Sh~2-301. We independently confirmed this distance by means of a histograms of parallaxes of red $1<B_P-R_P<4.5$~mag (with $RUWE<1.4$) stars found in GaiaDR3 catalogue within the radius of 3~arcmin from our target. The histogram peaks at 0.2-0.3~mas, suggesting 4~kpc, but the distance could be in the range 2-4~kpc. The median distance to the stars from this sample is 3.1~kpc, while the standard deviation (of the mean) is only 0.3~kpc. Based on the above, through this paper we will use the distance 3.3~kpc obtained for [DBS2003]~3 members, but to estimate more realistic uncertainties of physical properties computed later in this paper, we choose larger errorbars ($\pm$0.3~kpc) to accomodate the value determined by \citet{Pandey2022}.

We present and analyse new and archival data for Gaia20bdk and draw conclusions about its current nature. In Section~\ref{sec:observations}, we describe the reduction of the archival photometry, and our new spectroscopic and photometric data obtained during the outburst until early 2024. Results obtained from the analysis are presented and immediately discussed in Section~\ref{sec:analysis}, and summarised in Section~\ref{sec:summary}.

%--------------------------------------------------------------------
\section{Observations}
\label{sec:observations}

%-------------------------------------- Two column figure (place early!)
   \begin{figure*}
     \centering
     \includegraphics[width=2\columnwidth]{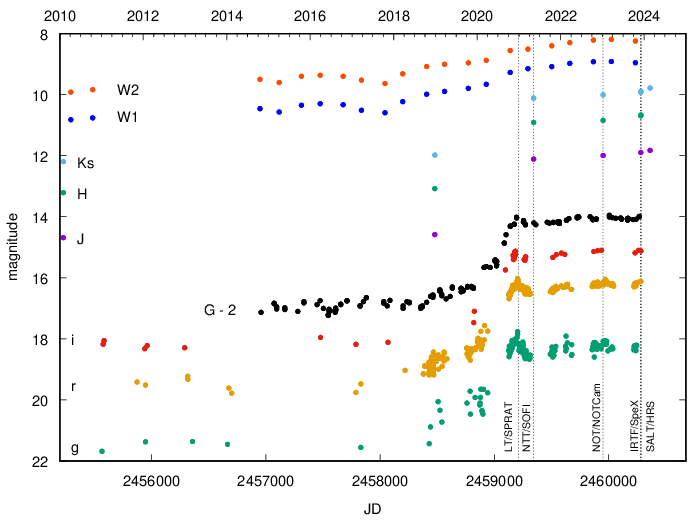}
     \caption{Combined $griJHK_S$ light curves of Gaia20bdk obtained by several ground-based surveys and telescopes, and $G$, $W1$, and $W2$ light curves obtained by Gaia and (NEO)WISE space telescopes, as described in Sections~\ref{sec:phot_pre} and \ref{sec:phot_out}. The first $JHK_S$ data points from 1999 Feb 1 obtained by 2MASS were shifted forward by exactly 11 years, so they appear on this plot on 2010 Feb 1.}
     \label{Fig.lc}
   \end{figure*}
%----------------------------------------------------------------------

\subsection{Pre-outburst photometry}
\label{sec:phot_pre}

In order to study the star during the quiescent phase, we collected archival public-domain visual and infrared photometry. The earliest multi-epoch 2010-2014 optical data were obtained by the Panoramic Survey Telescope \& Rapid Response System (PanSTaRRS) survey in $grizy$-bands
%\footnote{\url{http://ps1images.stsci.edu/cgi-bin/ps1cutouts}} 
\citep{2016arXiv161205560C,2020ApJS..251....3M,2020ApJS..251....4W}. 
We performed our own photometry of Gaia20bdk using the aperture of 11 pixels ($2\farcs75$) and computed the sky within 13-27 pixels annulus. Then, to calibrate the results to standard magnitudes (see in Table~\ref{tab:photometry_pre}), we used the magnitude zero points provided in the headers for each image and filter. We compare our results to APASS9 $gri$ magnitudes of stars in the same field determined by \citet{Henden2015, Henden2016} finding that their measurements are reproduced with our method within 0.05~mag, which is sufficient for our purposes. 

During December, 2014 -- September, 2021, observations in $griz$-filters were collected by the SkyMapper wide-field survey, conducted by a 1.35-m telescope located at Siding Spring Observatory, Australia \citep{2019PASA...36...33O}. In this paper we use the photometric data computed by the dedicated pipeline and available from the fourth data release \citep{Onken2024}.
In nearly the same time, observations in Sloan filters were made on two close epochs -- 2017 February 4 and March 19 -- by the VST Photometric H${\alpha}$ Survey of the Southern Galactic Plane and Bulge (VPHAS+; \citealt{Drew2014}). The survey was conducted by means of the 2.6-m ESO VLT Survey Telescope (VST) at Cerro Paranal, equipped with the OMEGACAM and $ugriH{\alpha}$ filters. We performed photometry on the calibrated fits images downloaded from the public ESO archive using {\sc DAOPHOT~II} procedures \citep{Stetson1987} distributed within the \texttt{astro-idl} library. To ensure optimal extraction (and to comply with our future observations), the aperture of 20.95 pixel ($4\farcs46$) was used for the target and four comparison stars. The inner and outer sky annuli were specifically adjusted for each star between 25-50 pixels to enable accurate sky level calculation. 
We took the $gri$ magnitudes of the four comparison stars from the APASS9 catalogue.
As no colour trends were found, we simply averaged the results obtained from all comparison stars, and computed standard deviation as the uncertainty measure (Tab.~\ref{tab:photometry_pre}). The target was not detected in the $u$-filter, therefore it was fainter than 21.86~mag, as inferred from the magnitude zero point provided in the image header. 

Starting in 2014, the Gaia spacecraft has been monitoring the sky more regularly. Here we show the public domain light curve obtained in the $G$-band until November, 2023, i.e. covering the rising branch and the outburst plateau.

The first pre-outburst near-infrared data were obtained in $JHK_S$ filters by the 2MASS survey on 1999 Feb 1 \citep{Skrutskie2006}, while the second on 2018 December 25-26, during the initial brightness rise \citep{Pandey2022}. On 2010 April 7 data were collected during the fully cryogenic phase of the Wide-field Infrared Survey Explorer \citep[WISE;][]{Wright2010} in $W1$, $W2$, $W3$ and $W4$ bands, centered at 3.4, 4.6, 12, and 22~$\mu$m respectively. Then in the post-cryogenic phase data were collected in $W1$ and $W2$-bands on 2010 October 15. From 2014 October, the NEOWISE mission continued to operate in $W1$ and $W2$-bands \citep{Mainzer2011}, providing long-term light curves covering the quiescent phase and the outburst. Usually over a dozen individual measurements per filter were obtained during each of two (typically 2 days long) pointings towards our target. We downloaded individual catalogue measurements, removed obvious outliers, and computed averages and their standard deviations per each epoch. Prior to this, to make sure that the measurements are unaffected by companions or nebulosity, we examined individual images. Saturation correction was not necessary.

\subsection{Outburst photometry}
\label{sec:phot_out}

We obtained our first observations of Gaia20bdk in $BVRI$ Bessell filters on three nights from December 2020 to January 2021 at the South African Astronomical Observatory (SAAO). We used the modern 1-m {\it Lesedi} telescope, equipped with a Sutherland High Speed Optical Camera (SHOC) providing $5\farcm7\times5\farcm7$ field of view. To match typical seeing conditions, we choose 2x2 binning resulting in 0$\farcs$666 pixel scale.

We continued optical monitoring at the Piszk\'estet\H{o} Mountain Station of Konkoly Observatory (Hungary) and at the Mount Suhora Observatory (MSO) of the University of the National Education Commission (Poland). In the first observatory we used the 80\,cm Ritchey-Chretien (RC80) telescope equipped with an FLI PL230 CCD camera, giving 0$\farcs$55 pixel scale and $18\farcm8\times18\farcm8$ field of view, and Johnson $BV$ and Sloan $r'i'$ filters, while in the second we used the 60\,cm Carl-Zeiss telescope equipped with Apogee Aspen-47 CG47-MB camera operating in the primary 2.4 meter focus, which results in the 1$\farcs$166 pixel scale and $19\farcm0 \times 19\farcm0$ field of view. A modern set of $BVRI$ interference filters manufactured by Custom Scientific was used. 

Observations were also obtained at the ESO-LaSilla observatory on 2021 May 10. We used the 3.58-m New Technology Telescope (NTT) equipped with ESO Faint Object Spectrograph and Camera version 2 (EFOSC2) and $BVR$ filters. With the $2\times2$ binning the angular resolution is 0$\farcs$24 per pixel. 

The original frames were corrected for {\it bias}, {\it dark} and {\it flat-field} in a standard way in {\small \sc IRAF}, while the RC80 data in {\small \sc IDL}. For the MSO data, the correction for {\it bias} was accomplished using the same exposure for {\it dark} as for science frames. For photometric reduction we used custom {\sc IDL} scripts, utilizing the \texttt{astro-idl} procedures from the {\small \sc DAOPHOT~II} package. %\citep{Stetson1987}. 
The aperture photometry was always made in the same aperture size $4\farcs46$, which translates to 4 pixels in MSO, 8.12 in RC80, 18.59 in EFOSC2, 6.66 in SHOC, and 20.95 at VST for pre-outburst photometry. 
The photometric calibration was done based on stable (to 0.02~mag) comparison stars, whose magnitudes were taken from the APASS9 catalogue.
As this catalogue provides Bessell $BV$ and Sloan $g'r'i'$ magnitudes, in the case of $RI$-filters the magnitudes were inferred from interpolation of their SEDs for effective wavelengths of those filters. The SEDs of these stars were constructed with the help of the 2MASS catalogue \citep{Cutri2003}. In the case of MSO and RC80 observations, performed usually between airmass 3--2.6, correction for effects introduced by differential and colour extinction between each comparison star and the target were included. Mean first- and second-order extinction coefficients determined for MSO were used for both observatories.
%, as shown in Table~\ref{Tab.coeff} in Appendix~\ref{ap1}. 
The MSO data were also transformed to the standard $BVRI$ Johnson-Cousins system using colour transformation equations. These corrections considerably decreased the scatter between individual measurements and comparison stars, especially in the $B$-filter. 
In all observatories, typically 3-6 frames per filter per night were obtained. After removal of obvious outliers (caused by satellite trails, guiding errors, cosmic rays) the results were averaged, and standard deviation was computed to estimate the uncertainty (Tab.~\ref{tab:photometry_out}).

Multi-colour coverage of the outburst is also provided by the Zwicky Transient Facility (ZTF, \citealt{Bellm2019}). 
ZTF monitoring is conducted by means of the 48-inch Samuel Oschin Schmidt Telescope at the Palomar Observatory, equipped with a custom-built mosaic CCD camera and Sloan filters. 
Regular monitoring of our target with a 2 day cadence in $gr$-filters started on 2018 September 17, just 1-2 months before the star showed the first signs of the major brightness rise. We downloaded the data provided by the dedicated pipeline \citep{masci2019} from data release no. 20. 

We also use the photometry obtained in the broad-band 'cyan' $c$ and 'orange' $o$ filters of the Asteroid Terrestrial-impact Last Alert System (ATLAS, \citealp{Tonry2018}, \citealp{Smith2020}, \citealp{Heinze2018}). The photometry was downloaded from the ATLAS Forced Photometry web service \citep{Shingles2021}. Although the data quality is low in pre-outburst and during the rising branch, it increases in the plateau. In combination with the 1~day sampling interrupted only by the seasonal breaks, this makes this dataset most suitable for studying the small-scale light variability during the plateau.

Our first near-infrared observations were obtained on 2021 May 9 using the ESO New Technology Telescope (NTT) equipped with the infrared spectrograph and imaging camera, Son of ISAAC (SOFI, \citealt{Sofi1998Msngr..91....9M}). We constructed the 'sky' frame by taking the median of the 5 dither positions after scaling them to the same level, as computed by 'mode' statistic. Then, as the resulting image was very uniform, we only subtracted the 'sky' frame from individual images and performed photometry on individual images in 4.88 pixel ($1\farcs404$) aperture and the sky ring in the range 15-30 pixels (Tab.~\ref{tab:photometry_out}). For this purpose, we chose seventeen stable comparison stars from the 2MASS catalogue. Results obtained from individual images (after removal of obvious outliers) were averaged, and standard deviation was computed to estimate the uncertainty (Tab.~\ref{tab:photometry_out}). 
%The same path was repeated later. 
%$0\farcs288$~pix$^{-1}$ ($4\farcm9\times4\farcm9$ FoV)

One and a half year later, on 2023 Jan 7, we performed observations in $JHK_s$ filters of the NOTCam, installed on the 2.56\,m Nordic Optical Telescope (NOT), located at the Roque de los Muchachos Observatory (La Palma, Canary Islands, Spain, Program ID: 66-103; PI: Nagy). 
%We used the NOTCam with $0\farcs234$~pix$^{-1}$ ($3\farcm9\times3\farcm9$ FoV). 
For each filter, we used a 5-point dither pattern with offsets of five pixels only. This small value and close vicinity of our target to the middle of the FoV, with intersecting bad column and line separating four chip quadrants, caused severe difficulties in obtaining proper photometry. Being unable to compute a 'sky' frame as for NTT/SOFI, we used the dark frames obtained on the same night with the same exposure times as our science frames, as well as the sky flats. The flatfield was constructed as a mean of a series of images, each obtained from subtraction a 'fainter' from a 'brighter' image. The resulting science images (after dark subtraction and division by flat frames) show horizontal nonlinear gradients of the background, strongest at the target position. To remove these gradients, we applied an empirical correction, by computing and subtracting the median value of each line. Then, we also removed the bad intersecting column and line by taking the average of the two neighbouring columns/lines. These operations resulted in a uniform background in the entire FoV, including the close vicinity of the target. Images where the target's point-spread fuction (PSF) was most seriously affected by the bad column and/or line, have been discarded from the scientific analysis. In other cases, we further limited potential distortion of target's PSFs by using fairly small 6 pixel aperture ($1\farcs404$). The sky was computed in the ring of 11-22 for the target , and 15-30 pixels for comparison stars, the same as used for NTT/SOFI. We also obtained photometry from the $J$-band acquisition image, taken on the same night during pointing setup for spectroscopic observations, in which the target was placed in the CCD region unaffected by the bad columns. We obtained the same magnitude (well within the errorbars), which proves that our results (Tab.~\ref{tab:photometry_out}) should not be impaired by the above surgeries. 

On 2023 December 3 and 4, and then on 2024 February 22, we performed photometry on mosaics obtained in $JHK$ filters, each composed from a 5-point dither, gathered by the guide camera of SpeX, which is a medium-resolution 0.7-5.3$\mu m$ spectrograph, mounted on the 3.2-m NASA InfraRed Telescope Facility (IRTF) on Mauna Kea. The photometric calibration was done using the 2MASS magnitudes of a few faint stars visible in the small, $1\farcm \times 1\farcm$ field of view (Tab.~\ref{tab:photometry_out}). 
%We note that effective wavelength of the SpeX's $K$-band filter is slightly longer ($\sim2.2 \mu$m) than in the case of 2MASS, but this is a very small difference.
%Due to low FoV of the camera, only a few comparison stars were used. In all three datasets, the nearby class 'A' sources from the 2MASS catalogue (16-17 in the case of SOFI and NOTCAM, only a few for SpeX) served as comparison stars.  Results obtained from individual images of SOFI and NOTCAM (after removal of obvious outliers) were averaged out, and standard deviation was computed to estimate the uncertainty (Tab.~\ref{tab:photometry_out}). 
We show pre and outburst photometry combined from the aforementioned instruments in Figure~\ref{Fig.lc}.

\subsection{Low-resolution optical spectrum}  
\begin{figure*}
    \centering
    \includegraphics[width=2.\columnwidth]{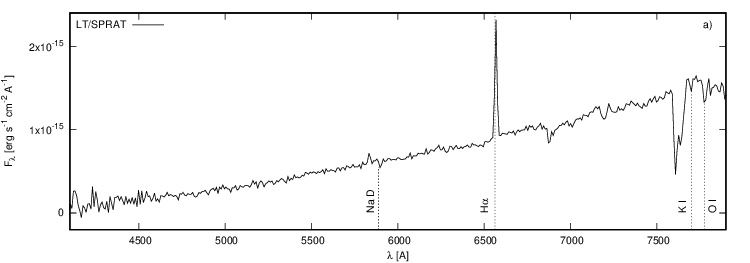}
    \includegraphics[width=2.\columnwidth]{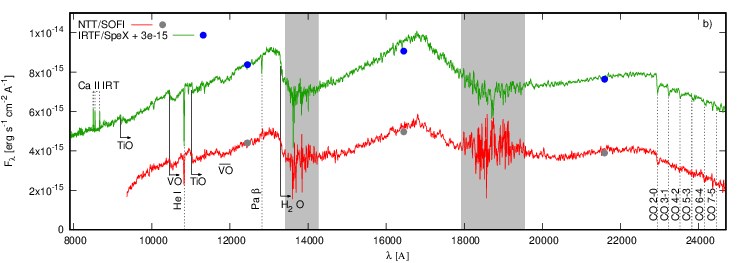}
    \caption{Low-resolution LT/SPRAT spectrum (panel a), and the NTT/SOFI (2021 May 9) and the IRTF/SpeX (2023 December 3) low- and medium-resolution flux-calibrated spectra of Gaia20bdk (panel b). The grey-shaded areas in this panel demarcate regions strongly affected by telluric lines. The IRTF spectrum and photometry were shifted up by $3\times10^{-15}$~erg cm$^{-1}$ s$^{-1}$ \AA$^{-1}$ for clarity. Only the major spectral lines and bands characteristic of FUors are indicated. Telluric absorption bands were not removed from the visual spectrum.}
    \label{fig:spectra}
\end{figure*}

On 2020 December 21/22 we observed Gaia20bdk by means of the SPectrograph for the Rapid Acquisition of Transients (SPRAT, \citealt{Piascik2014}), mounted on the Liverpool Telescope (LT; \citealt{LT}), under ProgID: XOL20B01 (PI: Pawel Zielinski). SPRAT provides low spectral resolution ($R=350$) in the range of $4020-7994$~\AA. 
The spectrum was reduced and calibrated to absolute flux units by a dedicated pipeline. The flux appears to be underestimated by 15-20\%, as obtained by comparison with the nearly-simultaneous ZTF and SAAO $griBVRI$ photometry. The low resolution of the spectrum makes impossible to perform more detailed analyses, however, it clearly shows the H$\alpha$ line in emission, while the \ion{Na}{i}~$\lambda\lambda$5890 and 5896,
%NaD (5889.95, 5895.924~\AA)
\ion{K}{i}~$\lambda$7699 
%KIb (7698.96~\AA) 
and \ion{O}{i}~$\lambda$7774 triplet in absorption (Fig.~\ref{fig:spectra}a).

\subsection{High-resolution optical spectrum}

On 2023 December 7 we used the High Resolution Spectrograph (HRS) mounted on the South African Large Telescope (SALT) under program 2021-2-LSP-001 (PI: Buckley). 
HRS is dual-beam, white-pupil, fibre-fed, échelle spectrograph, employing volume phase holographic (VPH) gratings as cross dispersers. Due to the low brightness of our target, we used the 'low resolution mode' ($R\approx14000$, \citealt{Crause2014}). The spectrum was extracted, wavelength calibrated and heliocentric-velocity corrected by means of the dedicated pipeline \citep{Kniazev2016, Kniazev2017}. Despite using the longest possible exposure time possible for this declination (2400~sec), the flux in the blue arm of the spectrum is almost zero. The flux increases only around 6000~\AA ~reaching signal-to-noise ratio (S/N) of 2-3 at the continuum, so that the H$\alpha$ line clearly shows the P-Cygni profile, typical for FUors. The \ion{Li}{i} $\lambda$6707 line is undefined, but all the spectra smoothed with gaussian function of 5-10 pixel width show depression at the expected wavelength. Except for the \ion{K}{i} $\lambda$7699 the spectra are too noisy to show any well-defined absorption lines which could be used for radial velocity determination and analysis of the disc's rotational profile. 
Another drawback is that the sky-fibre was most likely placed in the region with slightly brighter nebulae. Also, the wavelength calibration made by the pipeline resulted in shifts between the object- and the sky- fibres within each échelle order. Thus, in addition to the above problem, after the sky subtraction we observe strong artificial absorption lines. We identified those nebulae-related lines and distinguished them from terrestrial sky-lines by manual scaling of the sky-spectrum intensity and subtracting it from the target spectrum. At the same time, the discrepancies in wavelength calibrations were to some degree averted for individual targets lines by manual shift of the sky spectrum in wavelength to match position of sky lines observed in the target's spectrum. Only then subtraction of the corrected sky spectrum from target spectrum resulted in satisfactory results.

\subsection{Near-infrared spectroscopy}

On 2021 May 9, we obtained our first near-infrared spectrum by means of SOFI on NTT (Prog. ID 105.203T.002, PI Giannini). We used the low resolution blue and red grisms with GBF and GRF order sorting filters through the 0.6~arcsec wide slit, covering $0.95-1.64$ and $1.53-2.52$~$\mu$m with $R=930$ and 980, respectively. We reduced the spectra in {\sc IRAF}: first the thermal noise and the sky level were approximately removed by subtraction of images obtained in consecutive A and B nod positions. We skipped the flat-fielding as this operation slightly increased the background scatter of the otherwise fairly smooth differential images (i.e. consecutive A-B pairs), and also overall scatter of the extracted spectrum. Due to the field curvature, further steps were performed using the tasks under the \texttt{longslit} package: the xenon lamp spectra and respective line lists provided for SOFI users were used for the wavelength calibration, performed with the \texttt{(re)identify}, \texttt{fitcoords} and \texttt{transform} tasks, to transform the original $(x,y)$ image coordinates into the $(x,\lambda)$. The \texttt{apall} task, with variance weight extraction method and with sky-residuals removal by median, was used for spectrum extraction. Finally we removed atmospheric lines using the \texttt{telluric} task, in which we interactively scaled the normalised spectrum of the telluric standard star HD143520, obtained on the same night, but at different airmass and hour angle; however, it did not affect the cleaning process. Finally we combined the individual telluric-corrected images to form average 'blue' ($0.95-1.64$~$\mu$m) and 'red' ($1.53-2.52$~$\mu$m) spectra, and approximately flux-calibrated it to the simultaneously-obtained $JHK_S$ photometry. The response function was estimated by means of the Planck spectrum closely aligned to the $JHK_S$ points, with the help of nearly-simultaneous $rRiIW1W2$ photometric points (Fig.~\ref{fig:spectra}b). Only the blue edge of the response function was not perfectly estimated, so that the spectrum level is slightly underestimated below 10000~\AA.

On 2023 Jan 7 we obtained the second low-resolution near-infrared spectrum by means of NOTCam (Program ID: 66-103; PI: Nagy). We reduced the spectra in the same way as SOFI spectra. Argon arc lamp was used for $J$-band wavelength calibration, and xenon for the $H$ and $K$. We also decided to skip the flat-fielding operation as it only increased the noise in the extracted spectrum and/or disturbed its otherwise smooth shape. Telluric correction was achieved using HD56405 spectrum, observed at very similar airmass and hour angle as the target. Despite having the simultaneous $JHK_S$ photometry, we decided to skip flux-calibration of the spectrum due to 'rectangular' response function of all gratings, making the proper shape of the spectrum hard to recover.

The third and the best set of near-infrared spectra was obtained on 2023 December 2 and 3 (Program ID: 2023B037; PI: K\'osp\'al) by means of the IRTF equipped with SpeX \citep{SpeX2003PASP..115..362R}. SpeX is a medium-resolution infrared spectrograph, which we used in the Short XD (SXD) mode, providing spectra in the range $0.7-2.55$~$\mu$m. We used the $0.8\times15$~arcsec slit, providing a spectral resolution of $R = 800$. The integration time of the single spectrum was 180~sec. We collected 11 spectra in the first, and 12 spectra in the second night. After the spectrum extraction by the Spextool pipeline \citep{Vacca2003, Cushing2004}, telluric correction was performed using the A0V stars HD20995 on the first night, and HD65102 on the second. Due to rapidly increasing cloudiness during the end of the first night, only in the second night was the telluric standard observed in the close vicinity of the target. We calibrated these spectra to absolute flux by matching to the $JHK_S$ photometry obtained on the second night (Fig~\ref{fig:spectra}b).

%-------------------------------------------------------------
%  \begin{figure}
%  \centering
%  \includegraphics[width=1\linewidth]{figures/ttau_phase-spec.eps}
%  \caption{Light curve of T~Tau with moments of spectroscopic observations indicated by circles.}
%        \label{Fig.S}
%  \end{figure}
%-------------------------------------------------------------------------------------------

\begin{table}
\centering
\caption{Optical pre--outburst photometry of Gaia20bdk. Only our own photometric determinations are listed.}
\label{tab:photometry_pre}
\begin{tabular}{ccccc}
\hline \hline
JD             & Filter& Mag     & Unc   & Survey \\ \hline
%2451210.5341  & $J$  & 14.684 & 0.039 & 2MASS \\
%2451210.5341  & $H$  & 13.210 & 0.034 & 2MASS \\
%2451210.5341  & $K_S$& 12.195 & 0.027 & 2MASS \\ 
%2458478.5     & $J$  & 14.586 & 0.006 & \citealt{Pandey2022} \\
%2458478.5     & $H$  & 13.076 & 0.005 & \citealt{Pandey2022} \\
%2458478.5     & $K_S$& 11.978 & 0.004 & \citealt{Pandey2022} \\\hline\hline 
2455257.84488 & $y$  & 16.859 & 0.021 & PanSTARRS \\
2455283.76421 & $z$  & 17.380 & 0.069 & PanSTARRS \\
2455501.11667 & $y$  & 16.953 & 0.056 & PanSTARRS \\
2455566.90151 & $g$  & 21.684 & 0.237 & PanSTARRS \\
2455576.89514 & $i$  & 18.175 & 0.038 & PanSTARRS \\
2455584.89246 & $i$  & 18.059 & 0.012 & PanSTARRS \\ 
2455634.77410 & $z$  & 17.155 & 0.047 & PanSTARRS \\
2455635.77364 & $y$  & 16.761 & 0.173 & PanSTARRS \\
2455851.14883 & $y$  & 16.919 & 0.049 & PanSTARRS \\
2455874.08709 & $r$  & 19.413 & 0.005 & PanSTARRS \\
2455880.08244 & $z$  & 17.309 & 0.022 & PanSTARRS \\
2455940.91157 & $i$  & 18.321 & 0.034 & PanSTARRS \\
2455947.87062 & $r$  & 19.514 & 0.247 & PanSTARRS \\
2455947.89596 & $g$  & 21.374 & 0.381 & PanSTARRS \\
2455961.82170 & $i$  & 18.219 & 0.037 & PanSTARRS \\
2456016.73084 & $y$  & 16.902 & 0.054 & PanSTARRS \\
2456028.74594 & $z$  & 17.508 & 0.013 & PanSTARRS \\
2456209.11648 & $z$  & 17.241 & 0.024 & PanSTARRS \\
2456209.13848 & $y$  & 16.689 & 0.067 & PanSTARRS \\
2456290.02045 & $i$  & 18.290 & 0.057 & PanSTARRS \\
2456316.95608 & $r$  & 19.227 & 0.057 & PanSTARRS \\
2456317.90965 & $r$  & 19.326 & 0.041 & PanSTARRS \\
2456358.75987 & $g$  & 21.361 & 0.204 & PanSTARRS \\
2456641.98677 & $z$  & 17.420 & 0.044 & PanSTARRS \\
2456642.02794 & $y$  & 16.966 & 0.030 & PanSTARRS \\
2456664.92191 & $g$  & 21.455 & 0.081 & PanSTARRS \\
2456676.84724 & $r$  & 19.611 & 0.129 & PanSTARRS \\
2456701.85159 & $r$  & 19.778 & 0.269 & PanSTARRS \\
%2457012.00411 & $g$  & 22.462 & 0.411 & PanSTARRS \\ 
\hline
%2457006.08957 & $z$  & 17.068 & 0.083 & SkyMapper \\
%2457007.07777 & $z$  & 17.079 & 0.089 & SkyMapper \\
%2457008.05448 & $z$  & 17.233 & 0.108 & SkyMapper \\
%2457102.90240 & $z$  & 16.998 & 0.057 & SkyMapper \\
%2457478.87784 & $i$  & 17.771 & 0.025 & SkyMapper \\
%2457478.87923 & $z$  & 16.929 & 0.012 & SkyMapper \\
%2457735.15542 & $z$  & 17.089 & 0.093 & SkyMapper \\ 
%2458069.24697 & $i$  & 17.918 & 0.069 & SkyMapper \\
%2458771.22497 & $z$  & 16.455 & 0.065 & SkyMapper \\
%2458820.05699 & $i$  & 17.468 & 0.032 & SkyMapper \\
%2458820.05838 & $z$  & 16.494 & 0.027 & SkyMapper \\
%2458825.04814 & $r$  & 18.437 & 0.050 & SkyMapper \\
%2458825.05230 & $i$  & 17.097 & 0.028 & SkyMapper \\
%2458825.05369 & $z$  & 16.160 & 0.022 & SkyMapper \\
%2458899.97646 & $r$  & 18.151 & 0.035 & SkyMapper \\
%2459097.30028 & $i$  & 15.741 & 0.020 & SkyMapper \\
%2459097.30168 & $z$  & 14.826 & 0.017 & SkyMapper \\
%2459465.28638 & $g$  & 17.806 & 0.029 & SkyMapper \\ \hline
2457788.68586 & $r$  & 19.755 & 0.091 & VPHAS+ \\
2457788.69394 & $i$  & 18.177 & 0.051 & VPHAS+ \\
2457831.56270 & $u$  & $>21.86$&  --  & VPHAS+ \\
2457831.58004 & $g$  & 21.556 & 0.165 & VPHAS+ \\
2457831.59207 & $r$  & 19.475 & 0.075 & VPHAS+ \\ \hline
\hline
\end{tabular}
\end{table}

\section{Data analysis}
\label{sec:analysis}

\subsection{Further insights into the distance}

Rather than adopting the literature value (Sec.~\ref{sec:intro}) for the distance to the target, we sought to refine it using data from Gaia~DR3. A recent study for open clusters shows that Gaia20bdk shares common proper motions with the members of the open cluster [DBS2003]~3 \citep{DBS2003}, being the N-E part of the Sh 2-301 star forming region. This cluster can also be found in other catalogues \citep[e.g.:][]{Kharchenko2013,cantat2020,Pandey2022,Hunt2023}. The mean distance obtained by \citet{Hunt2023} for [DBS2003]~3 is 3.3~kpc, which is comparable to the distance others found. Analysing the distance distribution of the member stars we adapt an uncertainty of $\pm$0.3~kpc. Through this paper we will use these values.
%In addition to the distance known from the literature (Sec.~\ref{sec:intro}), we decided to take another approach to refine the value. Most recent study utilizing Gaia~DR3 data for open clusters shows that Gaia20bdk shares common proper motions with the members of the young open cluster [DBS2003]~3 \citep{Hunt2023}. This cluster overlaps with the ''NE-cluster'' distinguished in Sh~2-301 by \citet{Pandey2022}. GaiaDR3 catalogue provides very accurate $3.32\pm0.04$~kpc distance to the members of [DBS2003]~3, which in principle are young and therefore appear to be the more evolved members of the ''NE-cluster''. We additionally investigated 
%independently confirmed this distance by means of a histogram of 
%parallaxes of red $1<B_P-R_P<4.5$~mag (with $RUWE<1.4$) stars found in GaiaDR3 catalogue within the 3~arcmin radius from our target. 
%The histogram peaks at 0.2-0.3~mas, suggesting 4~kpc, but the distance could be in the range 2-4~kpc. 
%The median distance to the stars from this sample is 3.1~kpc, while the standard deviation (of the mean) is 0.3~kpc.
%Based on the above, through this paper we will use the mean distance 3.3~kpc obtained for [DBS2003]~3 members by \citet{Hunt2023}, but to estimate more realistic uncertainties of physical properties computed later in this paper, and to accommodate the value determined by \citet{Pandey2022}, we adapt the uncertainty given by our analysis ($\pm$0.3~kpc).

\begin{figure}
\centering
  \includegraphics[width=1\linewidth]{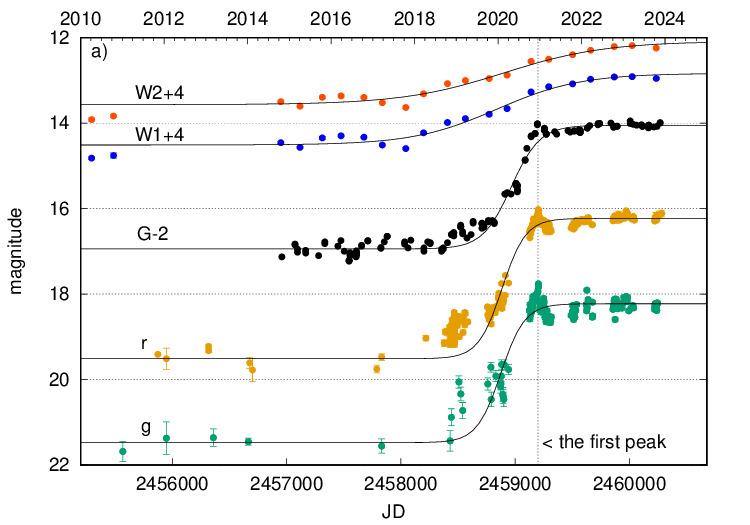}
  \includegraphics[width=1\linewidth]{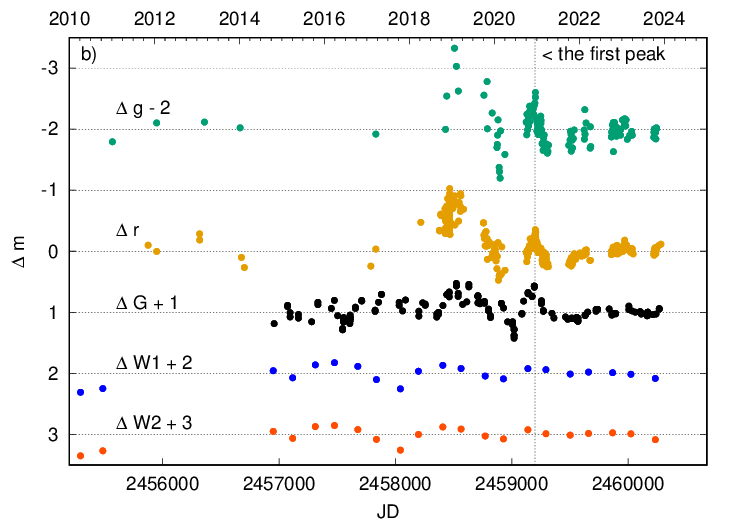}
\caption{Analytical fits of logistic function to the best sampled $grGW1W2$ light curves (
panel a). Second panel shows the residuals after subtraction of respective logistic functions.}
\label{Fig.lc.fit}
\end{figure}

\subsection{General description of the variability}
\label{sec:var_large_scale}

Based on the Gaia light curve displayed in Fig.~\ref{Fig.lc}, the object had quiescent brightness of $G=18.95\pm0.15$~mag, and this situation lasted until mid-November, 2018. Since then, until about mid-November, 2019, Gaia measurements started to show little brightening to $G=18.44\pm0.13$~mag. This initially slow brightening process accelerated sometime in December, 2019\footnote{Due to the break in Gaia measurements of our target, to estimate this moment we used the combined $r$-band light curve.}, which ultimately led to the Gaia Alert issued on 2020 March 2. Gaia20bdk continued the fast brightening until mid-December, 2020, when it peaked at 16.02~mag. In the next few months it slightly dropped to $\sim16.2$~mag, but immediately started slow rise with superimposed wave-like variability ($G=16.11\pm0.08$~mag) -- we will call this stage as 'plateau' later in this paper. The total brightness rise from the mean quiescent level to the first peak was $\Delta G = 2.93$~mag. Due to the slow brightening trend in the plateau and the wave-like variability, $\Delta G$ reached 3~mag for a short time in 2023. We note that this ''2-stages'' brightening behaviour is very similar to what was found in HBC~722, as described in Sec.~3.1 by \citet{Kospal2016}.

%To sum up, the lightening was not smooth but it had two phases, each about one year long; the first slow one (Nov 2018 -- Dec 2019), and the second faster (Dec 2019 -- Dec 2020). But this apparently is still unfinished process, as the mean brightness on the plateau clearly shows a growing tendency.

The same evolution is observed in $gri$ light curves, combined from the data collected by PanSTaRRS, SkyMap, VPHAS+ and ZTF, as well as at the Piszk\'estet\H{o} Mountain Station. The quiescent levels in these filters were $g=21.48\pm0.12$, $r=19.51\pm0.20$ and $i=18.16\pm0.12$~mag. After the initial (Nov, 2018 -- Nov, 2019) slow brightening phase, 
%first recognized above in the Gaia light curve
the brightening rate increased and in December, 2020, the light curves initially peaked for a few days at 17.76, 16.02 and  
%15.129 
$\sim$15.02 in $gri$-bands, respectively. Similarly as in the $G$-band, the brightness dropped slightly after the first peak, then rose and started to vary around $g=18.30\pm0.17$~mag, $r=16.28\pm0.11$~mag and $i=15.24\pm0.11$. The total brightness rise from the quiescence to ''the first peak'' was $\Delta g = 3.86$~mag, $\Delta r = 3.44$~mag, $\Delta i \approx 3.14$~mag. Due to the brightening trend, slower in $g$-band but faster in $ri$-bands these values may be crossed in the near future.

Information extracted from the $JHK_S$ bands is slightly less accurate due to the worse sampling. Based on 2MASS and NTT/SOFI observations, we got $\Delta J\approx 2.57$~mag, $\Delta H\approx2.30$~mag and $\Delta K_S \approx 2.08$~mag. The brightness is also slowly rising during the plateau. This is even more evident in NEOWISE data, where after the quiescent stage ($W1=10.51\pm0.20$ and $W2=9.57\pm0.21$~mag), in 2017/2018 the brightness started to rise with almost a constant rate until 2022. To estimate the amplitude at the same moment as for the optical bands (''the first peak''), we interpolated between the points obtained just before and just after December, 2020. We got $W1\approx9.21$ and $W2\approx8.53$~mag, which gives $\Delta W1\approx1.30$ and $\Delta W2\approx1.04$~mag. In October 2022, and March 2023, the values were by 0.3~mag higher.

In order to extract more accurate information about the outburst timing in different bands, we fitted the analytical logistic function in the form $m=m_0-\Delta m / (1+e^{-(t-t_{1/2})/\tau)})$ to the sufficiently well sampled light curves in $grGW1W2$ filters \citep{Hillenbrand2018_ApJ869146H,Kuhn2024,Guo2024MNRAS-L}. In this function $m_0$ is the quiescent magnitude, known from the observations and listed above. The adjustable parameters are $t_{1/2}$ -- the midpoint of the rising branch, $\Delta m$ -- the outburst amplitude\footnote {Although this parameter was determined above from our observations, we got much better results when we were fitting it, as the brightness of Gaia20bdk is still rising.}, and $\tau$ - the characteristic timescale of the brightness growth. We show the results in Figure~\ref{Fig.lc.fit}a.
However, the combination of the short time-scale secular variability superimposed on the major outburst -- especially strong in the optical bands during the brightness rise in 2018-2020 (Fig.~\ref{Fig.lc.fit}b) -- and the nonuniform sampling, caused a failure of the analytical method. 
%\footnote{Attempts to obey this situation by manually removing some measurements in $gr$-bands to make data coverage more uniform did not bring expected improvements, instead we observed human-dependent fits}. 
%Also, there is not clear whether the star reached the maximum in all bands, especially $W1W2$. 
There is especially no consensus in $t_{1/2}$, i.e. we don't observe monotonic wavelength-dependency as the above authors; In our case $t_{1/2}$ changes from $JD-2458000=875$, 885 and 965~d for $grG$ and then returns to 840 and 900 for $W1W2$, respectively.
We see, however, a clear trend in the characteristic brightness increase time $\tau_0$, which gradually progressed from $103\pm4$, $105\pm10$ and $125\pm6$~d in $grG$ to $370\pm55$ and $445\pm70$~d in $W1W2$.

At first sight the visual analysis and trend seen in $\tau_0$ may suggests that the outburst could start first in the $W1W2$ bands, i.e. around 2017/2018, which is one year earlier than noticed in the $grG$-filters. However, this is likely confusing due to the significant, about $\pm0.2$~mag, quiescent variability in the $W1W2$ bands (Fig.~\ref{Fig.lc.fit}b), as compared to the total brightness rise ($\Delta W1\approx1.30$ and $\Delta W2\approx1.04$~mag). Note that the local minimum visible in the end of 2017 (which on the first sight may look as the outburst start in infrared bands) is similar to that from the beginning of 2015, which suggests (quasi-)periodic event. Taking this fact into account we cannot exclude that the outburst could in fact start later in the $W1W2$ bands, maybe in the middle of 2018 or even together with the optical bands. We therefore conclude that there is no clear evidence of an 'outside-in' outburst propagation mechanism, like in Gaia17bpi and Gaia18dvy \citep{Hillenbrand2018_ApJ869146H, SzegediElek2020_ApJ899130S}.
Our data suggest instead that this outburst has been driven like in HBC~722, that is by a rapid increase in accretion caused by piled-up material from the inner disc leading to the first peak (and perhaps the earlier variation best seen in the 2019 residua), and a slower outward expansion of a hot component, as predicted by \citet{Bell1995}. We do not, however, have enough measurements to prove it later in Sec.~\ref{sec:sed_mod} analytically as accurately as \citet{Kospal2016} and most recently \citet{Carvalho2024}.

\begin{figure*}
\centering
    \includegraphics[width=2.\columnwidth]{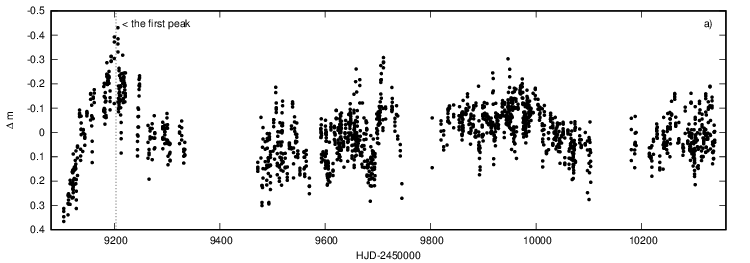}\\
    \includegraphics[width=0.95\columnwidth]{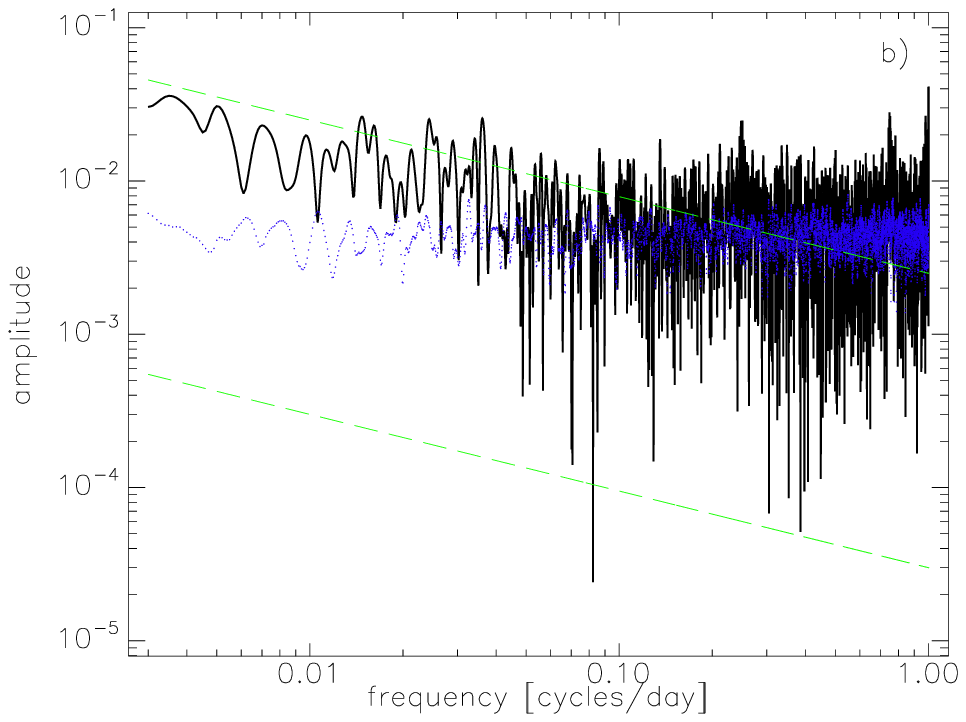}
    \includegraphics[width=1.05\columnwidth]{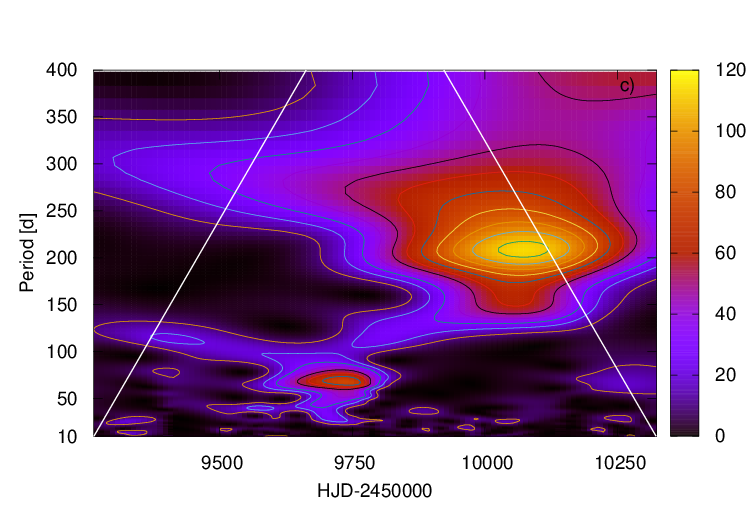} 
    \caption{ATLAS $o$-band light curve obtained during the plateau, with the slow brightening trend removed (panel a). Second panel shows Fourier spectrum (black line) in the log-log scale, calculated from the o-band data taken during the plateau and transformed to normalised flux unit. The flicker-noise nature ($a_f \sim f^{-1/2}$) of these oscillations is indicated by the two parallel green dashed lines, while the amplitude errors are represented by blue dots. Third panel (c) shows respective WWZ spectrum. The major edge effects lie outside of the white dashed lines. WWZ power is expressed in colours, as shown in the right label. }
    \label{fig:resid_o}
\end{figure*}

\subsection{Small-scale variability}
\label{sec:var_small_scale}

In addition to the major outburst, the sufficiently well sampled $grioGW1W2$ light curves show low-amplitude typically 0.01-0.2~mag light changes, some of them likely quasi-periodic oscillations (QPO). In the pre-outburst stage they are visible as 2~yr variation in the $W1W2$-bands after subtraction of the respective logistic functions (Fig.~\ref{Fig.lc.fit}b).\footnote{Although the logistic functions failed in their original application, they are good enough for removing the general brightness increase associated with the outburst.}
Until mid-2018 the variation remained invisible in (at the time best sampled optical) $rG$-bands. 
%However, one can notice {\bf possible} correlation between variability in optical and infrared bands right with the outburst start (2018/2019). 
The 2~yr variation first noticed in the $W1W2$-bands apparently continues during the plateau but with a smaller amplitude, which is probably due to the disc temperature increase. The variation became visible also in optical bands. Our measurements of the $G$-band light curve during the plateau refine the period to $1.8\pm0.2$~yr. In the case of the $g$-band, the 1.8~yr QPO appears to be less securely defined, which is probably caused by combination of the dominant inner disc high frequency variability and the lower photometric precision. The 1.8~yr QPO suggests 2~au location assuming that it is caused by Keplerian rotation of a disc inhomogeneity around 2.67~M$_{\sun}$ star. But the QPO is seen at shorter wavelengths during the plateau, as the disc temperature increased and the previously colder disc annuli dominating at infrared wavelengths during quiescence, are now more active in the optical ones. 
%Thus, assuming that the mechanism causing the variability is the same as before the outburst, this speaks for inhomogeneous wind as the most likely possibility \citep{Szabo2021}. 
A few more years of multi-colour monitoring supported by high resolution spectroscopy will shed more light on the physics behind the observed QPO.

The above low-frequency variations are not the only ones visible in the residuals. 
According to Fig.~\ref{Fig.lc.fit}b, the largest variations are noticed in 2019--2021 during the rising branch, when they reached 1~mag in the $gr$-bands. Amplitude of these oscillations gradually decreases towards longer wavelengths, but they are still visible by NEOWISE. One can notice likely positive correlation between variability in optical and infrared bands during the rising branch.
%although this finding may be slightly affected by the earlier mentioned difficulty with the proper logistic function fitting to the differently sampled light curves (Sec.~\ref{sec:var_large_scale}).} 
Similar conclusion can be drawn with regard to the ''first peak'', which we conventionally treat as the beginning of the plateau. We speculate that the ''first peak'' may have common driving mechanism with the preceding 1~mag variation from 2019.

Subsequently, the best sampled $o$-band ATLAS data show various light changes occurring in the time scales of days, weeks and months during the plateau (Fig.~\ref{fig:resid_o}a). These light changes are also visible in the $grG$ data, but in much less detail due to sparse sampling. Therefore the Sloan-filter data were not taken into account in the frequency analysis, but will be used in Sec.~\ref{sec:c-m} for investigation of colour properties of the small-scale light variations during the plateau.

In order to perform frequency analysis of the $o$-band time series, in the first step all outlier points were manually removed. The long-term trend in brightness was removed by a 2nd order polynomial fit (so that the further results are unaffected by uncertainty related with the logistic functions fits). The result is shown in Fig.~\ref{fig:resid_o}a. Then the magnitudes were transformed to flux units, normalised to unity at the mean brightness level. The Lomb-Scargle diagram \citep{Zechmeister2009} reveals a dozen of significant peaks (with the false alarm of probability of $10^{-10:-36}$), especially at 68 and 207~days. None of them, however, represents stable period or quasi-period. Instead, in accordance with visual inspection, they usually exhibited only two quickly disappearing oscillations. We also computed a Fourier spectrum, as described in \citet{Siwak2013}. It shows the flicker-noise nature of the light changes, as the spectrum amplitude $a_f$ scales with the frequency $f$ as $a_f \sim f^{-1/2}$ \citep{Press1978}. This relation is indicated by the two parallel green dashed lines in Fig.~\ref{fig:resid_o}b. This feature of the Fourier spectrum is typical in a variety of accreting objects \citep{Scaringi2015} and was for the first time studied in FU~Ori by \citet{Kenyon2000}. According to the authors, flickering is observed as a series of random brightness fluctuations with amplitudes of 0.01--1.0 mag that recur on dynamical timescales (a feature observed during Gaia20bdk's plateau), and are therefore believed to be a dynamical variation of the energy output from the disc.

Wavelet analysis allows to trace temporal changes and localise in time finite oscillatory packages. The goal of this analysis is to get deeper insight into the physical processes occurring at different disc annuli, which make up the entire flicker-noise spectrum \citep{Siwak2013,Siwak2018}.
Due to the nonuniform coverage from the ground, instead of wavelet analysis applied to uniformly sampled space-base data, we computed the weighted wavelet Z-transform (WWZ) proposed by \citet{Foster1996}. The major edge effects are indicated by the two white lines, which means that significant signals are contained only inside the trapezium region defined by these lines. We show only signals longer than 10~days, to avoid spurious results at the highest frequencies, most seriously affected by nonuniform sampling. We also decided to cut-off the first peak and start the analysis from $HJD=2459263$, as the strong variation caused by the ''first peak'' strongly re-scaled the spectrum's dynamical range, worsening visibility of the more subtle features. Examination of the WWZ spectrum (Fig.~\ref{fig:resid_o}c) suggests that the $68\pm10$~d signal could represent the final stage of a drifting QPO, which starts as a 115~d signal. However the significance of this drifting feature before it reached 68~d is low due to poor data coverage in the beginning of the plateau. After that the WWZ spectrum became dominated by the $207\pm25$~d signal. Further monitoring should reveal whether it is just a single oscillatory event, or it represents the beginning of a new QPO. The WWZ spectrum reveals a handful of isolated low-significance events at 15--40~d periods, which can in fact be identified directly in the light curve after more careful visual inspection. Perhaps some of them lasted for a few cycles, which would enable to classify them as ordinary QPOs, but it is impossible now due to unequal sampling from the ground. If true, they could be equivalents of the 11~day family in FU~Ori itself \citep{Siwak2018}, likely appearing at the sharp drop in the disc temperature \cite{Zhu2020}, but in Gaia20bdk slightly longer due to the higher stellar mass and stronger disc ionisation.

The one- and two-dimensional frequency analyses performed above did not firmly reveal time-coherent light variability patterns, like in FU~Ori \citep{Siwak2013, Siwak2018}, perhaps also in V2493~Cyg \citep{Green2013, Baek2015} and V646~Pup \citep{Siwak2020}.
%Indeed, it would be rather unlikely if the 68~d period would arise due to the rotation of a disc inhomogeneity due to its more outward location (0.45~au for 2.67~M$_{\sun}$ star), where the disc radiation dominates at infrared wavelengths. 
As for now, the available data suggest that Gaia20bdk shares common properties with FUors with larger $A_V$, like V1057~Cyg \citep{Szabo2021} and V1515~Cyg \citep{Szabo2022}, where TESS data revealed incoherent light changes of even stronger $a_f \sim f^{-1}$ stochastic (Brownian) nature, but evolving towards flicker noise ($a_f \sim f^{-1/2}$) in ground-based data, i.e. at lower frequencies like those investigated here. In the case of V1057~Cyg, \citet{Szabo2021} speculated that apparently the stochastic 'Brownian motion' of hypothetical clumpy condensations in the disc wind destroyed the coherence of the incoming light, i.e. information that the light from the innermost disc is carrying.
Longer, preferably space-based monitoring would resolve this ambiguity. We checked that angular resolution provided by TESS is unfortunately too small to study these issues in our target.

\begin{figure*}
    \centering
    \includegraphics[width=.9\columnwidth]{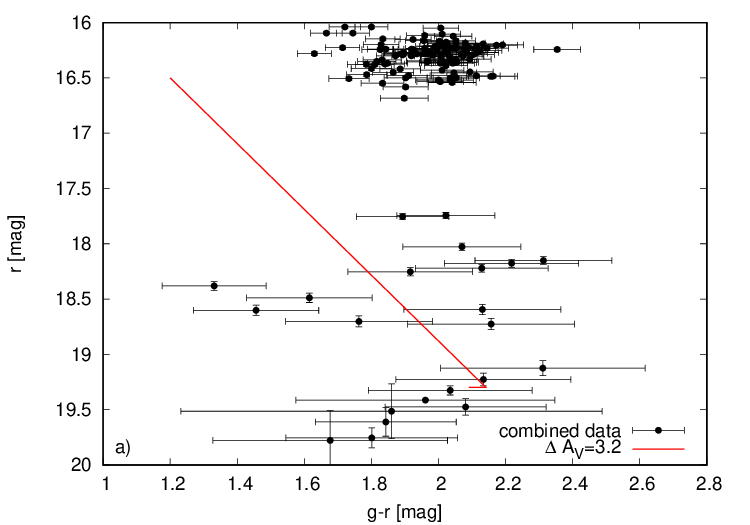}
    \includegraphics[width=.9\columnwidth]{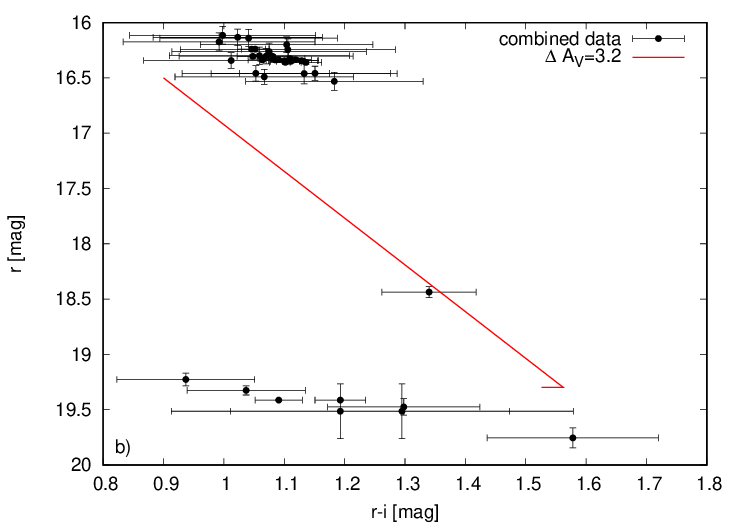}\\
    \includegraphics[width=.9\columnwidth]{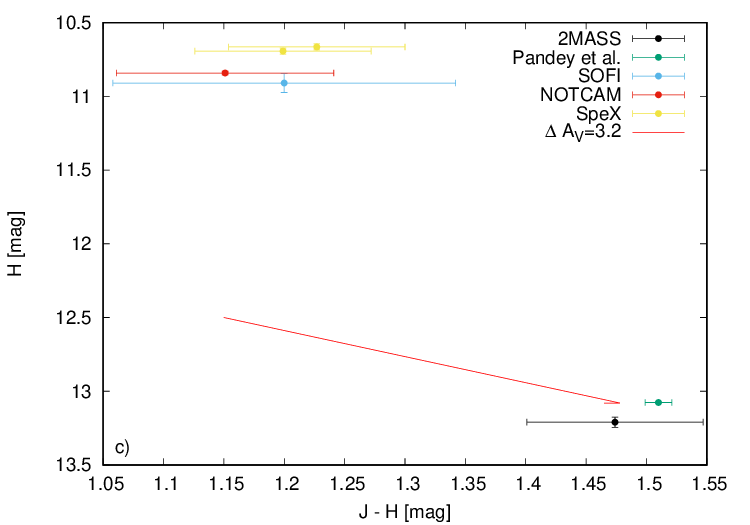}
    \includegraphics[width=.9\columnwidth]{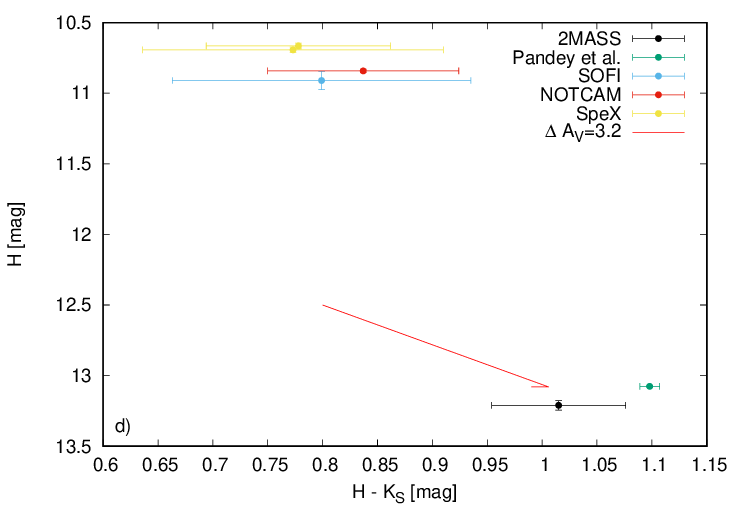}\\
    \includegraphics[width=.9\columnwidth]{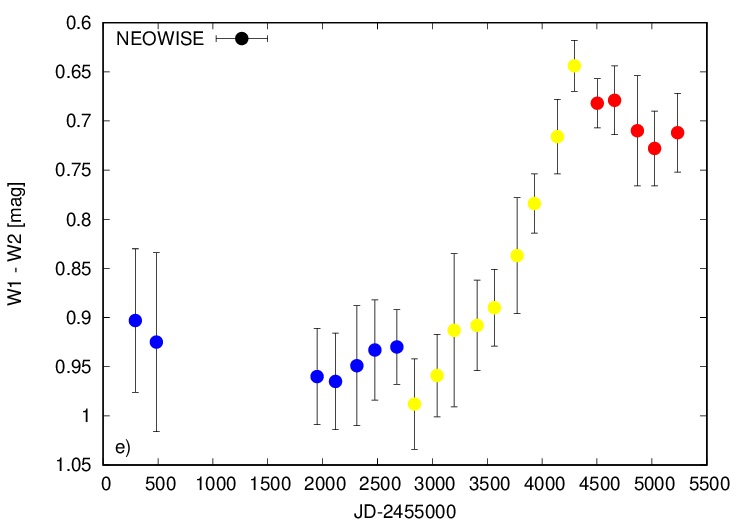}
    \includegraphics[width=.9\columnwidth]{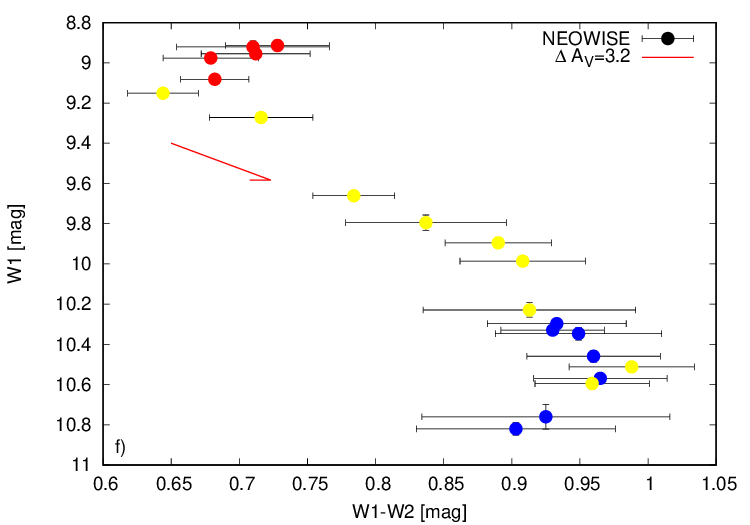}\\
    \includegraphics[width=.9\columnwidth]{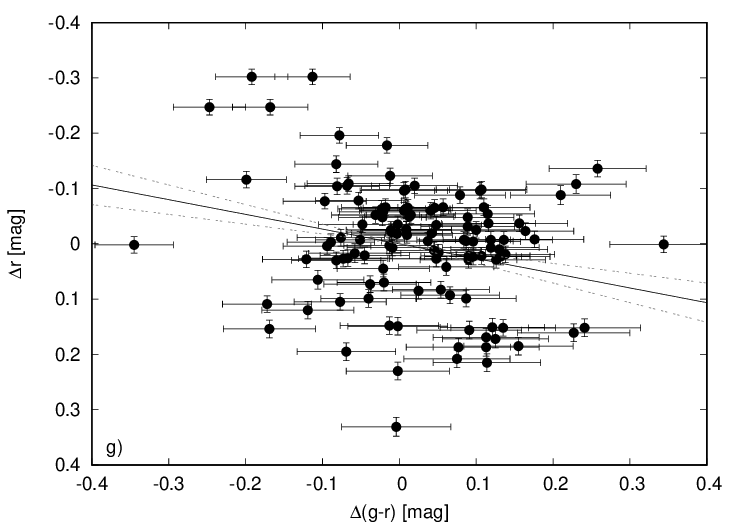}
    \includegraphics[width=.9\columnwidth]{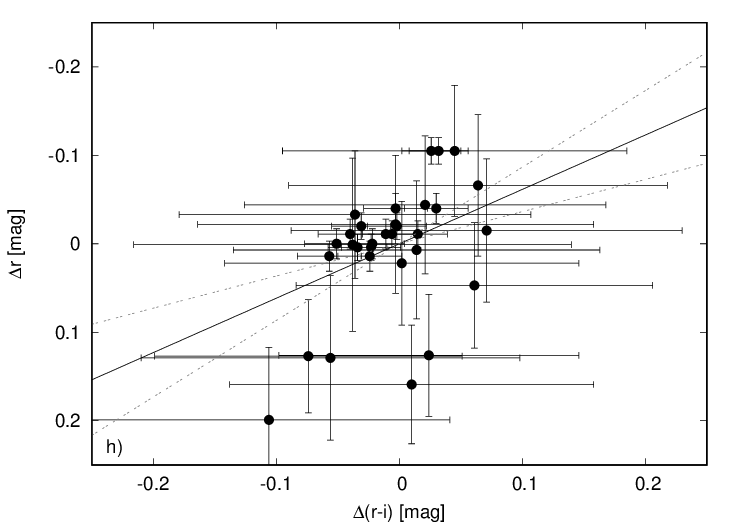}
    \caption{Colour--magnitude diagrams in optical and infrared part of spectrum. The (NEO)WISE data are colour-codded: blue represents the quiescence, yellow -- the rising branch and red -- the plateau. The bottom line shows two diagrams computed from the detrended ZTF and RC80 $gri$ data taken during the plateau only.}
    \label{fig:cmd}
\end{figure*}

\subsection{Colour--magnitude diagrams.}
\label{sec:c-m}

Colour--magnitude diagrams composed of $gri$-band data obtained in various observatories show that transition from the quiescence to the outburst stage did not occur due to the reduction in the extinction (Fig.~\ref{fig:cmd}a and b). The extinction path, computed for Sloan filters applying the \citet{Cardelli1989} law for total-to-selective extinction $R_V=3.1$ and $A_V=3.2$~mag, is indicated as the red semi-arrow. Due to the fact that observations in quiescence were never obtained simultaneously in all bands, we computed the colour indices by pairing data obtained $\pm$~90~days apart. Thus, these data cannot be used for detailed colour variability studies in the quiescence, but just as an indicator of the mean value of the quiescent colour index. Later, during the rising and the plateau, we computed colour indices by pairing points obtained during the same night.

Near-IR colour-–magnitude diagrams in $JHK_S$ filters, colour index versus time and colour-–magnitude diagram based on NEOWISE $W1W2$ photometry are shown in Fig.~\ref{fig:cmd}c, d, e and f, respectively. According to the colour–-magnitude diagrams composed of the ground-based $JHK_S$ data (Fig.~\ref{fig:cmd}c and d), the transition from the quiescent phase to the outburst cannot be explained by reduction in the extinction, which is similar as seen in Sloan bands. Even though the NEOWISE data obtained during the major brightness rise (marked by yellow) are distributed along the extinction path, the observed brightness and colour change is much higher than expected just by $\Delta A_V=3.2$~mag reduction. We note that although the brightness in the $W1W2$ bands was increasing until 2022/2023, the points obtained after the first optical peak (2020/2021, marked by red) started to show different trend, similar to that seen in quiescence (marked by blue).

Analysis of the colour--magnitude diagrams obtained during the plateau to understand colour properties of the flickering (Sec.~\ref{sec:var_small_scale}) is difficult due to the observing strategy adopted in the ZTF survey, providing the only source of nearly simultaneous multi-colour data. In that survey, data in distinct Sloan filters are usually obtained a few hours apart, so that the colour indices are of lower precision. The strategy was much more unfortunate in the case of ATLAS, where data in $co$-bands were usually obtained on different nights, and were therefore rejected from our colour analysis.
We performed least-squares weighted (by xy errors) fits of a linear function to the trends observed in the colour-magnitude diagrams composed of ZTF and Piszk\'estet\H{o} data obtained during the pleateau. The best fit is indicated by the black line and the 1-sigma errors in the slope are indicated by the grey-dashed lines. There is a very weak signature that the data points in the $(g-r)-r$ colour magnitude diagram show the star being bluer when brighter (Fig.~\ref{fig:cmd}g). Slightly more significant is the distribution observed in the $(r-i)-r$ diagram, showing that the star is redder when brighter (Fig.~\ref{fig:cmd}h). This result is similar to that obtained from fits to respective Johnson filters in FU~Ori by \citet{Kenyon2000}. The temperature of the flickering source obtained by the authors suggested the G0 supergiant atmosphere, that is the inner disc atmosphere. We decided not to investigate that issue in detail here, as the proper treatment of such colour--magnitude diagrams requires more accurate, preferably space based light curves to determine associated QPOs, if exists \citep{Siwak2018,Siwak2020,Szabo2021,Szabo2022}. We plan to return to this topic in a few years, once long uniformly sampled accurate light curves in (preferably) several filters will be collected.

\begin{figure}
%    \centering
    \includegraphics[width=1.\columnwidth]{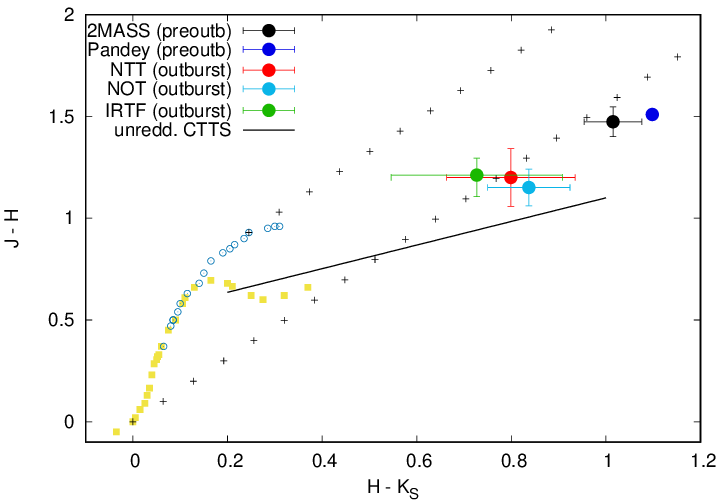}
    \caption{$J-H$ vs. $H-K_S$ colour--colour diagram prepared from pre-outburst (2MASS and \citealt{Pandey2022}) \& outburst (SOFI, NOT, IRTF) data of Gaia20bdk. The squares represent the zero age main sequence stars, while the open circles the giant branch \citep{Bessel1988}. The black line is the locus of unreddened CTTS \citep{Meyer1997}. The two parallel lines formed from pluses represent the reddening path; the step represents the reddening by additional 1~mag in $A_V$.}
    \label{fig_ccd}
\end{figure}

\subsection{Determination of $A_V$ during the quiescent phase from the infrared colour -- colour diagram}
\label{sec:c-c}

In Figure~\ref{fig_ccd} we present the $J-H$ vs. $H-K_S$ colour-colour diagram and the empirical position of unreddened CTTS established by \citet{Meyer1997}. Literature and our own studies (Sec.~\ref{sec:intro},  Sec.~\ref{sec:progenitor}) indicate not very embedded Class~I membership of our target, though the colour--colour diagram itself suggests that the progenitor of Gaia20bdk is a rather ordinary but substantially reddened CTTS. Assuming $R_V=3.1$ for \citet{Cardelli1989} extinction law, we obtained $A_V=5.9\pm0.1$~mag during the only epoch covered with infrared pre-outburst observations obtained by 2MASS. Slightly larger value is obtained by the data of \citet{Pandey2022}, but we skip this determination as the observations were obtained during the initial brightness rise of Gaia20bdk. %We will use other methods for $A_V$ determination in Sections~\ref{sec:Av_spectr} and \ref{sec:sed_mod}.

\subsection{Determination of $A_V$ during the outburst phase from the infrared spectra}
\label{sec:Av_spectr}

As the extinction towards FU~Ori is low and relatively well determined ($A_V=1.7\pm0.1$~mag; see in \citealt{Zhu2007, Siwak2018, Green2019ApJ...887...93G, Lykou2022}), we used another approach to determine the interstellar extinction towards our target, but now during the outburst. This method relies on a considerable similarity of all bona fide FUor spectra in near-infrared, and follows the idea of \citet{Connelley2018} about checking how much dereddening $\Delta A_V$ should be applied to our near-infrared flux calibrated spectra to match the calibrated spectrum of FU~Ori (but scaled to match the flux level of the target). For the single NTT/SOFI GB- and GR-band flux-calibrated spectra, we obtained the best match for $\Delta A_V=4.0\pm 0.1$~mag. In the case of the two IRTF/SpeX spectra, the best match was also found for $\Delta A_V=4.0-4.2$~mag.
This results in $A_V=5.7\pm0.2$~mag and $A_V=5.6-6.0$~mag towards Gaia20bdk during the outburst, respectively. These values are consistent with $A_V=5.9\pm0.1$~mag obtained in Section~\ref{sec:c-c} during quiescence. This means that the increased disc wind caused by the outburst itself did not lead to the substantial drop in extinction, as sometimes observed in young eruptive stars like Gaia19ajj \citep{Hillenbrand2019b}. 

%Finally we note that we did not perform similar procedure for NOTCam spectra due to the strong wavelength-dependency of the response functions of all separate $JHK_S$-grating spectra, and the lack of any overlap that would allow to perform at least approximate corrections on that effect.

\begin{figure}
    \includegraphics[width=0.99\columnwidth]{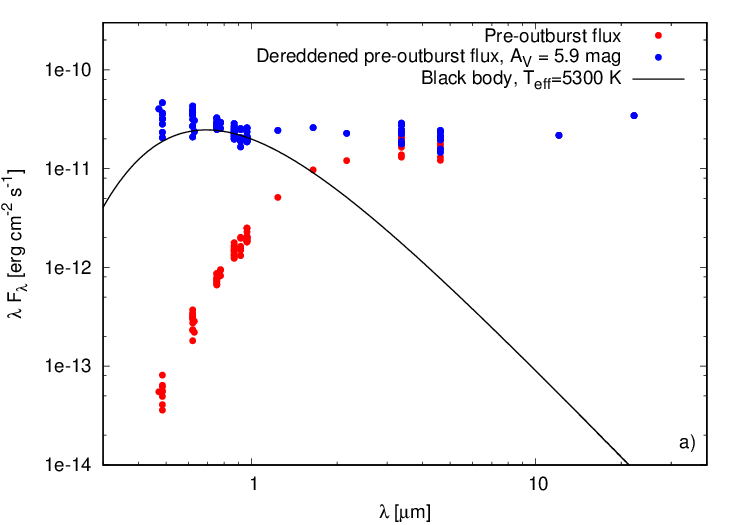}
    \includegraphics[width=0.99\columnwidth]{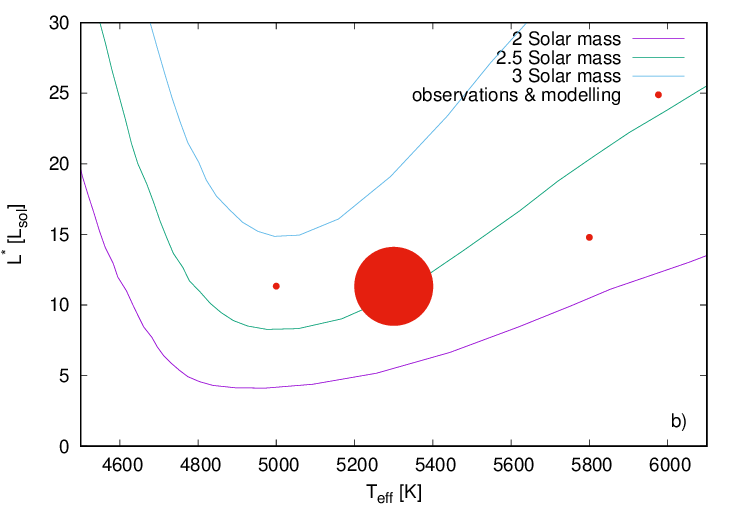}
    \includegraphics[width=0.99\columnwidth]{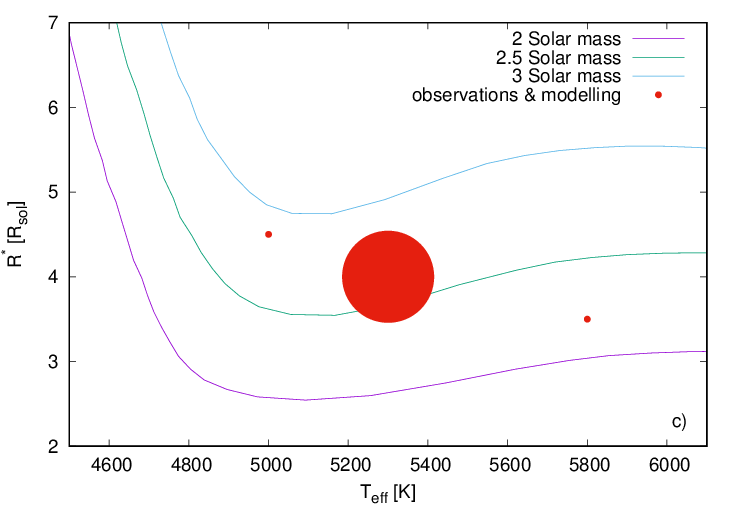}
    \caption{Pre-outburst SED suggesting either flat spectrum or Class~I membership (panel a). The same panel shows results of the dereddened quiescent data modelling. Position of the star in the H-R diagram based on modelling results (panels b\&c); the one obtained with the assumption of the optimal temperature (5300~K) and distance (3.3~kpc) is indicated by the largest dot. The size of the dot fairly well represent the uncertainty in $L_{*}$ and $R_{*}$ caused by uncertainty in distance (3.0 -- 3.6~kpc). (Similar size should be applied to the two small dots to indicate uncertainties).}
    \label{fig_modeling}
    \end{figure}

\subsection{Analysis of the quiescent SED. The progenitor}
\label{sec:progenitor}

Based on WISE magnitudes and using equation 8 in \citet{Kuhn2021}, we derived the spectral index $\alpha$ 
in the range of +0.3 to -0.6 for the pre-outburst SED (Fig.~\ref{fig_modeling}a). Assuming that the single $W4$ measurement is representative for the entire pre-outburst stage, this suggests either Class~I or flat-spectrum membership of Gaia20bdk. 
%The spectral index is not strong enough to indicate a dramatic difference from Class~II sources.

In addition to the above, using the pre-outburst dereddened data,
we derived physical parameters of the progenitor in order to supplement those presented by \citet{Pandey2022}.
For this purpose, we manually aligned a few dozens of Planck functions, computed for different effective temperatures $T_{\rm eff}$ and stellar radii $R_{\star}$ (with 100~K and 0.5~R$_{\sun}$ steps, respectively) to the lower part of the data from 4800--9200~\AA~ region (i.e. to 2-5 points with lowest fluxes per wavelength bin). This choice was dictated by the lack of obvious dips in the pre-outburst stage, which suggests that most of the variability came from the ordinary low-state accretion typical for CTTS. Bearing in mind that blue wavelengths are more enhanced by accretion effects, and that the disc radiation starts to dominate over the stellar one roughly from the $J$-band, we derived best fits for $T_{\rm eff}\approx5300^{+500}_{-300}$~K (Fig.~\ref{fig_modeling}a). For distance $d=3.3$~kpc this results in a stellar radius $R_{\star}\approx4.0^{-0.5}_{+0.5}$~R$_{\sun}$. The $\pm0.3$~kpc uncertainty in the distance introduces an extra $\pm0.5$~R$_{\sun}$ uncertainty to $R_{\star}$ in addition to the listed above ($\pm1$~R$_{\sun}$ in total).

We integrated the Planck function over the entire range of wavelengths, and assuming that the solar bolometric luminosity is $L_{\sun}=3.85\times10^{-33}$erg~s$^{-1}$, we derived a bolometric luminosity of the progenitor of $L_{\star}=11.3$~L$_{\sun}$ for 5000~K and 5300~K (SpT G7), and 12.4~L$_{\sun}$ for 5800~K. The more realistic uncertainty in $L_{\star}$ is related to uncertainty in distance, and for the most reliable $T_{\rm eff}=5300$~K the values are changing from $8.7$~L$_{\sun}$ for 3~kpc, to $14.5$~L$_{\sun}$ for 3.6~kpc. Our results superimposed on the evolutionary tracks of \citet{Siess2000} are in agreement with the mass $2.67\pm1.50$~M$_{\sun}$ derived by \citet{Pandey2022} (Fig.~\ref{fig_modeling}b and c), however, our results may suggest that the real mass uncertainty is smaller, about 0.5~M$_{\sun}$. We also explored the grid of evolutionary tracks prepared for different ages by \citet{Siess2000} in mass, radius, effective temperature and luminosity. Most of them indicate on 2-3~Myr, but 0.5-4~Myr are also possible.

\begin{table*}
\centering
\caption{Physical parameters of the disc during the outburst in three epochs. The disc inclination was set to 60~deg. The three-epoch data and the respective synthetic fluxes are plotted in Figure~\ref{fig_out_modeling}.}

\label{tab:outburst_sed}
\begin{tabular}{ccccccc}
\hline\hline
$d$~[kpc] & $R_{\rm in}$~[R$_{\sun}$] & $R_{out}$~[au] & $L_{d}$~[L$_{\odot}$] & $M\dot{M}$~[M$_{\odot}^{2}$~yr$^{-1}$] & $A_V$~[mag] & $\chi^2$ \\ \hline
{\it EPOCH 1}:& & & & \\ \hline
3.0 & 4.0$^{+0.5}_{-0.5}$& 0.3 &  97$^{-11}_{+14}$ & 2.92$^{+0.05}_{-0.05}\times10^{-5}$ & 4.50$^{-0.25}_{+0.25}$ & 1.76\\ [1ex]
3.3 & 4.0$^{+0.5}_{-0.5}$& 0.3 & 133$^{-16}_{+18}$ & 4.00$^{+0.11}_{-0.02}\times10^{-5}$ & 4.75$^{-0.25}_{+0.20}$ & 2.09\\ [1ex]
3.6 & 4.0$^{+0.5}_{-0.5}$& 0.3 & 177$^{-20}_{+19}$ & 5.34$^{+0.09}_{-0.27}\times10^{-5}$ & 4.95$^{-0.20}_{+0.15}$ & 2.46\\ \hline
{\it EPOCH 2}:& & & & \\ \hline
3.0 & 4.0$^{+0.5}_{-0.5}$& 0.8 &  92$^{-9}_{+12}$ & 2.50$^{+0.06}_{-0.07}\times10^{-5}$ & 3.95$^{-0.25}_{+0.25}$ & 9.13\\ [1ex]
3.3 & 4.0$^{+0.5}_{-0.5}$& 0.8 & 122$^{-11}_{+16}$ & 3.31$^{+0.12}_{-0.06}\times10^{-5}$ & 4.15$^{-0.20}_{+0.25}$ & 9.23\\ [1ex]
3.6 & 4.0$^{+0.5}_{-0.5}$& 0.8 & 155$^{-14}_{+24}$ & 4.22$^{+0.12}_{-0.00}\times10^{-5}$ & 4.30$^{-0.20}_{+0.25}$ & 9.34\\ \hline
{\it EPOCH 3}:& & & & \\ \hline
3.0 & 4.0$^{+0.5}_{-0.5}$& 0.4 & 124$^{-12}_{+18}$ & 3.57$^{+0.13}_{-0.04}\times10^{-5}$ & 4.50$^{-0.20}_{+0.25}$ &3.36\\ [1ex]
3.3 & 4.0$^{+0.5}_{-0.5}$& 0.4 & 165$^{-17}_{+22}$ & 4.76$^{+0.11}_{-0.07}\times10^{-5}$ & 4.70$^{-0.20}_{+0.20}$ & 3.60 \\ [1ex]
3.6 & 4.0$^{+0.5}_{-0.5}$& 0.4 & 220$^{-25}_{+25}$ & 6.35$^{+0.07}_{-0.25}\times10^{-5}$ & 4.90$^{-0.20}_{+0.15}$ &3.87\\ [1ex] \hline\hline
\end{tabular}
\end{table*}

\begin{figure}
    \includegraphics[width=1.\columnwidth]{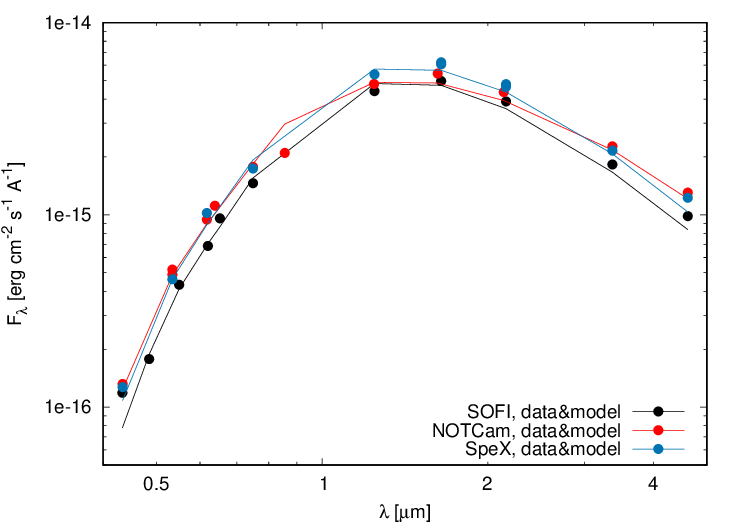}
    \caption{Result from disc modelling of SOFI, NOTCam and SpeX data ({\it EPOCH 1, 2} and {\it 3}, respectively), as shown in Tab.~\ref{tab:outburst_sed}.} 
    \label{fig_out_modeling}
    \end{figure}

\subsection{Analysis of the outburst SED. Determination of the mass accretion rate.}
\label{sec:sed_mod}

Accretion disc model introduced by \citet{Kospal2016} for HBC~722, and then applied to other FUors \citep{Szabo2021, Szabo2022, Nagy2023, Siwak2023} was used here to extract more information about the outburst physics. 
For this purpose first we constructed three separate outburst SEDs from the data gathered in $JHK_S$ filters on 2021 May 9 ({\it EPOCH 1}), 2023 Jan 7 ({\it EPOCH 2}), and 2023 December 3 ({\it EPOCH 3}) and the $griBVRIW1W2$-band data, usually interpolated from the closest data points. The errors of all data points were conservatively assumed to be 5\%. As no information about system geometry is available, we set the disc inclination to $i=60$~deg because $\cos i = 0.5$.\footnote{We also considered other values, but we always came to the conclusion that they do not lead to more satisfactory results.} The inner disc radius was assumed as equal to the stellar radius $R_{\rm in}=4\pm0.5$~R$_{\odot}$. Errors of the remaining parameters are derived by means of the models computed for the upper (4.5) and the lower (3.5) value of $R_{\rm in}$. For each epoch, we considered three values of the distance: 3.0, 3.3 and 3.6~kpc. The passive disc component (e.g. \citealt{Liu2022,Carvalho2024}) was roughly estimated and taken into account during the search of the outer disc radius $R_{out}$ for each epoch.
%; in other words the fit was forced to go slightly below the W1 and W2 data points.

For {\it EPOCH~1} our best fit model for the most likely distance 3.3~kpc resulted in an outer disc radius 0.3~au (0.4~au with passive disc emission component unaccounted), disc bolometric luminosity $L_{d}=133$~L$_{\odot}$, $M\dot{M}= 4.0\times10^{-5}$~M$^2_{\odot}$~yr$^{-1}$ (where $\dot{M}$ is the mass accretion rate), and $A_V=4.75$~mag. 
The last parameter is 1.2~mag lower than determined in Sections~\ref{sec:Av_spectr} and \ref{sec:c-c}. 
We speculate that this may be due to the model imprefections, and maybe also by the fact in all cases our extinction estimates are based on photometric and spectroscopic data gathered in $JHK_S$-bands.

The fit to the {\it EPOCH 2} data is less accurate ($\chi^2=9.23$). The best fit resulted in higher outer disc radii, about 0.8~au (1-1.1~au with passive disc emission component unaccounted), and slightly lower $L_d$ and $M\dot{M}$. We also encountered problems with the proper reproduction of $A_V$, which resulted in 4.15~mag only.

We encountered the same dilemmas with model-returned $A_V=4.70$~mag while modelling the {\it EPOCH~3} dataset. The model resulting in the most realistic parameters gave $\chi^2=3.60$. As in EPOCH 1, we obtained 0.4~au for the outer disc radius (0.5-0.6~au with passive disc emission component unaccounted). The disc luminosity is higher (165~L$_{\odot}$) than in earlier epochs and the same is true for $M\dot{M}$. We show the modelling results for $d=3.3$~kpc in Figure~\ref{fig_out_modeling}.

Taking $d=3.3$~kpc and the stellar mass 2.67~M$_{\odot}$, we calculate a mass accretion rate $\dot{M}=1.50\times10^{-5}$~M$_{\odot}$~yr$^{-1}$ during the first epoch, $\dot{M}=1.24\times10^{-5}$~M$_{\odot}$~yr$^{-1}$ during the second, and $\dot{M}=1.78\times10^{-5}$~M$_{\odot}$~yr$^{-1}$ during the third. Based on results listed in Table~\ref{tab:outburst_sed}, slightly larger $\dot{M}$ can be derived for $d=3.6$~kpc, and smaller for 3.0~kpc.  Uncertainty in the stellar mass given by \citet{Pandey2022} affects the results more seriously (to $\pm 60\%$), but even then the mass accretion rate remains in the range typical for FUors. We think, however, that the real uncertainty in mass is much smaller based on evolutionary tracks (Fig.~\ref{fig_modeling}b and c) and it does not exceed 0.5~M$_{\odot}$, so that the $\dot{M}$ computed from individual models are accurate to about 30\%.

The ''variations'' of the outer disc radius in different {\it EPOCHs} are probably caused by the non-negligible 
small-scale variability observed in all bands. We found that {\it EPOCH} 1 and 3 SEDs were composed of measurements obtained close to the local light minima, whereas {\it EPOCH}~2 during the local light maximum during the plateau. Future observations may let us determine the mechanism staying behind these small-scale light variations, and to determine $R_{out}$ in an unambiguous way.

\subsection{Spectral features and their variability}

Our spectra of Gaia20bdk contain all features commonly found in spectra of classical FUors, as summarised by \citet{Connelley2018}. The most pronounced are the strong molecular $\Delta\nu=2$ CO band heads absorptions (starting from 2.29$\mu$m in the $K$-band) and the triangular shape of the entire $H$-band, caused by water vapour absorption typically observed in low-gravity young stars (Fig.~\ref{fig:spectra}b). The spectra also show the two major absorption lines in the $J$-band, \ion{He}{i} and Pa$\beta$ (though in 2021 May 9 the second line was likely still in emission or equal to the continuum). Both 2021 and 2023 spectra show TiO and VO molecular absorption bands. The first dominate the red part of late-type stars spectra, while the second are typical of higher luminosity M7 stars, but absent in M-dwarfs. We mark these sharply starting bands by continuous black lines and arrows in Fig.~\ref{fig:spectra}b, similarly like the beginning of the water vapour absorption (H$_2$O). 

\begin{figure*}
    \centering
    \includegraphics[width=2.\columnwidth]{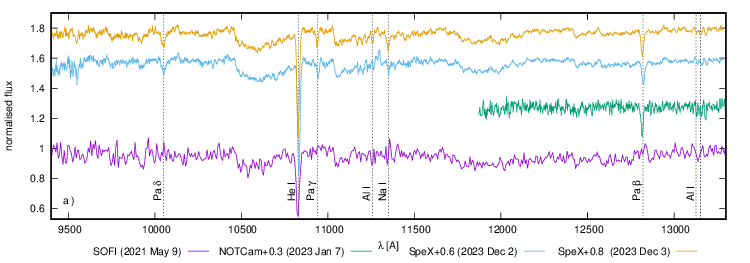}
    \includegraphics[width=2.\columnwidth]{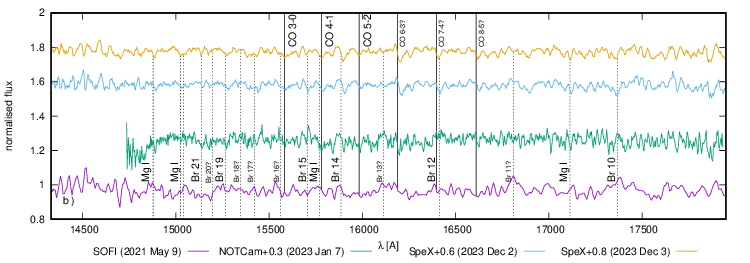}
    \includegraphics[width=2.\columnwidth]{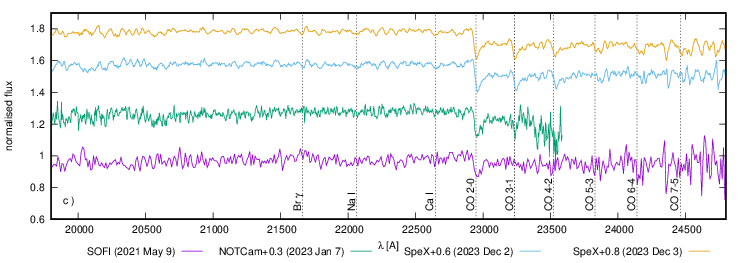}
    \caption{Lines identified in the near-infrared spectra. In the case of $H$\&$K$ bands, unidentified lines (or with uncertain identification) from the bracket and CO series, are labelled with a smaller font and a question mark. Not that the large fraction of positive results we owe to the IRTF/SpeX spectra. In the $H$-band, the position of CO series is marked by solid black lines, while the bracket series and metal lines by the dashed lines.} 
    \label{fig:spectra_norm}
\end{figure*}

In addition to Pa$\beta$, the more detailed look into the $J$-band spectra reveals Pa$\gamma$ and Pa$\delta$ (Fig.~\ref{fig:spectra_norm}a). They were likely in emission or undetected in 2021 May 9, but are very well defined in spectra from 2023 December. Attempts to find Pa$\alpha$ (1.875$\mu$m) failed even in the best SpeX spectrum due to strong contamination of this region by telluric lines.

We identified five lines from the Bracket series in the $H$-band (Fig.~\ref{fig:spectra_norm}b). They are typically weak in FUors, thus they are best visible in the SpeX spectra, which have the best signal-to-noise ratio among the spectra we obtained. Positions and names of six undetected or tentatively detected lines from the Bracket series are indicated for completeness, but these are marked with question mark. In the $K$-band SpeX spectra we also detected Br$\gamma$ showing P-Cygni profile (Fig.~\ref{fig:spectra_norm}c), as typically observed in classical FUors by \citet{Connelley2018}.

In addition to the major CO bands visible in the $K$-band, in the $H$-band we also identified three (out of six) lines belonging to the $\Delta \nu=3$ band head. We mark their positions by solid black lines. The first two lines closely coincide with positions of other lines, but detailed look reveals stronger broadening (caused by lines merging/blending) and two separate bottoms (local minima) at the expected positions of each line. 
Other young eruptive stars also show this CO series, e.g. Gaia19ajj \citep{Hillenbrand2019b}, PGIR20dci \citep{Hillenbrand2021}, V960~Mon and RNO54 \citep{Hillenbrand2023}.

We also detected metals commonly found in other FUors by \citet{Connelley2018}. Many of these positive detections we owe to the SpeX spectra, which reveal \ion{Na}{i} and \ion{Ca}{i} in the $K$-band, \ion{Mg}{i} in the $H$-band, and \ion{Al}{i} and \ion{Na}{i} in the $J$-band.
Following the authors, we measured equivalent widths of \ion{Na}{i}, \ion{Ca}{i} and the first $\Delta\nu=2$ CO band overtone (measured from 2.292 to 2.320$\mu$m), obtaining EW(\ion{Na}{i}+\ion{Ca}{i})$\approx4.4$ and EW(CO)$\approx29.4$~\AA. This places Gaia20bdk exactly in the region occupied by the bona fide FUors and FUor-like objects in figures 9 and 10 of \citet{Connelley2018}. More specifically, Gaia20bdk in practise overlaps with FU~Ori itself on those diagrams.

\begin{figure}
%    \centering
    \includegraphics[width=1.\columnwidth]{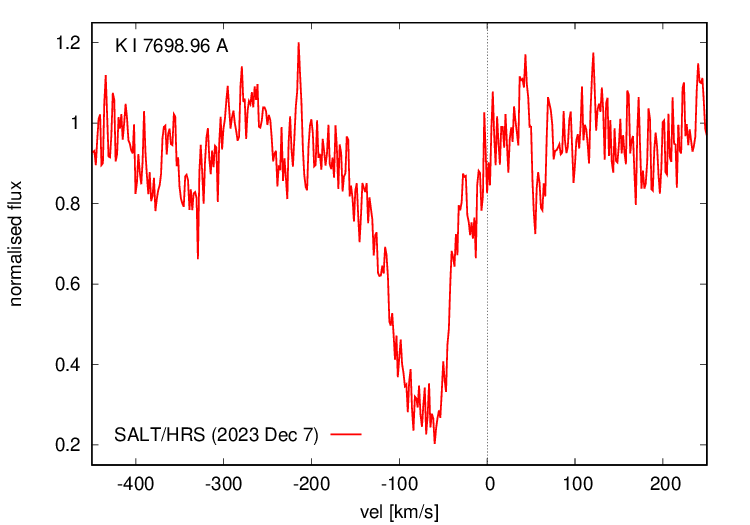}
    \includegraphics[width=1.\columnwidth]{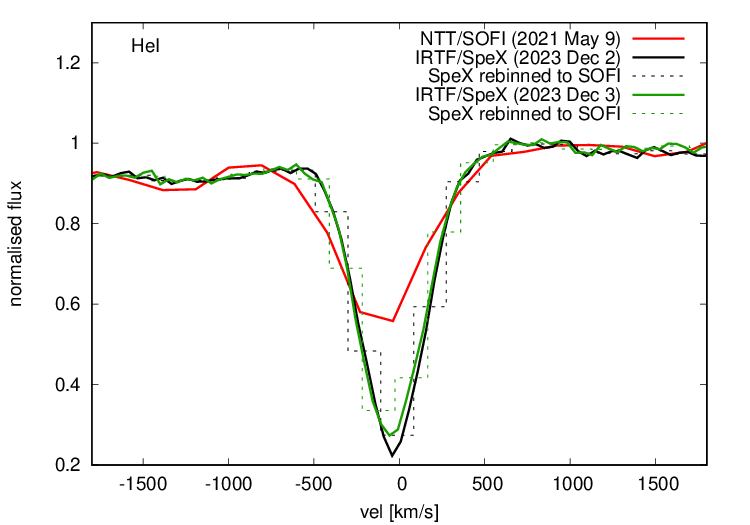}
    \includegraphics[width=1.\columnwidth]{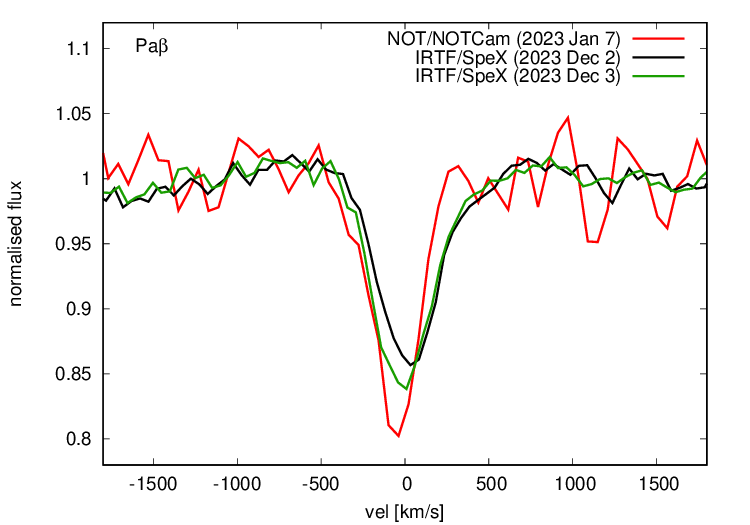}
    \caption{Single snapshot of \ion{K}{i}~$\lambda$7699, and the evolution of \ion{He}{i} and Pa$\beta$ lines in velocity space.} 
    \label{fig:spectra_abs}
\end{figure}

Three metal elements, commonly found in FUors were also identified in absorption in our LT/SPRAT spectrum, that is Na~D, \ion{K}{i}~$\lambda$7699 and \ion{O}{i}~$\lambda$7774 triplet (Fig.~\ref{fig:spectra}a). The SALT/HRS spectrum obtained three years later firmly confirms the absorption status of the \ion{K}{i}~$\lambda\lambda$7665 and 7699; though the first one is strongly contaminated by telluric lines) and the \ion{O}{i} triplet. In the case of the \ion{K}{i}~$\lambda$7699, in addition to the depression likely intrinsic to the disc (i.e. near the expected wavelength), the HRS spectrum reveals stronger and blueshifted absorption components (at heliocentric radial velocity $-75\pm2$~km~s$^{-1}$, Fig.~\ref{fig:spectra_abs}). Assuming that systemic radial velocity of Gaia20bdk is not larger than $\pm30$~km~s$^{-1}$, this may indicate a wind outflow or a shell feature, like those observed in V1057~Cyg by \citet{Herbig2003} and \citet{Szabo2021}.
% and Gaia19ajj \citep{Hillenbrand2019}.
According to the authors of the first paper, who decomposed such lines into several separate components, they are probably due to condensations in the expanding wind passing in front of the star-disc system.

Although the HRS spectrum is too noisy to directly reveal the Na~D and \ion{Li}{i}~$\lambda$6707 lines, spectra (especially those from the target fibre, i.e. without the sky subtraction) smoothed with gaussian functions of different widths, always show flux depression near the expected wavelength. However, similarly as observed by \citet{Herbig2003} and \citet{Szabo2021}, the lithium line appears to be blueshifted, thus we likely detected the shell component dominating the entire line profile like in \ion{K}{i}~$\lambda$7699.

\begin{figure*}
    \centering
    \includegraphics[width=1.\columnwidth]{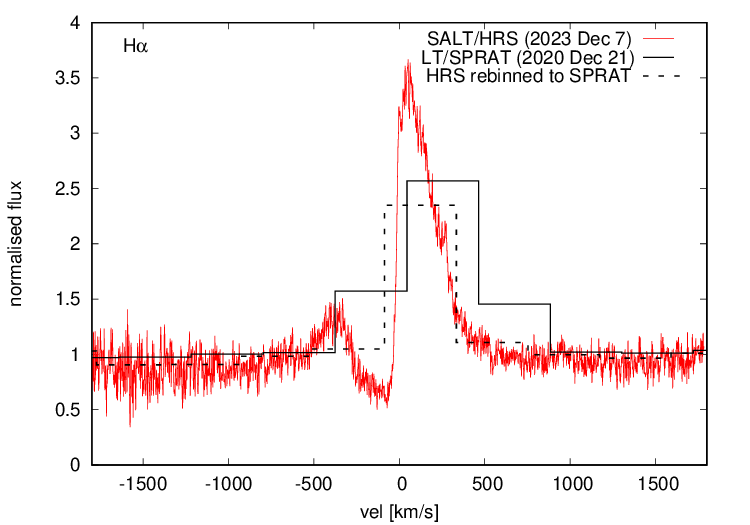}
    \includegraphics[width=1.\columnwidth]{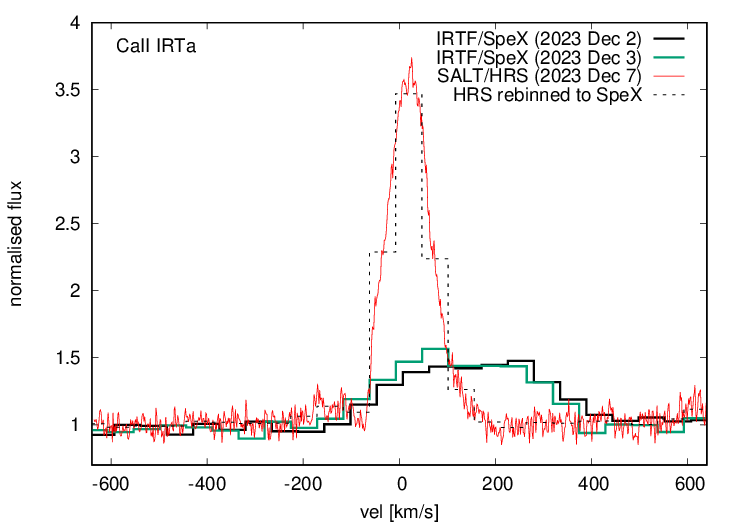}\\
    \includegraphics[width=1.\columnwidth]{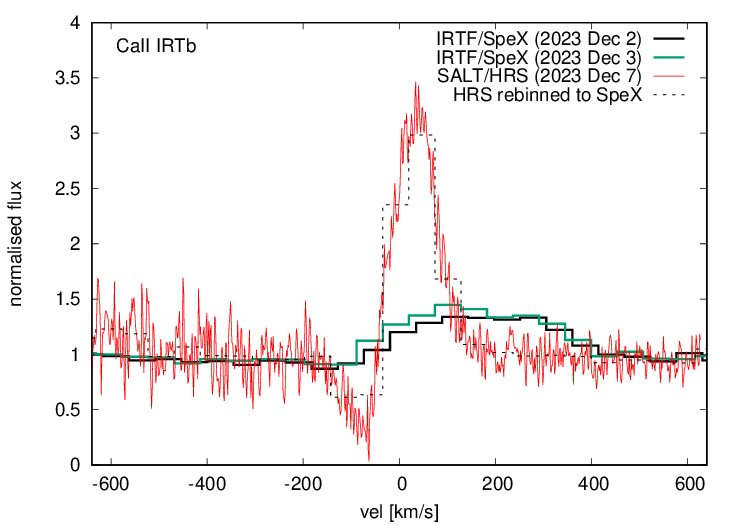}
    \includegraphics[width=1.\columnwidth]{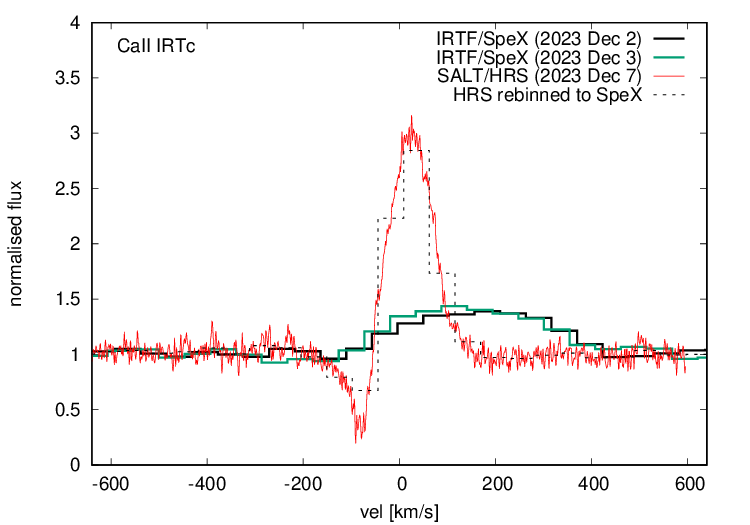}
    \caption{Evolution of H$\alpha$ and \ion{Ca}{ii}~IRT in velocity space.} 
    \label{fig:spectra_pcyg}
\end{figure*}

The visual and near-infrared spectra contain only a few emission lines. The strongest is H$\alpha$ and slightly weaker are those from the calcium infrared triplet (\ion{Ca}{ii}~IRT), as shown in Fig.~\ref{fig:spectra}ab. However, high-resolution SALT/HRS spectra revealed their P-Cygni nature (Fig.~\ref{fig:spectra_pcyg}), commonly found in FUors. The highest wind velocities inferred from the blueshifted absorption components reach $-400$~km~s$^{-1}$ (H$\alpha$), and almost twice less in the calcium triplet.

The structure of H$\alpha$ is likely variable, as inferred from comparison of the line profiles obtained by SPRAT and HRS. 
As these instruments have very different spectral resolutions, to enable direct comparison we downgraded resolution of the HRS spectrum (0.04~\AA~pix$^{-1}$) to that of SPRAT (9.2~\AA~pix$^{-1}$). 
The result shown in Figure~\ref{fig:spectra_pcyg} reveals that three years earlier (2020 December 21) the H$\alpha$ line most likely showed the emission component only. 
This could be true if the absorption component was relatively weak or just moderate in comparison to the emission one, as observed in the spectrum of Gaia21bty (likely a FUor), obtained at the beginning of the outburst \citep{Siwak2023}.

Obtained within five days three spectra covering the calcium infrared triplet reveal strong variability of all three components. Both SpeX spectra show just a strongly redshifted emission. But a more careful look reveals similar variations of all the three components within one day, which may be interpreted as a gradual evolution towards the shape observed four days later by HRS. The HRS spectrum reveals stronger emission component but also previously unseen absorption component (both leading to the P-Cygni profile) in all but the first lines (Fig.~\ref{fig:spectra_pcyg}). Based on the HRS spectrum resampled to the SpeX's resolution, the absorption component was likely absent in the SpeX spectra. Thus, it appears that P-Cygni profile is not always present in CaII~IRT. Time-series observations obtained with a daily cadence are necessary to study this phenomenon, and the same conclusion also applies to the H$\alpha$ (see also in \citealt{Powell2012}).

We made similar comparisons for the He~I~10830 (for this purpose the SpeX spectra were resampled to the lower SOFI's resolution), where the line turned out to be the strongest in 2023 (Fig.~\ref{fig:spectra_abs}). The variability of Pa$\beta$ is more complex, as in the first spectrum taken by SOFI (2021 May 9) the line was undefined in the noisy continuum, but only later (2023) seen in absorption. In Gaia21bty, the two infrared spectra taken early in the outburst showed the line first in emission and then (48 days later) probably equal to the continuum \citep{Siwak2023}. Note that Pa$\beta$ depth in the Gaia20bdk spectra is also variable, as inferred from NOTCam and SpeX spectra, both providing practically the same resolutions. We also note that both in the He~I and Pa$\beta$ their blue and read wings extend to $\pm500$~km~s$^{-1}$, and that the blue ones are more affected by the extended wind absorption.

\section{Conclusions}
\label{sec:summary}

Gaia20bdk -- a Class~I member of relatively distant (3-3.6~kpc) Sh~2-301 Star Forming Region in Canis Major -- started systematic brightening in late 2018. It became quickly obvious that both the brightening amplitude (from 3.9 in blue to 1.3~mag in infrared) and the rate may indicate a new FU Ori-type eruptive event. Literature study shows that the brightening process was similar to that observed in HBC~722. The colour--magnitude diagrams show that the brightening did not occur along the extinction path, therefore it can not be explained by extinction reduction in the line of sight. Our near-infrared spectra show all features typical for FUors, especially CO-bandhead in absorption, triangular shape of $H$-band spectrum, and most of hydrogen and heavier elements in absorption, definitely confirming the FUor classification of the object. The H${\alpha}$ and \ion{Ca}{ii}~IRT lines are showing P-Cygni profiles, with stronger emission components accompanied by weaker absorptions, indicative on high-velocity (up to -500~km~s$^{-1}$) disc wind. The shape of the \ion{Ca}{ii}~IRT varies on a time scale of days, which is probably caused by non uniform density of the inner disc wind.

The light curve plateau shows small-scale photometric variability. It is usually associated with disc flickering observed in other FUors and accreting objects. The longest variability component shows 1.8~yr quasi-periodicity, and it was probably also present in the quiescence. If so, it would be caused by variable extinction or a wind, which means that it originates in the outer disc environments rather than in the inner, and would probably not be related with the eruptive event. The shorter 68--210~d significant signals are likely single oscillatory events, thus not meeting the QPO definition. But the wavelet spectrum suggests possible family of 10--40~d QPOs, which could originate in the inner disc at the place of the sharp temperature drop, like numerically predicted at 0.15~au in FU~Ori by \citet{Zhu2020}. 

From the data gathered during the plateau, we constructed three SEDs. Models assuming the distance of 3.3~kpc, the inner disc radius 4~R$_{\sun}$ and the stellar mass 2.67~M$_{\sun}$ result in the disc luminosity of 133--165~L$_{\sun}$ and mass accretion rate $1.2-1.8\times10^{-5}$~M$_{\sun}$~yr$^{-1}$, which is typical for FUors. In practice, these values are invariant on the assumed distance (3--3.6~kpc) and the stellar mass (2.2--3.2~M$_\sun$). Our best-fit models obtained for {\it EPOCHs 1 and 3}, suggest rather small outer 'active' disc radius (0.3--0.4~au). Unfortunately, $A_V$ of 5.6-6.0~mag obtained from the near-infrared observations, is not well reproduced by our disc models.

Further observing efforts are necessary to monitor the outburst behaviour and to resolve several inconsistencies raised above. The target is bright enough to make time-series spectroscopic and photometric observations for a variety of further analyses, especially Doppler imaging of the inner disc atmosphere and the wind structure.

\begin{acknowledgements}
%ESA PRODEX
We acknowledge support from the ESA PRODEX contract nr. 4000132054.
We acknowledge the Hungarian National Research, Development and Innovation Office grant OTKA FK 146023.
G. M. and Zs. N. were supported by the J\'anos Bolyai Research Scholarship of the Hungarian Academy of Sciences. G.M. acknowledges support from the European Union’s Horizon 2020 research and innovation programme under grant agreement No. 101004141.

%ERC - SACCRED
This project has received funding from the European Research Council (ERC) under the European Union's Horizon 2020 research and innovation programme under grant agreement No 716155 (SACCRED). 

%AK
This work was also supported by the NKFIH excellence grant TKP2021-NKTA-64.

%Ele
E.F. acknowledges financial support from the project PRIN-INAF 2019 'Spectroscopically Tracing the Disc Dispersal Evolution (STRADE)'.

%Zsof Sz.
Zs.M.Sz. acknowledges funding from a St Leonards scholarship from the University of St Andrews. For the purpose of open access, the author has applied a Creative Commons Attribution (CC BY) licence to any Author Accepted Manuscript version arising.
Member of the International Max Planck Research School (IMPRS) for Astronomy and Astrophysics at the Universities of Bonn and Cologne.

%NOT and OPTICON acknowledgement
Based on observations made with the Nordic Optical Telescope, owned in collaboration by the University of Turku and Aarhus University, and operated jointly by Aarhus University, the University of Turku and the University of Oslo, representing Denmark, Finland and Norway, the University of Iceland and Stockholm University at the Observatorio del Roque de los Muchachos, La Palma, Spain, of the Instituto de Astrofisica de Canarias.
The research leading to these results has received funding from the European Community's Horizon 2020 Programme (H2020/2021-2024) under grant agreement number 101004719 (ORP).

%ESO SOFI EFOSC VPHAS+ VVV VVVX
Based on observations made with ESO Telescopes at the La Silla and Paranal Observatories under programme ID: 105.203T.001. Based on data obtained from the ESO Science Archive Facility.

%IRTF SpeX
This work is based in part on data obtained with the NASA Infrared Telescope Facility, which is operated by the University of Hawaii under a contract with the National Aeronautics and Space Administration.
The authors wish to recognize and acknowledge the very significant cultural role and reverence that the summit of Maunakea has always had within the indigenous Hawaiian community.  We are most fortunate to have the opportunity to conduct observations from this mountain. 

%SAAO
This paper uses observations made at the South African Astronomical Observatory (SAAO).
Some of the observations reported in this paper were obtained with the Southern African Large Telescope (SALT). Polish participation in SALT is funded by grant No. MNiSW DIR/WK/2016/07.
%No. MEiN nr 2021/WK/01.

%2MASS
This publication makes use of data products from the Two Micron All Sky Survey, which is a joint project of the University of Massachusetts and the Infrared Processing and Analysis Center/California Institute of Technology, funded by the National Aeronautics and Space Administration and the National Science Foundation.

%WISE
This publication makes use of data products from the Wide-field Infrared Survey Explorer, which is a joint project of the University of California, Los Angeles, and the Jet Propulsion Laboratory/California Institute of Technology, funded by the National Aeronautics and Space Administration.

%IRSA
This research has made use of the NASA/IPAC Infrared Science Archive, which is funded by the National Aeronautics and Space Administration and operated by the California Institute of Technology.

%Gaia
We acknowledge ESA Gaia, DPAC and the Photometric Science Alerts Team (\url{http://gsaweb.ast.cam.ac.uk/alerts}).

%PanStarrs
The Pan-STARRS1 Surveys (PS1) and the PS1 public science archive have been made possible through contributions by the Institute for Astronomy, the University of Hawaii, the Pan-STARRS Project Office, the Max-Planck Society and its participating institutes, the Max Planck Institute for Astronomy, Heidelberg and the Max Planck Institute for Extraterrestrial Physics, Garching, The Johns Hopkins University, Durham University, the University of Edinburgh, the Queen's University Belfast, the Harvard-Smithsonian Center for Astrophysics, the Las Cumbres Observatory Global Telescope Network Incorporated, the National Central University of Taiwan, the Space Telescope Science Institute, the National Aeronautics and Space Administration under Grant No. NNX08AR22G issued through the Planetary Science Division of the NASA Science Mission Directorate, the National Science Foundation Grant No. AST-1238877, the University of Maryland, Eotvos Lorand University (ELTE), the Los Alamos National Laboratory, and the Gordon and Betty Moore Foundation.

%SkyMapper
The national facility capability for SkyMapper has been funded through ARC LIEF grant LE130100104 from the Australian Research Council, awarded to the University of Sydney, the Australian National University, Swinburne University of Technology, the University of Queensland, the University of Western Australia, the University of Melbourne, Curtin University of Technology, Monash University and the Australian Astronomical Observatory. SkyMapper is owned and operated by The Australian National University's Research School of Astronomy and Astrophysics. The survey data were processed and provided by the SkyMapper Team at ANU. The SkyMapper node of the All-Sky Virtual Observatory (ASVO) is hosted at the National Computational Infrastructure (NCI). Development and support of the SkyMapper node of the ASVO has been funded in part by Astronomy Australia Limited (AAL) and the Australian Government through the Commonwealth's Education Investment Fund (EIF) and National Collaborative Research Infrastructure Strategy (NCRIS), particularly the National eResearch Collaboration Tools and Resources (NeCTAR) and the Australian National Data Service Projects (ANDS).

%SVO
This research has made use of the Spanish Virtual Observatory (\url{http://svo.cab.inta-csic.es}) supported from Ministerio de Ciencia e Innovación through grant PID2020-112949GB-I00..

%MSO
This paper uses observations made at the Mount Suhora Astronomical Observatory, Poland. 

%RC80
The operation of the RC80 telescope at Konkoly Observatory has been supported by the GINOP
2.3.2-15-2016-00033 grant of the National Research, Development and Innovation Office (NKFIH) funded by
the European Union.

%LT
Based on the observations made with the Liverpool Telescope operated on the island of La Palma by Liverpool John Moores University in the Spanish Observatorio del Roque de los Muchachos of the Instituto de Astrofisica de Canarias with financial support from the UK Science and Technology Facilities Council. The Liverpool Telescope is operated on the island of La Palma by Liverpool John Moores University in the Spanish Observatorio del Roque de los Muchachos of the Instituto de Astrofisica de Canarias with financial support from the UK Science and Technology Facilities Council.

%ADS
This paper made use of NASA's Astrophysics Data System (ADS) Bibliographic Services. 
\end{acknowledgements}
\bibliographystyle{aa}
\bibliography{gaia20bdk.bbl}

\begin{thebibliography}{115}
\expandafter\ifx\csname natexlab\endcsname\relax\def\natexlab#1{#1}\fi

\bibitem[{{{\'A}brah{\'a}m} {et~al.}(2019){{\'A}brah{\'a}m}, {Chen},
  {K{\'o}sp{\'a}l}, {Bouwman}, {Carmona}, {Haas}, {Sicilia-Aguilar}, {Sobrino
  Figaredo}, {van Boekel}, \& {Varga}}]{Abraham2019_ApJ887156A}
{{\'A}brah{\'a}m}, P., {Chen}, L., {K{\'o}sp{\'a}l}, {\'A}., {et~al.} 2019,
  \apj, 887, 156

\bibitem[{{{\'A}brah{\'a}m} {et~al.}(2009){{\'A}brah{\'a}m}, {Juh{\'a}sz},
  {Dullemond}, {K{\'o}sp{\'a}l}, {van Boekel}, {Bouwman}, {Henning},
  {Mo{\'o}r}, {Mosoni}, {Sicilia-Aguilar}, \&
  {Sipos}}]{Abraham2009_Natur459224A}
{{\'A}brah{\'a}m}, P., {Juh{\'a}sz}, A., {Dullemond}, C.~P., {et~al.} 2009,
  \nat, 459, 224

\bibitem[{{Acosta-Pulido} {et~al.}(2007){Acosta-Pulido}, {Kun},
  {{\'A}brah{\'a}m}, {K{\'o}sp{\'a}l}, {Csizmadia}, {Kiss}, {Mo{\'o}r},
  {Szabados}, {Benk{\H{o}}}, {Barrena Delgado}, {Charcos-Llorens}, {Eredics},
  {Kiss}, {Manchado}, {R{\'a}cz}, {Ramos Almeida}, {Sz{\'e}kely}, \&
  {Vidal-N{\'u}{\~n}ez}}]{Acosta2007}
{Acosta-Pulido}, J.~A., {Kun}, M., {{\'A}brah{\'a}m}, P., {et~al.} 2007, \aj,
  133, 2020

\bibitem[{{Andrews} {et~al.}(2004){Andrews}, {Rothberg}, \&
  {Simon}}]{Andrews2004}
{Andrews}, S.~M., {Rothberg}, B., \& {Simon}, T. 2004, \apjl, 610, L45

\bibitem[{{Ashraf} {et~al.}(2024){Ashraf}, {Jose}, {Lee}, {Contreras Pe{\~n}a},
  {Herczeg}, {Liu}, {Johnstone}, \& {Lee}}]{Ashraf2024MNRAS.527.11651}
{Ashraf}, M., {Jose}, J., {Lee}, H.-G., {et~al.} 2024, \mnras, 527, 11651

\bibitem[{{Audard} {et~al.}(2014){Audard}, {{\'A}brah{\'a}m}, {Dunham},
  {Green}, {Grosso}, {Hamaguchi}, {Kastner}, {K{\'o}sp{\'a}l}, {Lodato},
  {Romanova}, {Skinner}, {Vorobyov}, \& {Zhu}}]{Audard_2014prpl.conf387A}
{Audard}, M., {{\'A}brah{\'a}m}, P., {Dunham}, M.~M., {et~al.} 2014, in
  Protostars and Planets VI, ed. H.~{Beuther}, R.~S. {Klessen}, C.~P.
  {Dullemond}, \& T.~{Henning}, 387

\bibitem[{{Baek} {et~al.}(2015){Baek}, {Pak}, {Green}, {Meschiari}, {Lee},
  {Jeon}, {Choi}, {Im}, {Sung}, \& {Park}}]{Baek2015}
{Baek}, G., {Pak}, S., {Green}, J.~D., {et~al.} 2015, \aj, 149, 11

\bibitem[{{Bailer-Jones} {et~al.}(2021){Bailer-Jones}, {Rybizki}, {Fouesneau},
  {Demleitner}, \& {Andrae}}]{Bailer2021}
{Bailer-Jones}, C.~A.~L., {Rybizki}, J., {Fouesneau}, M., {Demleitner}, M., \&
  {Andrae}, R. 2021, \aj, 161, 147

\bibitem[{{Bell} {et~al.}(1995){Bell}, {Lin}, {Hartmann}, \&
  {Kenyon}}]{Bell1995}
{Bell}, K.~R., {Lin}, D.~N.~C., {Hartmann}, L.~W., \& {Kenyon}, S.~J. 1995,
  \apj, 444, 376

\bibitem[{{Bellm} {et~al.}(2019){Bellm}, {Kulkarni}, {Graham}, {Dekany},
  {Smith}, {Riddle}, {Masci}, {Helou}, {Prince}, {Adams}, {Barbarino},
  {Barlow}, {Bauer}, {Beck}, {Belicki}, {Biswas}, {Blagorodnova}, {Bodewits},
  {Bolin}, {Brinnel}, {Brooke}, {Bue}, {Bulla}, {Burruss}, {Cenko}, {Chang},
  {Connolly}, {Coughlin}, {Cromer}, {Cunningham}, {De}, {Delacroix}, {Desai},
  {Duev}, {Eadie}, {Farnham}, {Feeney}, {Feindt}, {Flynn}, {Franckowiak},
  {Frederick}, {Fremling}, {Gal-Yam}, {Gezari}, {Giomi}, {Goldstein},
  {Golkhou}, {Goobar}, {Groom}, {Hacopians}, {Hale}, {Henning}, {Ho}, {Hover},
  {Howell}, {Hung}, {Huppenkothen}, {Imel}, {Ip}, {Ivezi{\'c}}, {Jackson},
  {Jones}, {Juric}, {Kasliwal}, {Kaspi}, {Kaye}, {Kelley}, {Kowalski},
  {Kramer}, {Kupfer}, {Landry}, {Laher}, {Lee}, {Lin}, {Lin}, {Lunnan},
  {Giomi}, {Mahabal}, {Mao}, {Miller}, {Monkewitz}, {Murphy}, {Ngeow},
  {Nordin}, {Nugent}, {Ofek}, {Patterson}, {Penprase}, {Porter}, {Rauch},
  {Rebbapragada}, {Reiley}, {Rigault}, {Rodriguez}, {van Roestel}, {Rusholme},
  {van Santen}, {Schulze}, {Shupe}, {Singer}, {Soumagnac}, {Stein}, {Surace},
  {Sollerman}, {Szkody}, {Taddia}, {Terek}, {Van Sistine}, {van Velzen},
  {Vestrand}, {Walters}, {Ward}, {Ye}, {Yu}, {Yan}, \& {Zolkower}}]{Bellm2019}
{Bellm}, E.~C., {Kulkarni}, S.~R., {Graham}, M.~J., {et~al.} 2019, \pasp, 131,
  018002

\bibitem[{{Bessell} \& {Brett}(1988)}]{Bessel1988}
{Bessell}, M.~S. \& {Brett}, J.~M. 1988, \pasp, 100, 1134

\bibitem[{{Cantat-Gaudin} {et~al.}(2020){Cantat-Gaudin}, {Anders},
  {Castro-Ginard}, {Jordi}, {Romero-G{\'o}mez}, {Soubiran}, {Casamiquela},
  {Tarricq}, {Moitinho}, {Vallenari}, {Bragaglia}, {Krone-Martins}, \&
  {Kounkel}}]{cantat2020}
{Cantat-Gaudin}, T., {Anders}, F., {Castro-Ginard}, A., {et~al.} 2020, \aap,
  640, A1

\bibitem[{{Cardelli} {et~al.}(1989){Cardelli}, {Clayton}, \&
  {Mathis}}]{Cardelli1989}
{Cardelli}, J.~A., {Clayton}, G.~C., \& {Mathis}, J.~S. 1989, \apj, 345, 245

\bibitem[{{Carvalho} {et~al.}(2024){Carvalho}, {Hillenbrand}, {Seebeck}, \&
  {Covey}}]{Carvalho2024}
{Carvalho}, A., {Hillenbrand}, L., {Seebeck}, J., \& {Covey}, K. 2024, \apj,
  971, 44

\bibitem[{{Chambers} {et~al.}(2016){Chambers}, {Magnier}, {Metcalfe},
  {Flewelling}, {Huber}, {Waters}, {Denneau}, {Draper}, {Farrow}, {Finkbeiner},
  {Holmberg}, {Koppenhoefer}, {Price}, {Rest}, {Saglia}, {Schlafly}, {Smartt},
  {Sweeney}, {Wainscoat}, {Burgett}, {Chastel}, {Grav}, {Heasley}, {Hodapp},
  {Jedicke}, {Kaiser}, {Kudritzki}, {Luppino}, {Lupton}, {Monet}, {Morgan},
  {Onaka}, {Shiao}, {Stubbs}, {Tonry}, {White}, {Ba{\~n}ados}, {Bell},
  {Bender}, {Bernard}, {Boegner}, {Boffi}, {Botticella}, {Calamida},
  {Casertano}, {Chen}, {Chen}, {Cole}, {Deacon}, {Frenk}, {Fitzsimmons},
  {Gezari}, {Gibbs}, {Goessl}, {Goggia}, {Gourgue}, {Goldman}, {Grant},
  {Grebel}, {Hambly}, {Hasinger}, {Heavens}, {Heckman}, {Henderson}, {Henning},
  {Holman}, {Hopp}, {Ip}, {Isani}, {Jackson}, {Keyes}, {Koekemoer}, {Kotak},
  {Le}, {Liska}, {Long}, {Lucey}, {Liu}, {Martin}, {Masci}, {McLean}, {Mindel},
  {Misra}, {Morganson}, {Murphy}, {Obaika}, {Narayan}, {Nieto-Santisteban},
  {Norberg}, {Peacock}, {Pier}, {Postman}, {Primak}, {Rae}, {Rai}, {Riess},
  {Riffeser}, {Rix}, {R{\"o}ser}, {Russel}, {Rutz}, {Schilbach}, {Schultz},
  {Scolnic}, {Strolger}, {Szalay}, {Seitz}, {Small}, {Smith}, {Soderblom},
  {Taylor}, {Thomson}, {Taylor}, {Thakar}, {Thiel}, {Thilker}, {Unger},
  {Urata}, {Valenti}, {Wagner}, {Walder}, {Walter}, {Watters}, {Werner},
  {Wood-Vasey}, \& {Wyse}}]{2016arXiv161205560C}
{Chambers}, K.~C., {Magnier}, E.~A., {Metcalfe}, N., {et~al.} 2016, arXiv
  e-prints, arXiv:1612.05560

\bibitem[{{Cieza} {et~al.}(2016){Cieza}, {Casassus}, {Tobin}, {Bos},
  {Williams}, {Perez}, {Zhu}, {Caceres}, {Canovas}, {Dunham}, {Hales},
  {Prieto}, {Principe}, {Schreiber}, {Ruiz-Rodriguez}, \& {Zurlo}}]{Cieza2016}
{Cieza}, L.~A., {Casassus}, S., {Tobin}, J., {et~al.} 2016, \nat, 535, 258

\bibitem[{{Connelley} \& {Reipurth}(2018)}]{Connelley2018}
{Connelley}, M.~S. \& {Reipurth}, B. 2018, \apj, 861, 145

\bibitem[{{Contreras Pe{\~n}a} {et~al.}(2023){Contreras Pe{\~n}a}, {Ashraf},
  {Lee}, {Herczeg}, {Lucas}, {Guo}, {Johnstone}, {Lee}, \&
  {Jose}}]{Contreras-Pena2023JKAS...56..253C}
{Contreras Pe{\~n}a}, C., {Ashraf}, M., {Lee}, J.-E., {et~al.} 2023, Journal of
  Korean Astronomical Society, 56, 253

\bibitem[{{Contreras Pe{\~n}a} {et~al.}(2024){Contreras Pe{\~n}a}, {Lucas},
  {Guo}, \& {Smith}}]{Contreras-Pena2024}
{Contreras Pe{\~n}a}, C., {Lucas}, P.~W., {Guo}, Z., \& {Smith}, L. 2024,
  \mnras, 528, 1823

\bibitem[{{Contreras Pe{\~n}a} {et~al.}(2017){Contreras Pe{\~n}a}, {Lucas},
  {Kurtev}, {Minniti}, {Caratti o Garatti}, {Marocco}, {Thompson}, {Froebrich},
  {Kumar}, {Stimson}, {Navarro Molina}, {Borissova}, {Gledhill}, \&
  {Terzi}}]{Contreras-Pena2017}
{Contreras Pe{\~n}a}, C., {Lucas}, P.~W., {Kurtev}, R., {et~al.} 2017, \mnras,
  465, 3039

\bibitem[{{Contreras Pe{\~n}a} {et~al.}(2019){Contreras Pe{\~n}a}, {Naylor}, \&
  {Morrell}}]{Contreras-Pena2019}
{Contreras Pe{\~n}a}, C., {Naylor}, T., \& {Morrell}, S. 2019, \mnras, 486,
  4590

\bibitem[{{Crause} {et~al.}(2014){Crause}, {Sharples}, {Bramall}, {Schmoll},
  {Clark}, {Younger}, {Tyas}, {Ryan}, {Brink}, {Strydom}, {Buckley},
  {Wilkinson}, {Crawford}, \& {Depagne}}]{Crause2014}
{Crause}, L.~A., {Sharples}, R.~M., {Bramall}, D.~G., {et~al.} 2014, in Society
  of Photo-Optical Instrumentation Engineers (SPIE) Conference Series, Vol.
  9147, Ground-based and Airborne Instrumentation for Astronomy V, ed. S.~K.
  {Ramsay}, I.~S. {McLean}, \& H.~{Takami}, 91476T

\bibitem[{{Cushing} {et~al.}(2004){Cushing}, {Vacca}, \&
  {Rayner}}]{Cushing2004}
{Cushing}, M.~C., {Vacca}, W.~D., \& {Rayner}, J.~T. 2004, \pasp, 116, 362

\bibitem[{{Cutri} {et~al.}(2003){Cutri}, {Skrutskie}, {van Dyk}, {Beichman},
  {Carpenter}, {Chester}, {Cambresy}, {Evans}, {Fowler}, {Gizis}, {Howard},
  {Huchra}, {Jarrett}, {Kopan}, {Kirkpatrick}, {Light}, {Marsh}, {McCallon},
  {Schneider}, {Stiening}, {Sykes}, {Weinberg}, {Wheaton}, {Wheelock}, \&
  {Zacarias}}]{Cutri2003}
{Cutri}, R.~M., {Skrutskie}, M.~F., {van Dyk}, S., {et~al.} 2003, VizieR Online
  Data Catalog, II/246

\bibitem[{{Drew} {et~al.}(2014){Drew}, {Gonzalez-Solares}, {Greimel}, {Irwin},
  {K{\"u}pc{\"u} Yoldas}, {Lewis}, {Barentsen}, {Eisl{\"o}ffel}, {Farnhill},
  {Martin}, {Walsh}, {Walton}, {Mohr-Smith}, {Raddi}, {Sale}, {Wright},
  {Groot}, {Barlow}, {Corradi}, {Drake}, {Fabregat}, {Frew}, {G{\"a}nsicke},
  {Knigge}, {Mampaso}, {Morris}, {Naylor}, {Parker}, {Phillipps}, {Ruhland},
  {Steeghs}, {Unruh}, {Vink}, {Wesson}, \& {Zijlstra}}]{Drew2014}
{Drew}, J.~E., {Gonzalez-Solares}, E., {Greimel}, R., {et~al.} 2014, \mnras,
  440, 2036

\bibitem[{{Dutra} {et~al.}(2003){Dutra}, {Bica}, {Soares}, \&
  {Barbuy}}]{DBS2003}
{Dutra}, C.~M., {Bica}, E., {Soares}, J., \& {Barbuy}, B. 2003, \aap, 400, 533

\bibitem[{{Fiorellino} {et~al.}(2024){Fiorellino}, {{\'A}brah{\'a}m},
  {K{\'o}sp{\'a}l}, {Kun}, {Alcal{\'a}}, {Caratti o Garatti}, {Cruz-S{\'a}enz
  de Miera}, {Garc{\'\i}a-{\'A}lvarez}, {Giannini}, {Park}, {Siwak},
  {Szil{\'a}gyi}, {Covino}, {Marton}, {Nagy}, {Nisini}, {Marianna Szab{\'o}},
  {Bora}, {Cseh}, {Kalup}, {Krezinger}, {Kriskovics}, {Og{\l}oza}, {P{\'a}l},
  {S{\'o}dor}, {Sonbas}, {Szak{\'a}ts}, {Vida}, {Vink{\'o}}, {Wyrzykowski}, \&
  {Zielinski}}]{Fiorellino2024}
{Fiorellino}, E., {{\'A}brah{\'a}m}, P., {K{\'o}sp{\'a}l}, {\'A}., {et~al.}
  2024, \aap, 686, A160

\bibitem[{{Fiorellino} {et~al.}(2023){Fiorellino}, {Tychoniec}, {Cruz-S{\'a}enz
  de Miera}, {Antoniucci}, {K{\'o}sp{\'a}l}, {Manara}, {Nisini}, \&
  {Rosotti}}]{Fiorellino2023}
{Fiorellino}, E., {Tychoniec}, {\L}., {Cruz-S{\'a}enz de Miera}, F., {et~al.}
  2023, \apj, 944, 135

\bibitem[{{Fischer} {et~al.}(2023){Fischer}, {Hillenbrand}, {Herczeg},
  {Johnstone}, {Kospal}, \& {Dunham}}]{fischer2023}
{Fischer}, W.~J., {Hillenbrand}, L.~A., {Herczeg}, G.~J., {et~al.} 2023, in
  Astronomical Society of the Pacific Conference Series, Vol. 534, Protostars
  and Planets VII, ed. S.~{Inutsuka}, Y.~{Aikawa}, T.~{Muto}, K.~{Tomida}, \&
  M.~{Tamura}, 355

\bibitem[{{Foster}(1996)}]{Foster1996}
{Foster}, G. 1996, \aj, 112, 1709

\bibitem[{{Gaia Collaboration} {et~al.}(2023){Gaia Collaboration}, {Vallenari},
  {Brown}, {Prusti}, {de Bruijne}, {Arenou}, {Babusiaux}, {Biermann},
  {Creevey}, {Ducourant}, {Evans}, {Eyer}, {Guerra}, {Hutton}, {Jordi},
  {Klioner}, {Lammers}, {Lindegren}, {Luri}, {Mignard}, {Panem}, {Pourbaix},
  {Randich}, {Sartoretti}, {Soubiran}, {Tanga}, {Walton}, {Bailer-Jones},
  {Bastian}, {Drimmel}, {Jansen}, {Katz}, {Lattanzi}, {van Leeuwen}, {Bakker},
  {Cacciari}, {Casta{\~n}eda}, {De Angeli}, {Fabricius}, {Fouesneau},
  {Fr{\'e}mat}, {Galluccio}, {Guerrier}, {Heiter}, {Masana}, {Messineo},
  {Mowlavi}, {Nicolas}, {Nienartowicz}, {Pailler}, {Panuzzo}, {Riclet}, {Roux},
  {Seabroke}, {Sordo}, {Th{\'e}venin}, {Gracia-Abril}, {Portell}, {Teyssier},
  {Altmann}, {Andrae}, {Audard}, {Bellas-Velidis}, {Benson}, {Berthier},
  {Blomme}, {Burgess}, {Busonero}, {Busso}, {C{\'a}novas}, {Carry}, {Cellino},
  {Cheek}, {Clementini}, {Damerdji}, {Davidson}, {de Teodoro}, {Nu{\~n}ez
  Campos}, {Delchambre}, {Dell'Oro}, {Esquej}, {Fern{\'a}ndez-Hern{\'a}ndez},
  {Fraile}, {Garabato}, {Garc{\'\i}a-Lario}, {Gosset}, {Haigron}, {Halbwachs},
  {Hambly}, {Harrison}, {Hern{\'a}ndez}, {Hestroffer}, {Hodgkin}, {Holl},
  {Jan{\ss}en}, {Jevardat de Fombelle}, {Jordan}, {Krone-Martins}, {Lanzafame},
  {L{\"o}ffler}, {Marchal}, {Marrese}, {Moitinho}, {Muinonen}, {Osborne},
  {Pancino}, {Pauwels}, {Recio-Blanco}, {Reyl{\'e}}, {Riello}, {Rimoldini},
  {Roegiers}, {Rybizki}, {Sarro}, {Siopis}, {Smith}, {Sozzetti}, {Utrilla},
  {van Leeuwen}, {Abbas}, {{\'A}brah{\'a}m}, {Abreu Aramburu}, {Aerts},
  {Aguado}, {Ajaj}, {Aldea-Montero}, {Altavilla}, {{\'A}lvarez}, {Alves},
  {Anders}, {Anderson}, {Anglada Varela}, {Antoja}, {Baines}, {Baker},
  {Balaguer-N{\'u}{\~n}ez}, {Balbinot}, {Balog}, {Barache}, {Barbato},
  {Barros}, {Barstow}, {Bartolom{\'e}}, {Bassilana}, {Bauchet}, {Becciani},
  {Bellazzini}, {Berihuete}, {Bernet}, {Bertone}, {Bianchi}, {Binnenfeld},
  {Blanco-Cuaresma}, {Blazere}, {Boch}, {Bombrun}, {Bossini}, {Bouquillon},
  {Bragaglia}, {Bramante}, {Breedt}, {Bressan}, {Brouillet}, {Brugaletta},
  {Bucciarelli}, {Burlacu}, {Butkevich}, {Buzzi}, {Caffau}, {Cancelliere},
  {Cantat-Gaudin}, {Carballo}, {Carlucci}, {Carnerero}, {Carrasco},
  {Casamiquela}, {Castellani}, {Castro-Ginard}, {Chaoul}, {Charlot}, {Chemin},
  {Chiaramida}, {Chiavassa}, {Chornay}, {Comoretto}, {Contursi}, {Cooper},
  {Cornez}, {Cowell}, {Crifo}, {Cropper}, {Crosta}, {Crowley}, {Dafonte},
  {Dapergolas}, {David}, {David}, {de Laverny}, {De Luise}, {De March}, {De
  Ridder}, {de Souza}, {de Torres}, {del Peloso}, {del Pozo}, {Delbo},
  {Delgado}, {Delisle}, {Demouchy}, {Dharmawardena}, {Di Matteo}, {Diakite},
  {Diener}, {Distefano}, {Dolding}, {Edvardsson}, {Enke}, {Fabre}, {Fabrizio},
  {Faigler}, {Fedorets}, {Fernique}, {Fienga}, {Figueras}, {Fournier},
  {Fouron}, {Fragkoudi}, {Gai}, {Garcia-Gutierrez}, {Garcia-Reinaldos},
  {Garc{\'\i}a-Torres}, {Garofalo}, {Gavel}, {Gavras}, {Gerlach}, {Geyer},
  {Giacobbe}, {Gilmore}, {Girona}, {Giuffrida}, {Gomel}, {Gomez},
  {Gonz{\'a}lez-N{\'u}{\~n}ez}, {Gonz{\'a}lez-Santamar{\'\i}a},
  {Gonz{\'a}lez-Vidal}, {Granvik}, {Guillout}, {Guiraud},
  {Guti{\'e}rrez-S{\'a}nchez}, {Guy}, {Hatzidimitriou}, {Hauser}, {Haywood},
  {Helmer}, {Helmi}, {Sarmiento}, {Hidalgo}, {Hilger}, {H{\l}adczuk}, {Hobbs},
  {Holland}, {Huckle}, {Jardine}, {Jasniewicz}, {Jean-Antoine Piccolo},
  {Jim{\'e}nez-Arranz}, {Jorissen}, {Juaristi Campillo}, {Julbe}, {Karbevska},
  {Kervella}, {Khanna}, {Kontizas}, {Kordopatis}, {Korn}, {K{\'o}sp{\'a}l},
  {Kostrzewa-Rutkowska}, {Kruszy{\'n}ska}, {Kun}, {Laizeau}, {Lambert},
  {Lanza}, {Lasne}, {Le Campion}, {Lebreton}, {Lebzelter}, {Leccia}, {Leclerc},
  {Lecoeur-Taibi}, {Liao}, {Licata}, {Lindstr{\o}m}, {Lister}, {Livanou},
  {Lobel}, {Lorca}, {Loup}, {Madrero Pardo}, {Magdaleno Romeo}, {Managau},
  {Mann}, {Manteiga}, {Marchant}, {Marconi}, {Marcos}, {Marcos Santos},
  {Mar{\'\i}n Pina}, {Marinoni}, {Marocco}, {Marshall}, {Martin Polo},
  {Mart{\'\i}n-Fleitas}, {Marton}, {Mary}, {Masip}, {Massari},
  {Mastrobuono-Battisti}, {Mazeh}, {McMillan}, {Messina}, {Michalik}, {Millar},
  {Mints}, {Molina}, {Molinaro}, {Moln{\'a}r}, {Monari}, {Mongui{\'o}},
  {Montegriffo}, {Montero}, {Mor}, {Mora}, {Morbidelli}, {Morel}, {Morris},
  {Muraveva}, {Murphy}, {Musella}, {Nagy}, {Noval}, {Oca{\~n}a}, {Ogden},
  {Ordenovic}, {Osinde}, {Pagani}, {Pagano}, {Palaversa}, {Palicio},
  {Pallas-Quintela}, {Panahi}, {Payne-Wardenaar}, {Pe{\~n}alosa Esteller},
  {Penttil{\"a}}, {Pichon}, {Piersimoni}, {Pineau}, {Plachy}, {Plum}, {Poggio},
  {Pr{\v{s}}a}, {Pulone}, {Racero}, {Ragaini}, {Rainer}, {Raiteri}, {Rambaux},
  {Ramos}, {Ramos-Lerate}, {Re Fiorentin}, {Regibo}, {Richards}, {Rios Diaz},
  {Ripepi}, {Riva}, {Rix}, {Rixon}, {Robichon}, {Robin}, {Robin}, {Roelens},
  {Rogues}, {Rohrbasser}, {Romero-G{\'o}mez}, {Rowell}, {Royer}, {Ruz Mieres},
  {Rybicki}, {Sadowski}, {S{\'a}ez N{\'u}{\~n}ez}, {Sagrist{\`a} Sell{\'e}s},
  {Sahlmann}, {Salguero}, {Samaras}, {Sanchez Gimenez}, {Sanna},
  {Santove{\~n}a}, {Sarasso}, {Schultheis}, {Sciacca}, {Segol}, {Segovia},
  {S{\'e}gransan}, {Semeux}, {Shahaf}, {Siddiqui}, {Siebert}, {Siltala},
  {Silvelo}, {Slezak}, {Slezak}, {Smart}, {Snaith}, {Solano}, {Solitro},
  {Souami}, {Souchay}, {Spagna}, {Spina}, {Spoto}, {Steele},
  {Steidelm{\"u}ller}, {Stephenson}, {S{\"u}veges}, {Surdej}, {Szabados},
  {Szegedi-Elek}, {Taris}, {Taylor}, {Teixeira}, {Tolomei}, {Tonello}, {Torra},
  {Torra}, {Torralba Elipe}, {Trabucchi}, {Tsounis}, {Turon}, {Ulla}, {Unger},
  {Vaillant}, {van Dillen}, {van Reeven}, {Vanel}, {Vecchiato}, {Viala},
  {Vicente}, {Voutsinas}, {Weiler}, {Wevers}, {Wyrzykowski}, {Yoldas}, {Yvard},
  {Zhao}, {Zorec}, {Zucker}, \& {Zwitter}}]{GaiaDR3}
{Gaia Collaboration}, {Vallenari}, A., {Brown}, A.~G.~A., {et~al.} 2023, \aap,
  674, A1

\bibitem[{{Gaia Collaboration \& Prusti} {et~al.}(2016){Gaia Collaboration \&
  Prusti}, {de Bruijne}, {Brown}, {Vallenari}, {Babusiaux}, {Bailer-Jones},
  {Bastian}, {Biermann}, {Evans}, {Eyer}, {Jansen}, {Jordi}, {Klioner},
  {Lammers}, {Lindegren}, {Luri}, {Mignard}, {Milligan}, {Panem}, {Poinsignon},
  {Pourbaix}, {Randich}, {Sarri}, {Sartoretti}, {Siddiqui}, {Soubiran},
  {Valette}, {van Leeuwen}, {Walton}, {Aerts}, {Arenou}, {Cropper}, {Drimmel},
  {H{\o}g}, {Katz}, {Lattanzi}, {O'Mullane}, {Grebel}, {Holland}, {Huc},
  {Passot}, {Bramante}, {Cacciari}, {Casta{\~n}eda}, {Chaoul}, {Cheek}, {De
  Angeli}, {Fabricius}, {Guerra}, {Hern{\'a}ndez}, {Jean-Antoine-Piccolo},
  {Masana}, {Messineo}, {Mowlavi}, {Nienartowicz}, {Ord{\'o}{\~n}ez-Blanco},
  {Panuzzo}, {Portell}, {Richards}, {Riello}, {Seabroke}, {Tanga},
  {Th{\'e}venin}, {Torra}, {Els}, {Gracia-Abril}, {Comoretto},
  {Garcia-Reinaldos}, {Lock}, {Mercier}, {Altmann}, {Andrae}, {Astraatmadja},
  {Bellas-Velidis}, {Benson}, {Berthier}, {Blomme}, {Busso}, {Carry},
  {Cellino}, {Clementini}, {Cowell}, {Creevey}, {Cuypers}, {Davidson}, {De
  Ridder}, {de Torres}, {Delchambre}, {Dell'Oro}, {Ducourant}, {Fr{\'e}mat},
  {Garc{\'\i}a-Torres}, {Gosset}, {Halbwachs}, {Hambly}, {Harrison}, {Hauser},
  {Hestroffer}, {Hodgkin}, {Huckle}, {Hutton}, {Jasniewicz}, {Jordan},
  {Kontizas}, {Korn}, {Lanzafame}, {Manteiga}, {Moitinho}, {Muinonen},
  {Osinde}, {Pancino}, {Pauwels}, {Petit}, {Recio-Blanco}, {Robin}, {Sarro},
  {Siopis}, {Smith}, {Smith}, {Sozzetti}, {Thuillot}, {van Reeven}, {Viala},
  {Abbas}, {Abreu Aramburu}, {Accart}, {Aguado}, {Allan}, {Allasia},
  {Altavilla}, {{\'A}lvarez}, {Alves}, {Anderson}, {Andrei}, {Anglada Varela},
  {Antiche}, {Antoja}, {Ant{\'o}n}, {Arcay}, {Atzei}, {Ayache}, {Bach},
  {Baker}, {Balaguer-N{\'u}{\~n}ez}, {Barache}, {Barata}, {Barbier}, {Barblan},
  {Baroni}, {Barrado y Navascu{\'e}s}, {Barros}, {Barstow}, {Becciani},
  {Bellazzini}, {Bellei}, {Bello Garc{\'\i}a}, {Belokurov}, {Bendjoya},
  {Berihuete}, {Bianchi}, {Bienaym{\'e}}, {Billebaud}, {Blagorodnova},
  {Blanco-Cuaresma}, {Boch}, {Bombrun}, {Borrachero}, {Bouquillon}, {Bourda},
  {Bouy}, {Bragaglia}, {Breddels}, {Brouillet}, {Br{\"u}semeister},
  {Bucciarelli}, {Budnik}, {Burgess}, {Burgon}, {Burlacu}, {Busonero}, {Buzzi},
  {Caffau}, {Cambras}, {Campbell}, {Cancelliere}, {Cantat-Gaudin}, {Carlucci},
  {Carrasco}, {Castellani}, {Charlot}, {Charnas}, {Charvet}, {Chassat},
  {Chiavassa}, {Clotet}, {Cocozza}, {Collins}, {Collins}, {Costigan}, {Crifo},
  {Cross}, {Crosta}, {Crowley}, {Dafonte}, {Damerdji}, {Dapergolas}, {David},
  {David}, {De Cat}, {de Felice}, {de Laverny}, {De Luise}, {De March}, {de
  Martino}, {de Souza}, {Debosscher}, {del Pozo}, {Delbo}, {Delgado},
  {Delgado}, {di Marco}, {Di Matteo}, {Diakite}, {Distefano}, {Dolding}, {Dos
  Anjos}, {Drazinos}, {Dur{\'a}n}, {Dzigan}, {Ecale}, {Edvardsson}, {Enke},
  {Erdmann}, {Escolar}, {Espina}, {Evans}, {Eynard Bontemps}, {Fabre},
  {Fabrizio}, {Faigler}, {Falc{\~a}o}, {Farr{\`a}s Casas}, {Faye}, {Federici},
  {Fedorets}, {Fern{\'a}ndez-Hern{\'a}ndez}, {Fernique}, {Fienga}, {Figueras},
  {Filippi}, {Findeisen}, {Fonti}, {Fouesneau}, {Fraile}, {Fraser}, {Fuchs},
  {Furnell}, {Gai}, {Galleti}, {Galluccio}, {Garabato}, {Garc{\'\i}a-Sedano},
  {Gar{\'e}}, {Garofalo}, {Garralda}, {Gavras}, {Gerssen}, {Geyer}, {Gilmore},
  {Girona}, {Giuffrida}, {Gomes}, {Gonz{\'a}lez-Marcos},
  {Gonz{\'a}lez-N{\'u}{\~n}ez}, {Gonz{\'a}lez-Vidal}, {Granvik}, {Guerrier},
  {Guillout}, {Guiraud}, {G{\'u}rpide}, {Guti{\'e}rrez-S{\'a}nchez}, {Guy},
  {Haigron}, {Hatzidimitriou}, {Haywood}, {Heiter}, {Helmi}, {Hobbs},
  {Hofmann}, {Holl}, {Holland}, {Hunt}, {Hypki}, {Icardi}, {Irwin}, {Jevardat
  de Fombelle}, {Jofr{\'e}}, {Jonker}, {Jorissen}, {Julbe}, {Karampelas},
  {Kochoska}, {Kohley}, {Kolenberg}, {Kontizas}, {Koposov}, {Kordopatis},
  {Koubsky}, {Kowalczyk}, {Krone-Martins}, {Kudryashova}, {Kull}, {Bachchan},
  {Lacoste-Seris}, {Lanza}, {Lavigne}, {Le Poncin-Lafitte}, {Lebreton},
  {Lebzelter}, {Leccia}, {Leclerc}, {Lecoeur-Taibi}, {Lemaitre}, {Lenhardt},
  {Leroux}, {Liao}, {Licata}, {Lindstr{\o}m}, {Lister}, {Livanou}, {Lobel},
  {L{\"o}ffler}, {L{\'o}pez}, {Lopez-Lozano}, {Lorenz}, {Loureiro},
  {MacDonald}, {Magalh{\~a}es Fernandes}, {Managau}, {Mann}, {Mantelet},
  {Marchal}, {Marchant}, {Marconi}, {Marie}, {Marinoni}, {Marrese},
  {Marschalk{\'o}}, {Marshall}, {Mart{\'\i}n-Fleitas}, {Martino}, {Mary},
  {Matijevi{\v{c}}}, {Mazeh}, {McMillan}, {Messina}, {Mestre}, {Michalik},
  {Millar}, {Miranda}, {Molina}, {Molinaro}, {Molinaro}, {Moln{\'a}r},
  {Moniez}, {Montegriffo}, {Monteiro}, {Mor}, {Mora}, {Morbidelli}, {Morel},
  {Morgenthaler}, {Morley}, {Morris}, {Mulone}, {Muraveva}, {Musella},
  {Narbonne}, {Nelemans}, {Nicastro}, {Noval}, {Ord{\'e}novic},
  {Ordieres-Mer{\'e}}, {Osborne}, {Pagani}, {Pagano}, {Pailler}, {Palacin},
  {Palaversa}, {Parsons}, {Paulsen}, {Pecoraro}, {Pedrosa}, {Pentik{\"a}inen},
  {Pereira}, {Pichon}, {Piersimoni}, {Pineau}, {Plachy}, {Plum}, {Poujoulet},
  {Pr{\v{s}}a}, {Pulone}, {Ragaini}, {Rago}, {Rambaux}, {Ramos-Lerate},
  {Ranalli}, {Rauw}, {Read}, {Regibo}, {Renk}, {Reyl{\'e}}, {Ribeiro},
  {Rimoldini}, {Ripepi}, {Riva}, {Rixon}, {Roelens}, {Romero-G{\'o}mez},
  {Rowell}, {Royer}, {Rudolph}, {Ruiz-Dern}, {Sadowski}, {Sagrist{\`a}
  Sell{\'e}s}, {Sahlmann}, {Salgado}, {Salguero}, {Sarasso}, {Savietto},
  {Schnorhk}, {Schultheis}, {Sciacca}, {Segol}, {Segovia}, {Segransan},
  {Serpell}, {Shih}, {Smareglia}, {Smart}, {Smith}, {Solano}, {Solitro},
  {Sordo}, {Soria Nieto}, {Souchay}, {Spagna}, {Spoto}, {Stampa}, {Steele},
  {Steidelm{\"u}ller}, {Stephenson}, {Stoev}, {Suess}, {S{\"u}veges}, {Surdej},
  {Szabados}, {Szegedi-Elek}, {Tapiador}, {Taris}, {Tauran}, {Taylor},
  {Teixeira}, {Terrett}, {Tingley}, {Trager}, {Turon}, {Ulla}, {Utrilla},
  {Valentini}, {van Elteren}, {Van Hemelryck}, {van Leeuwen}, {Varadi},
  {Vecchiato}, {Veljanoski}, {Via}, {Vicente}, {Vogt}, {Voss}, {Votruba},
  {Voutsinas}, {Walmsley}, {Weiler}, {Weingrill}, {Werner}, {Wevers},
  {Whitehead}, {Wyrzykowski}, {Yoldas}, {{\v{Z}}erjal}, {Zucker}, {Zurbach},
  {Zwitter}, {Alecu}, {Allen}, {Allende Prieto}, {Amorim},
  {Anglada-Escud{\'e}}, {Arsenijevic}, {Azaz}, {Balm}, {Beck}, {Bernstein},
  {Bigot}, {Bijaoui}, {Blasco}, {Bonfigli}, {Bono}, {Boudreault}, {Bressan},
  {Brown}, {Brunet}, {Bunclark}, {Buonanno}, {Butkevich}, {Carret}, {Carrion},
  {Chemin}, {Ch{\'e}reau}, {Corcione}, {Darmigny}, {de Boer}, {de Teodoro}, {de
  Zeeuw}, {Delle Luche}, {Domingues}, {Dubath}, {Fodor}, {Fr{\'e}zouls},
  {Fries}, {Fustes}, {Fyfe}, {Gallardo}, {Gallegos}, {Gardiol}, {Gebran},
  {Gomboc}, {G{\'o}mez}, {Grux}, {Gueguen}, {Heyrovsky}, {Hoar}, {Iannicola},
  {Isasi Parache}, {Janotto}, {Joliet}, {Jonckheere}, {Keil}, {Kim},
  {Klagyivik}, {Klar}, {Knude}, {Kochukhov}, {Kolka}, {Kos}, {Kutka}, {Lainey},
  {LeBouquin}, {Liu}, {Loreggia}, {Makarov}, {Marseille}, {Martayan},
  {Martinez-Rubi}, {Massart}, {Meynadier}, {Mignot}, {Munari}, {Nguyen},
  {Nordlander}, {Ocvirk}, {O'Flaherty}, {Olias Sanz}, {Ortiz}, {Osorio},
  {Oszkiewicz}, {Ouzounis}, {Palmer}, {Park}, {Pasquato}, {Peltzer}, {Peralta},
  {P{\'e}turaud}, {Pieniluoma}, {Pigozzi}, {Poels}, {Prat}, {Prod'homme},
  {Raison}, {Rebordao}, {Risquez}, {Rocca-Volmerange}, {Rosen}, {Ruiz-Fuertes},
  {Russo}, {Sembay}, {Serraller Vizcaino}, {Short}, {Siebert}, {Silva},
  {Sinachopoulos}, {Slezak}, {Soffel}, {Sosnowska}, {Strai{\v{z}}ys}, {ter
  Linden}, {Terrell}, {Theil}, {Tiede}, {Troisi}, {Tsalmantza}, {Tur},
  {Vaccari}, {Vachier}, {Valles}, {Van Hamme}, {Veltz}, {Virtanen}, {Wallut},
  {Wichmann}, {Wilkinson}, {Ziaeepour}, \& {Zschocke}}]{Gaia2016}
{Gaia Collaboration \& Prusti}, T., {de Bruijne}, J.~H.~J., {Brown}, A.~G.~A.,
  {et~al.} 2016, \aap, 595, A1

\bibitem[{{Green} {et~al.}(2019){Green}, {Schlafly}, {Zucker}, {Speagle}, \&
  {Finkbeiner}}]{Green2019ApJ...887...93G}
{Green}, G.~M., {Schlafly}, E., {Zucker}, C., {Speagle}, J.~S., \&
  {Finkbeiner}, D. 2019, \apj, 887, 93

\bibitem[{{Green} {et~al.}(2013){Green}, {Robertson}, {Baek}, {Pooley}, {Pak},
  {Im}, {Lee}, {Jeon}, {Choi}, \& {Meschiari}}]{Green2013}
{Green}, J.~D., {Robertson}, P., {Baek}, G., {et~al.} 2013, \apj, 764
  [\eprint[arXiv]{1212.2610}]

\bibitem[{{Guo} {et~al.}(2020){Guo}, {Lucas}, {Contreras Pe{\~n}a}, {Kurtev},
  {Smith}, {Borissova}, {Alonso-Garc{\'\i}a}, {Minniti}, {Caratti o Garatti},
  \& {Froebrich}}]{Guo2020}
{Guo}, Z., {Lucas}, P.~W., {Contreras Pe{\~n}a}, C., {et~al.} 2020, \mnras,
  492, 294

\bibitem[{{Guo} {et~al.}(2021){Guo}, {Lucas}, {Contreras Pena}, {Smith},
  {Kurtev}, {Borissova}, {Alonso-García}, {Minniti}, {Chené}, {Kumar},
  {Caratti o Garatti}, \& {Froebrich}}]{Guo2021}
{Guo}, Z., {Lucas}, P.~W., {Contreras Pena}, C., {et~al.} 2021, \mnras, 504,
  830–856

\bibitem[{{Guo} {et~al.}(2024{\natexlab{a}}){Guo}, {Lucas}, {Kurtev},
  {Borissova}, {Contreras Pe{\~n}a}, {Yurchenko}, {Smith}, {Minniti}, {Saito},
  {Bayo}, {Catelan}, {Alonso-Garc{\'\i}a}, {Caratti o Garatti}, {Morris},
  {Froebrich}, {Tennyson}, {Mauc{\'o}}, {Aguayo}, {Miller}, \&
  {Muthu}}]{Guo2024MNRAS}
{Guo}, Z., {Lucas}, P.~W., {Kurtev}, R., {et~al.} 2024{\natexlab{a}}, \mnras,
  528, 1769

\bibitem[{{Guo} {et~al.}(2024{\natexlab{b}}){Guo}, {Lucas}, {Kurtev},
  {Borissova}, {Elbakyan}, {Morris}, {Bayo}, {Smith}, {Caratti o Garatti},
  {Contreras Pe{\~n}a}, {Minniti}, {Jose}, {Ashraf}, {Alonso-Garc{\'\i}a},
  {Miller}, \& {Muthu}}]{Guo2024MNRAS-L}
{Guo}, Z., {Lucas}, P.~W., {Kurtev}, R.~G., {et~al.} 2024{\natexlab{b}},
  \mnras, 529, L115

\bibitem[{{Hanaoka} {et~al.}(2019){Hanaoka}, {Kaneda}, {Suzuki}, {Kokusho},
  {Oyabu}, {Ishihara}, {Kohno}, {Furuta}, {Tsuchikawa}, \&
  {Saito}}]{Hanaoka2019}
{Hanaoka}, M., {Kaneda}, H., {Suzuki}, T., {et~al.} 2019, \pasj, 71, 6

\bibitem[{{Hartmann} \& {Kenyon}(1996)}]{Hartmann1996}
{Hartmann}, L. \& {Kenyon}, S.~J. 1996, \araa, 34, 207

\bibitem[{{Heinze} {et~al.}(2018){Heinze}, {Tonry}, {Denneau}, {Flewelling},
  {Stalder}, {Rest}, {Smith}, {Smartt}, \& {Weiland}}]{Heinze2018}
{Heinze}, A.~N., {Tonry}, J.~L., {Denneau}, L., {et~al.} 2018, \aj, 156, 241

\bibitem[{{Henden} {et~al.}(2015){Henden}, {Levine}, {Terrell}, \&
  {Welch}}]{Henden2015}
{Henden}, A.~A., {Levine}, S., {Terrell}, D., \& {Welch}, D.~L. 2015, in
  American Astronomical Society Meeting Abstracts, Vol. 225, American
  Astronomical Society Meeting Abstracts \#225, 336.16

\bibitem[{{Henden} {et~al.}(2016){Henden}, {Templeton}, {Terrell}, {Smith},
  {Levine}, \& {Welch}}]{Henden2016}
{Henden}, A.~A., {Templeton}, M., {Terrell}, D., {et~al.} 2016, VizieR Online
  Data Catalog, II/336

\bibitem[{{Herbig}(1977)}]{Herbig1977}
{Herbig}, G.~H. 1977, \apj, 217, 693

\bibitem[{{Herbig}(1989)}]{Herbig1989}
{Herbig}, G.~H. 1989, in European Southern Observatory Conference and Workshop
  Proceedings, Vol.~33, European Southern Observatory Conference and Workshop
  Proceedings, 233--246

\bibitem[{{Herbig} {et~al.}(2003){Herbig}, {Petrov}, \&
  {Duemmler}}]{Herbig2003}
{Herbig}, G.~H., {Petrov}, P.~P., \& {Duemmler}, R. 2003, \apj, 595, 384

\bibitem[{{Herczeg} {et~al.}(2016){Herczeg}, {Dong}, {Shappee}, {Chen},
  {Hillenbrand}, {Jose}, {Kochanek}, {Prieto}, {Stanek}, {Kaplan}, {Holoien},
  {Mairs}, {Johnstone}, {Gully-Santiago}, {Zhu}, {Smith}, {Bersier}, {Mulders},
  {Filippenko}, {Ayani}, {Brimacombe}, {Brown}, {Connelley}, {Harmanen},
  {Itoh}, {Kawabata}, {Maehara}, {Takata}, {Yuk}, \& {Zheng}}]{Herczeg2016}
{Herczeg}, G.~J., {Dong}, S., {Shappee}, B.~J., {et~al.} 2016, \apj, 831, 133

\bibitem[{{Hillenbrand} {et~al.}(2023){Hillenbrand}, {Carvalho}, {van Roestel},
  \& {De}}]{Hillenbrand2023}
{Hillenbrand}, L.~A., {Carvalho}, A., {van Roestel}, J., \& {De}, K. 2023,
  \apjl, 958, L27

\bibitem[{{Hillenbrand} {et~al.}(2018){Hillenbrand}, {Contreras Pe{\~n}a},
  {Morrell}, {Naylor}, {Kuhn}, {Cutri}, {Rebull}, {Hodgkin}, {Froebrich}, \&
  {Mainzer}}]{Hillenbrand2018_ApJ869146H}
{Hillenbrand}, L.~A., {Contreras Pe{\~n}a}, C., {Morrell}, S., {et~al.} 2018,
  \apj, 869, 146

\bibitem[{{Hillenbrand} {et~al.}(2021){Hillenbrand}, {De}, {Hankins},
  {Kasliwal}, {Rebull}, {Lau}, {Cutri}, {Ashley}, {Karambelkar}, {Moore},
  {Travouillon}, \& {Mainzer}}]{Hillenbrand2021}
{Hillenbrand}, L.~A., {De}, K., {Hankins}, M., {et~al.} 2021, \aj, 161, 220

\bibitem[{{Hillenbrand} {et~al.}(2019{\natexlab{a}}){Hillenbrand}, {Miller},
  {Carpenter}, {Kasliwal}, {Isaacson}, {Tang}, {Joshi}, {Banerjee}, \&
  {Cutri}}]{Hillenbrand2019a}
{Hillenbrand}, L.~A., {Miller}, A., {Carpenter}, J.~M., {et~al.}
  2019{\natexlab{a}}, \apj, 874, 82

\bibitem[{{Hillenbrand} {et~al.}(2019{\natexlab{b}}){Hillenbrand}, {Reipurth},
  {Connelley}, {Cutri}, \& {Isaacson}}]{Hillenbrand2019b}
{Hillenbrand}, L.~A., {Reipurth}, B., {Connelley}, M., {Cutri}, R.~M., \&
  {Isaacson}, H. 2019{\natexlab{b}}, \aj, 158, 240

\bibitem[{{Hodapp} {et~al.}(2020){Hodapp}, {Denneau}, {Tucker}, {Shappee},
  {Huber}, {Payne}, {Do}, {Lin}, {Connelley}, {Varricatt}, {Tonry}, {Chambers},
  \& {Magnier}}]{Hodapp2020}
{Hodapp}, K.~W., {Denneau}, L., {Tucker}, M., {et~al.} 2020, \aj, 160, 164

\bibitem[{{Hodgkin} {et~al.}(2021){Hodgkin}, {Harrison}, {Breedt}, {Wevers},
  {Rixon}, {Delgado}, {Yoldas}, {Kostrzewa-Rutkowska}, {Wyrzykowski}, {van
  Leeuwen}, {Blagorodnova}, {Campbell}, {Eappachen}, {Fraser}, {Ihanec},
  {Koposov}, {Kruszy{\'n}ska}, {Marton}, {Rybicki}, {Brown}, {Burgess},
  {Busso}, {Cowell}, {De Angeli}, {Diener}, {Evans}, {Gilmore}, {Holland},
  {Jonker}, {van Leeuwen}, {Mignard}, {Osborne}, {Portell}, {Prusti},
  {Richards}, {Riello}, {Seabroke}, {Walton}, {{\'A}brah{\'a}m}, {Altavilla},
  {Baker}, {Bastian}, {O'Brien}, {de Bruijne}, {Butterley}, {Carrasco},
  {Casta{\~n}eda}, {Clark}, {Clementini}, {Copperwheat}, {Cropper},
  {Damljanovic}, {Davidson}, {Davis}, {Dennefeld}, {Dhillon}, {Dolding},
  {Dominik}, {Esquej}, {Eyer}, {Fabricius}, {Fridman}, {Froebrich}, {Garralda},
  {Gomboc}, {Gonz{\'a}lez-Vidal}, {Guerra}, {Hambly}, {Hardy}, {Holl},
  {Hourihane}, {Japelj}, {Kann}, {Kiss}, {Knigge}, {Kolb}, {Komossa},
  {K{\'o}sp{\'a}l}, {Kov{\'a}cs}, {Kun}, {Leto}, {Lewis}, {Littlefair},
  {Mahabal}, {Mundell}, {Nagy}, {Padeletti}, {Palaversa}, {Pigulski},
  {Pretorius}, {van Reeven}, {Ribeiro}, {Roelens}, {Rowell}, {Schartel},
  {Scholz}, {Schwope}, {Sip{\H{o}}cz}, {Smartt}, {Smith}, {Serraller},
  {Steeghs}, {Sullivan}, {Szabados}, {Szegedi-Elek}, {Tisserand}, {Tomasella},
  {van Velzen}, {Whitelock}, {Wilson}, \& {Young}}]{Hodgkin2021}
{Hodgkin}, S.~T., {Harrison}, D.~L., {Breedt}, E., {et~al.} 2021, \aap, 652,
  A76

\bibitem[{{Holoien} {et~al.}(2014){Holoien}, {Prieto}, {Stanek}, {Kochanek},
  {Shappee}, {Zhu}, {Sicilia-Aguilar}, {Grupe}, {Croxall}, {Adams}, {Simon},
  {Morrell}, {McGraw}, {Wagner}, {Basu}, {Beacom}, {Bersier}, {Brimacombe},
  {Jencson}, {Pojmanski}, {Starrfield}, {Szczygie{\l}}, \&
  {Woodward}}]{Holoien2014}
{Holoien}, T. W.~S., {Prieto}, J.~L., {Stanek}, K.~Z., {et~al.} 2014, \apjl,
  785, L35

\bibitem[{{Hou} \& {Han}(2014)}]{Hou2014}
{Hou}, L.~G. \& {Han}, J.~L. 2014, \aap, 569, A125

\bibitem[{{Hubbard}(2017{\natexlab{a}})}]{Hubbard2017}
{Hubbard}, A. 2017{\natexlab{a}}, \apj, 840, 6

\bibitem[{{Hubbard}(2017{\natexlab{b}})}]{Hubbard2017ApJ...840L...5H}
{Hubbard}, A. 2017{\natexlab{b}}, \apjl, 840, L5

\bibitem[{{Hunt} \& {Reffert}(2023)}]{Hunt2023}
{Hunt}, E.~L. \& {Reffert}, S. 2023, \aap, 673, A114

\bibitem[{{Kenyon} {et~al.}(2000){Kenyon}, {Kolotilov}, {Ibragimov}, \&
  {Mattei}}]{Kenyon2000}
{Kenyon}, S.~J., {Kolotilov}, E.~A., {Ibragimov}, M.~A., \& {Mattei}, J.~A.
  2000, \apj, 531, 1028

\bibitem[{{Kharchenko} {et~al.}(2013){Kharchenko}, {Piskunov}, {Schilbach},
  {R{\"o}ser}, \& {Scholz}}]{Kharchenko2013}
{Kharchenko}, N.~V., {Piskunov}, A.~E., {Schilbach}, E., {R{\"o}ser}, S., \&
  {Scholz}, R.~D. 2013, \aap, 558, A53

\bibitem[{{Kniazev} {et~al.}(2016){Kniazev}, {Gvaramadze}, \&
  {Berdnikov}}]{Kniazev2016}
{Kniazev}, A.~Y., {Gvaramadze}, V.~V., \& {Berdnikov}, L.~N. 2016, \mnras, 459,
  3068

\bibitem[{{Kniazev} {et~al.}(2017){Kniazev}, {Gvaramadze}, \&
  {Berdnikov}}]{Kniazev2017}
{Kniazev}, A.~Y., {Gvaramadze}, V.~V., \& {Berdnikov}, L.~N. 2017, in
  Astronomical Society of the Pacific Conference Series, Vol. 510, Stars: From
  Collapse to Collapse, ed. Y.~Y. {Balega}, D.~O. {Kudryavtsev}, I.~I.
  {Romanyuk}, \& I.~A. {Yakunin}, 480

\bibitem[{{K{\'o}sp{\'a}l} {et~al.}(2016){K{\'o}sp{\'a}l}, {{\'A}brah{\'a}m},
  {Acosta-Pulido}, {Dunham}, {Garc{\'\i}a-{\'A}lvarez}, {Hogerheijde}, {Kun},
  {Mo{\'o}r}, {Farkas}, {Hajdu}, {Hodos{\'a}n}, {Kov{\'a}cs}, {Kriskovics},
  {Marton}, {Moln{\'a}r}, {P{\'a}l}, {S{\'a}rneczky}, {S{\'o}dor},
  {Szak{\'a}ts}, {Szalai}, {Szegedi-Elek}, {Szing}, {T{\'o}th}, {Vida}, \&
  {Vink{\'o}}}]{Kospal2016}
{K{\'o}sp{\'a}l}, {\'A}., {{\'A}brah{\'a}m}, P., {Acosta-Pulido}, J.~A.,
  {et~al.} 2016, \aap, 596, A52

\bibitem[{{K{\'o}sp{\'a}l} {et~al.}(2020){K{\'o}sp{\'a}l}, {{\'A}brah{\'a}m},
  {Carmona}, {Chen}, {Green}, {van Boekel}, \& {White}}]{Kospal2020}
{K{\'o}sp{\'a}l}, {\'A}., {{\'A}brah{\'a}m}, P., {Carmona}, A., {et~al.} 2020,
  \apjl, 895, L48

\bibitem[{{K{\'o}sp{\'a}l} {et~al.}(2023){K{\'o}sp{\'a}l}, {{\'A}brah{\'a}m},
  {Diehl}, {Banzatti}, {Bouwman}, {Chen}, {Cruz-S{\'a}enz de Miera}, {Green},
  {Henning}, \& {Rab}}]{Kospal2023}
{K{\'o}sp{\'a}l}, {\'A}., {{\'A}brah{\'a}m}, P., {Diehl}, L., {et~al.} 2023,
  \apjl, 945, L7

\bibitem[{{Kuhn} {et~al.}(2021){Kuhn}, {de Souza}, {Krone-Martins},
  {Castro-Ginard}, {Ishida}, {Povich}, {Hillenbrand}, \& {COIN
  Collaboration}}]{Kuhn2021}
{Kuhn}, M.~A., {de Souza}, R.~S., {Krone-Martins}, A., {et~al.} 2021, \apjs,
  254, 33

\bibitem[{{Kuhn} {et~al.}(2024){Kuhn}, {Hillenbrand}, {Connelley}, {Rich},
  {Staels}, {Carvalho}, {Lucas}, {Fremling}, {Karambelkar}, {Lee}, {Ahumada},
  {Ishida}, {De}, {de Souza}, \& {Kasliwal}}]{Kuhn2024}
{Kuhn}, M.~A., {Hillenbrand}, L.~A., {Connelley}, M.~S., {et~al.} 2024, \mnras
  [\eprint[arXiv]{2401.09522}]

\bibitem[{{Kun} {et~al.}(2019){Kun}, {{\'A}brah{\'a}m}, {Acosta Pulido},
  {Mo{\'o}r}, \& {Prusti}}]{Kun2019MNRAS.483.4424K}
{Kun}, M., {{\'A}brah{\'a}m}, P., {Acosta Pulido}, J.~A., {Mo{\'o}r}, A., \&
  {Prusti}, T. 2019, \mnras, 483, 4424

\bibitem[{{Liu} {et~al.}(2022){Liu}, {Herczeg}, {Johnstone},
  {Contreras-Pe{\~n}a}, {Lee}, {Yang}, {Zhou}, {Yoon}, {Lee}, {Kunitomo}, \&
  {Jose}}]{Liu2022}
{Liu}, H., {Herczeg}, G.~J., {Johnstone}, D., {et~al.} 2022, \apj, 936, 152

\bibitem[{{Lykou} {et~al.}(2022){Lykou}, {{\'A}brah{\'a}m}, {Chen}, {Varga},
  {K{\'o}sp{\'a}l}, {Matter}, {Siwak}, {Szab{\'o}}, {Zhu}, {Liu}, {Lopez},
  {Allouche}, {Augereau}, {Berio}, {Cruzal{\`e}bes}, {Dominik}, {Henning},
  {Hofmann}, {Hogerheijde}, {Jaffe}, {Kokoulina}, {Lagarde}, {Meilland},
  {Millour}, {Pantin}, {Petrov}, {Robbe-Dubois}, {Schertl}, {Scheuck}, {van
  Boekel}, {Waters}, {Weigelt}, \& {Wolf}}]{Lykou2022}
{Lykou}, F., {{\'A}brah{\'a}m}, P., {Chen}, L., {et~al.} 2022, arXiv e-prints,
  arXiv:2205.10173

\bibitem[{{Magakian} {et~al.}(2023){Magakian}, {Movsessian}, {Andreasyan},
  {Moiseev}, \& {Uklein}}]{Magakian2023A&A...675A..79M}
{Magakian}, T.~Y., {Movsessian}, T.~A., {Andreasyan}, H.~R., {Moiseev}, A.~V.,
  \& {Uklein}, R.~I. 2023, \aap, 675, A79

\bibitem[{{Magnier} {et~al.}(2020){Magnier}, {Chambers}, {Flewelling},
  {Hoblitt}, {Huber}, {Price}, {Sweeney}, {Waters}, {Denneau}, {Draper},
  {Hodapp}, {Jedicke}, {Kaiser}, {Kudritzki}, {Metcalfe}, {Stubbs}, \&
  {Wainscoat}}]{2020ApJS..251....3M}
{Magnier}, E.~A., {Chambers}, K.~C., {Flewelling}, H.~A., {et~al.} 2020, \apjs,
  251, 3

\bibitem[{{Mainzer} {et~al.}(2011){Mainzer}, {Bauer}, {Grav}, {Masiero},
  {Cutri}, {Dailey}, {Eisenhardt}, {McMillan}, {Wright}, {Walker}, {Jedicke},
  {Spahr}, {Tholen}, {Alles}, {Beck}, {Brandenburg}, {Conrow}, {Evans},
  {Fowler}, {Jarrett}, {Marsh}, {Masci}, {McCallon}, {Wheelock}, {Wittman},
  {Wyatt}, {DeBaun}, {Elliott}, {Elsbury}, {Gautier}, {Gomillion}, {Leisawitz},
  {Maleszewski}, {Micheli}, \& {Wilkins}}]{Mainzer2011}
{Mainzer}, A., {Bauer}, J., {Grav}, T., {et~al.} 2011, \apj, 731, 53

\bibitem[{{Marton} {et~al.}(2016){Marton}, {T{\'o}th}, {Paladini}, {Kun},
  {Zahorecz}, {McGehee}, \& {Kiss}}]{Marton2016}
{Marton}, G., {T{\'o}th}, L.~V., {Paladini}, R., {et~al.} 2016, \mnras, 458,
  3479

\bibitem[{{Masci} {et~al.}(2019){Masci}, {Laher}, {Rusholme}, {Shupe}, {Groom},
  {Surace}, {Jackson}, {Monkewitz}, {Beck}, {Flynn}, {Terek}, {Landry},
  {Hacopians}, {Desai}, {Howell}, {Brooke}, {Imel}, {Wachter}, {Ye}, {Lin},
  {Cenko}, {Cunningham}, {Rebbapragada}, {Bue}, {Miller}, {Mahabal}, {Bellm},
  {Patterson}, {Juri{\'c}}, {Golkhou}, {Ofek}, {Walters}, {Graham}, {Kasliwal},
  {Dekany}, {Kupfer}, {Burdge}, {Cannella}, {Barlow}, {Van Sistine}, {Giomi},
  {Fremling}, {Blagorodnova}, {Levitan}, {Riddle}, {Smith}, {Helou}, {Prince},
  \& {Kulkarni}}]{masci2019}
{Masci}, F.~J., {Laher}, R.~R., {Rusholme}, B., {et~al.} 2019, \pasp, 131,
  018003

\bibitem[{{Meyer} {et~al.}(1997){Meyer}, {Calvet}, \&
  {Hillenbrand}}]{Meyer1997}
{Meyer}, M.~R., {Calvet}, N., \& {Hillenbrand}, L.~A. 1997, \aj, 114, 288

\bibitem[{{Molyarova} {et~al.}(2018){Molyarova}, {Akimkin}, {Semenov},
  {\'Abrah\'am}, {Henning}, {K\'osp\'al}, {Vorobyov}, \&
  {Wiebe}}]{Molyarova2018}
{Molyarova}, T., {Akimkin}, V., {Semenov}, D., {et~al.} 2018, \apj, 866, 46

\bibitem[{{Moorwood} {et~al.}(1998){Moorwood}, {Cuby}, \&
  {Lidman}}]{Sofi1998Msngr..91....9M}
{Moorwood}, A., {Cuby}, J.~G., \& {Lidman}, C. 1998, The Messenger, 91, 9

\bibitem[{{Muzerolle} {et~al.}(2005){Muzerolle}, {Megeath}, {Flaherty},
  {Gordon}, {Rieke}, {Young}, \& {Lada}}]{Muzerolle2015}
{Muzerolle}, J., {Megeath}, S.~T., {Flaherty}, K.~M., {et~al.} 2005, \apjl,
  620, L107

\bibitem[{{Nagy} {et~al.}(2023){Nagy}, {Park}, {{\'A}brah{\'a}m},
  {K{\'o}sp{\'a}l}, {Cruz-S{\'a}enz de Miera}, {Kun}, {Siwak}, {Szab{\'o}},
  {Szil{\'a}gyi}, {Fiorellino}, {Giannini}, {Lee}, {Lee}, {Marton}, {Szabados},
  {Vitali}, {Andrzejewski}, {Gromadzki}, {Hodgkin}, {Jab{\l}o{\'n}ska},
  {Mendez}, {Merc}, {Michniewicz}, {Miko{\l}ajczyk}, {Pylypenko}, {Ratajczak},
  {Wyrzykowski}, {Zejmo}, \& {Zieli{\'n}ski}}]{Nagy2023}
{Nagy}, Z., {Park}, S., {{\'A}brah{\'a}m}, P., {et~al.} 2023, \mnras, 524, 3344

\bibitem[{{Ninan} {et~al.}(2015){Ninan}, {Ojha}, {Baug}, {Bhatt}, {Mohan},
  {Ghosh}, {Men'shchikov}, {Anupama}, {Tamura}, \& {Henning}}]{Ninan2015}
{Ninan}, J.~P., {Ojha}, D.~K., {Baug}, T., {et~al.} 2015, \apj, 815, 4

\bibitem[{{Onken} {et~al.}(2019){Onken}, {Wolf}, {Bessell}, {Chang}, {Da
  Costa}, {Luvaul}, {Mackey}, {Schmidt}, \& {Shao}}]{2019PASA...36...33O}
{Onken}, C.~A., {Wolf}, C., {Bessell}, M.~S., {et~al.} 2019, \pasa, 36, e033

\bibitem[{{Onken} {et~al.}(2024){Onken}, {Wolf}, {Bessell}, {Chang}, {Luvaul},
  {Tonry}, {White}, \& {Da Costa}}]{Onken2024}
{Onken}, C.~A., {Wolf}, C., {Bessell}, M.~S., {et~al.} 2024, arXiv e-prints,
  arXiv:2402.02015

\bibitem[{{Pandey} {et~al.}(2022){Pandey}, {Sharma}, {Dewangan}, {Ojha},
  {Panwar}, {Das}, {Bisen}, {Ghosh}, \& {Sinha}}]{Pandey2022}
{Pandey}, R., {Sharma}, S., {Dewangan}, L.~K., {et~al.} 2022, \apj, 926, 25

\bibitem[{{Park} {et~al.}(2021){Park}, {K{\'o}sp{\'a}l}, {Cruz-S{\'a}enz de
  Miera}, {Siwak}, {Dr{\'o}{\.z}d{\.z}}, {Ign{\'a}cz}, {Jaffe},
  {K{\"o}nyves-T{\'o}th}, {Kriskovics}, {Lee}, {Lee}, {Mace}, {Og{\l}oza},
  {P{\'a}l}, {Potter}, {Szab{\'o}}, {Sefako}, \& {Worters}}]{Park2021}
{Park}, S., {K{\'o}sp{\'a}l}, {\'A}., {Cruz-S{\'a}enz de Miera}, F., {et~al.}
  2021, \apj, 923, 171

\bibitem[{{Piascik} {et~al.}(2014){Piascik}, {Steele}, {Bates}, {Mottram},
  {Smith}, {Barnsley}, \& {Bolton}}]{Piascik2014}
{Piascik}, A.~S., {Steele}, I.~A., {Bates}, S.~D., {et~al.} 2014, in Society of
  Photo-Optical Instrumentation Engineers (SPIE) Conference Series, Vol. 9147,
  Ground-based and Airborne Instrumentation for Astronomy V, ed. S.~K.
  {Ramsay}, I.~S. {McLean}, \& H.~{Takami}, 91478H

\bibitem[{{Powell} {et~al.}(2012){Powell}, {Irwin}, {Bouvier}, \&
  {Clarke}}]{Powell2012}
{Powell}, S.~L., {Irwin}, M., {Bouvier}, J., \& {Clarke}, C.~J. 2012, \mnras,
  426, 3315

\bibitem[{{Press}(1978)}]{Press1978}
{Press}, W.~H. 1978, Comments on Modern Physics, Part C - Comments on
  Astrophysics, 7, 103

\bibitem[{{Rayner} {et~al.}(2003){Rayner}, {Toomey}, {Onaka}, {Denault},
  {Stahlberger}, {Vacca}, {Cushing}, \& {Wang}}]{SpeX2003PASP..115..362R}
{Rayner}, J.~T., {Toomey}, D.~W., {Onaka}, P.~M., {et~al.} 2003, \pasp, 115,
  362

\bibitem[{{Scaringi} {et~al.}(2015){Scaringi}, {Maccarone}, {Kording},
  {Knigge}, {Vaughan}, {Marsh}, {Aranzana}, {Dhillon}, \&
  {Barros}}]{Scaringi2015}
{Scaringi}, S., {Maccarone}, T.~J., {Kording}, E., {et~al.} 2015, Science
  Advances, 1, e1500686

\bibitem[{{Shingles} {et~al.}(2021){Shingles}, {Smith}, {Young}, {Smartt},
  {Tonry}, {Denneau}, {Heinze}, {Weiland}, {Flewelling}, {Stalder},
  {Clocchiatti}, {F{\"o}rster}, {Pignata}, {Rest}, {Anderson}, {Stubbs}, \&
  {Erasmus}}]{Shingles2021}
{Shingles}, L., {Smith}, K.~W., {Young}, D.~R., {et~al.} 2021, Transient Name
  Server AstroNote, 7, 1

\bibitem[{{Sicilia-Aguilar} {et~al.}(2017){Sicilia-Aguilar}, {Oprandi},
  {Froebrich}, {Fang}, {Prieto}, {Stanek}, {Scholz}, {Kochanek}, {Henning},
  {Gredel}, {Holoien}, {Rabus}, {Shappee}, {Billington}, {Campbell-White}, \&
  {Zegmott}}]{Sicilia2017}
{Sicilia-Aguilar}, A., {Oprandi}, A., {Froebrich}, D., {et~al.} 2017, \aap,
  607, A127

\bibitem[{{Siess} {et~al.}(2000){Siess}, {Dufour}, \& {Forestini}}]{Siess2000}
{Siess}, L., {Dufour}, E., \& {Forestini}, M. 2000, \aap, 358, 593

\bibitem[{{Siwak} {et~al.}(2023){Siwak}, {Hillenbrand}, {K{\'o}sp{\'a}l},
  {{\'A}brah{\'a}m}, {Giannini}, {De}, {Mo{\'o}r}, {Szil{\'a}gyi},
  {Jan{\'\i}k}, {Koen}, {Park}, {Nagy}, {Cruz-S{\'a}enz de Miera},
  {Fiorellino}, {Marton}, {Kun}, {Lucas}, {Udalski}, \&
  {Szab{\'o}}}]{Siwak2023}
{Siwak}, M., {Hillenbrand}, L.~A., {K{\'o}sp{\'a}l}, {\'A}., {et~al.} 2023,
  \mnras, 524, 5548

\bibitem[{{Siwak} {et~al.}(2020){Siwak}, {Og{\l}oza}, \&
  {Krzesi{\'n}ski}}]{Siwak2020}
{Siwak}, M., {Og{\l}oza}, W., \& {Krzesi{\'n}ski}, J. 2020, \aap, 644
  [\eprint[arXiv]{2011.00254}]

\bibitem[{{Siwak} {et~al.}(2013){Siwak}, {Ruci\'nski}, {Matthews}, {Kuschnig},
  {Guenther}, {Moffat}, {Rowe}, {Sasselov}, \& {Weiss}}]{Siwak2013}
{Siwak}, M., {Ruci\'nski}, S.~M., {Matthews}, J.~M., {et~al.} 2013, \mnras,
  432, 194

\bibitem[{{Siwak} {et~al.}(2018){Siwak}, {Winiarski}, {Og{\l}oza},
  {Dr{\'o}{\.z}d{\.z}}, {Zo{\l}a}, {Moffat}, {Stachowski}, {Rucinski},
  {Cameron}, {Matthews}, {Weiss}, {Kuschnig}, {Rowe}, {Guenther}, \&
  {Sasselov}}]{Siwak2018}
{Siwak}, M., {Winiarski}, M., {Og{\l}oza}, W., {et~al.} 2018, \aap, 618, A79

\bibitem[{{Skrutskie} {et~al.}(2006){Skrutskie}, {Cutri}, {Stiening},
  {Weinberg}, {Schneider}, {Carpenter}, {Beichman}, {Capps}, {Chester},
  {Elias}, {Huchra}, {Liebert}, {Lonsdale}, {Monet}, {Price}, {Seitzer},
  {Jarrett}, {Kirkpatrick}, {Gizis}, {Howard}, {Evans}, {Fowler}, {Fullmer},
  {Hurt}, {Light}, {Kopan}, {Marsh}, {McCallon}, {Tam}, {Van Dyk}, \&
  {Wheelock}}]{Skrutskie2006}
{Skrutskie}, M.~F., {Cutri}, R.~M., {Stiening}, R., {et~al.} 2006, \aj, 131,
  1163

\bibitem[{{Smith} {et~al.}(2020){Smith}, {Smartt}, {Young}, {Tonry}, {Denneau},
  {Flewelling}, {Heinze}, {Weiland}, {Stalder}, {Rest}, {Stubbs}, {Anderson},
  {Chen}, {Clark}, {Do}, {F{\"o}rster}, {Fulton}, {Gillanders}, {McBrien},
  {O'Neill}, {Srivastav}, \& {Wright}}]{Smith2020}
{Smith}, K.~W., {Smartt}, S.~J., {Young}, D.~R., {et~al.} 2020, \pasp, 132,
  085002

\bibitem[{{Steele} {et~al.}(2004){Steele}, {Smith}, {Rees}, {Baker}, {Bates},
  {Bode}, {Bowman}, {Carter}, {Etherton}, {Ford}, {Fraser}, {Gomboc}, {Lett},
  {Mansfield}, {Marchant}, {Medrano-Cerda}, {Mottram}, {Raback}, {Scott},
  {Tomlinson}, \& {Zamanov}}]{LT}
{Steele}, I.~A., {Smith}, R.~J., {Rees}, P.~C., {et~al.} 2004, in Society of
  Photo-Optical Instrumentation Engineers (SPIE) Conference Series, Vol. 5489,
  Ground-based Telescopes, ed. J.~{Oschmann}, Jacobus~M., 679--692

\bibitem[{{Stetson}(1987)}]{Stetson1987}
{Stetson}, P.~B. 1987, \pasp, 99, 191

\bibitem[{{Szab{\'o}} {et~al.}(2021){Szab{\'o}}, {K{\'o}sp{\'a}l},
  {{\'A}brah{\'a}m}, {Park}, {Siwak}, {Green}, {Mo{\'o}r}, {P{\'a}l},
  {Acosta-Pulido}, {Lee}, {Cseh}, {Cs{\"o}rnyei}, {Hanyecz},
  {K{\"o}nyves-T{\'o}th}, {Krezinger}, {Kriskovics}, {Ordasi}, {S{\'a}rneczky},
  {Seli}, {Szak{\'a}ts}, {Szing}, \& {Vida}}]{Szabo2021}
{Szab{\'o}}, Z.~M., {K{\'o}sp{\'a}l}, {\'A}., {{\'A}brah{\'a}m}, P., {et~al.}
  2021, \apj, 917, 80

\bibitem[{{Szab{\'o}} {et~al.}(2022){Szab{\'o}}, {K{\'o}sp{\'a}l},
  {{\'A}brah{\'a}m}, {Park}, {Siwak}, {Green}, {P{\'a}l}, {Acosta-Pulido},
  {Lee}, {Ibrahimov}, {Grankin}, {Kov{\'a}cs}, {Bora}, {B{\'o}di}, {Cseh},
  {Cs{\"o}rnyei}, {Dr{\'o}{\.z}d{\.z}}, {Hanyecz}, {Ign{\'a}cz}, {Kalup},
  {K{\"o}nyves-T{\'o}th}, {Krezinger}, {Kriskovics}, {Og{\l}oza}, {Ordasi},
  {S{\'a}rneczky}, {Seli}, {Szak{\'a}ts}, {S{\'o}dor}, {Szing}, {Vida}, \&
  {Vink{\'o}}}]{Szabo2022}
{Szab{\'o}}, Z.~M., {K{\'o}sp{\'a}l}, {\'A}., {{\'A}brah{\'a}m}, P., {et~al.}
  2022, \apj, 936, 64

\bibitem[{{Szegedi-Elek} {et~al.}(2020){Szegedi-Elek}, {{\'A}brah{\'a}m},
  {Wyrzykowski}, {Kun}, {K{\'o}sp{\'a}l}, {Chen}, {Marton}, {Mo{\'o}r}, {Kiss},
  {P{\'a}l}, {Szabados}, {Varga}, {Varga-Vereb{\'e}lyi}, {Andreas}, {Bachelet},
  {Bischoff}, {B{\'o}di}, {Breedt}, {Burgaz}, {Butterley}, {Carrasco},
  {{\v{C}}epas}, {Damljanovic}, {Gezer}, {Godunova}, {Gromadzki}, {Gurgul},
  {Hardy}, {Hildebrandt}, {Hoffmann}, {Hundertmark}, {Ihanec}, {Janulis},
  {Kalup}, {Kaczmarek}, {K{\"o}nyves-T{\'o}th}, {Krezinger}, {Kruszy{\'n}ska},
  {Littlefair}, {Maskoli{\={u}}nas}, {M{\'e}sz{\'a}ros}, {Miko{\l}ajczyk},
  {Mugrauer}, {Netzel}, {Ordasi}, {Pak{\v{s}}tien{\.{e}}}, {Rybicki},
  {S{\'a}rneczky}, {Seli}, {Simon}, {{\v{S}}i{\v{s}}kauskait{\.{e}}},
  {S{\'o}dor}, {Sokolovsky}, {Stenglein}, {Street}, {Szak{\'a}ts}, {Tomasella},
  {Tsapras}, {Vida}, {Zdanavi{\v{c}}ius}, {Zieli{\'n}ski}, {Zieli{\'n}ski}, \&
  {Zi{\'o}{\l}kowska}}]{SzegediElek2020_ApJ899130S}
{Szegedi-Elek}, E., {{\'A}brah{\'a}m}, P., {Wyrzykowski}, {\L}., {et~al.} 2020,
  \apj, 899, 130

\bibitem[{{Tonry} {et~al.}(2018){Tonry}, {Denneau}, {Heinze}, {Stalder},
  {Smith}, {Smartt}, {Stubbs}, {Weiland}, \& {Rest}}]{Tonry2018}
{Tonry}, J.~L., {Denneau}, L., {Heinze}, A.~N., {et~al.} 2018, \pasp, 130,
  064505

\bibitem[{{Tran} {et~al.}(2024){Tran}, {De}, \& {Hillenbrand}}]{Tran2023}
{Tran}, V., {De}, K., \& {Hillenbrand}, L. 2024, \mnras, 530, 2076

\bibitem[{{Vacca} {et~al.}(2003){Vacca}, {Cushing}, \& {Rayner}}]{Vacca2003}
{Vacca}, W.~D., {Cushing}, M.~C., \& {Rayner}, J.~T. 2003, \pasp, 115, 389

\bibitem[{{Waters} {et~al.}(2020){Waters}, {Magnier}, {Price}, {Chambers},
  {Burgett}, {Draper}, {Flewelling}, {Hodapp}, {Huber}, {Jedicke}, {Kaiser},
  {Kudritzki}, {Lupton}, {Metcalfe}, {Rest}, {Sweeney}, {Tonry}, {Wainscoat},
  \& {Wood-Vasey}}]{2020ApJS..251....4W}
{Waters}, C.~Z., {Magnier}, E.~A., {Price}, P.~A., {et~al.} 2020, \apjs, 251, 4

\bibitem[{{Wiebe} {et~al.}(2019){Wiebe}, {Molyarova}, V., {Vorobyov}, \&
  {Semenov}}]{Wiebe2019}
{Wiebe}, D.~S., {Molyarova}, T.~S., V., A.~V., {Vorobyov}, E.~I., \& {Semenov},
  D.~A. 2019, \mnras, 485, 1843

\bibitem[{{Wright} {et~al.}(2010){Wright}, {Eisenhardt}, {Mainzer}, {Ressler},
  {Cutri}, {Jarrett}, {Kirkpatrick}, {Padgett}, {McMillan}, {Skrutskie},
  {Stanford}, {Cohen}, {Walker}, {Mather}, {Leisawitz}, {Gautier}, {McLean},
  {Benford}, {Lonsdale}, {Blain}, {Mendez}, {Irace}, {Duval}, {Liu}, {Royer},
  {Heinrichsen}, {Howard}, {Shannon}, {Kendall}, {Walsh}, {Larsen}, {Cardon},
  {Schick}, {Schwalm}, {Abid}, {Fabinsky}, {Naes}, \& {Tsai}}]{Wright2010}
{Wright}, E.~L., {Eisenhardt}, P. R.~M., {Mainzer}, A.~K., {et~al.} 2010, \aj,
  140, 1868

\bibitem[{{Yoon} {et~al.}(2021){Yoon}, {Lee}, {Lee}, {Herczeg}, {Park}, {Mace},
  {Lee}, \& {Jaffe}}]{Yoon2021}
{Yoon}, S.-Y., {Lee}, J.-E., {Lee}, S., {et~al.} 2021, \apj, 919, 116

\bibitem[{{Zechmeister} \& {K{\"u}rster}(2009)}]{Zechmeister2009}
{Zechmeister}, M. \& {K{\"u}rster}, M. 2009, \aap, 496, 577

\bibitem[{{Zhu} {et~al.}(2007){Zhu}, {Hartmann}, {Calvet}, {Hernandez},
  {Muzerolle}, \& {Tannirkulam}}]{Zhu2007}
{Zhu}, Z., {Hartmann}, L., {Calvet}, N., {et~al.} 2007, \apj, 669, 483

\bibitem[{{Zhu} {et~al.}(2020){Zhu}, {Jiang}, \& {Stone}}]{Zhu2020}
{Zhu}, Z., {Jiang}, Y.-F., \& {Stone}, J.~M. 2020, \mnras, 495, 3494

\end{thebibliography}
%\begin{thebibliography}{}
%\end{thebibliography}
%\begin{appendix} %First appendix
%\section{Photometric observations during the outburst.}
%\label{ap1}
\onecolumn
%\longtab{
\begin{longtable}{ccccc}
\caption{\label{tab:photometry_out} Results from our own near-infrared and optical photometry of Gaia20bdk, gathered during the outburst.}\\
\hline\hline
JD            & Filter & Mag     & Unc   & Telescope \\ \hline
\endfirsthead
\caption{continued.}\\
\hline\hline
JD            & Filter & Mag     & Unc   & Telescope \\ \hline
\endhead 
\hline
\endfoot
2459343.50551 & $J$  & 12.110 & 0.078 & NTT   \\ 
2459343.50826 & $H$  & 10.910 & 0.064 & NTT   \\
2459343.50691 & $K_S$& 10.111 & 0.072 & NTT   \\ \hline
2459951.55187 & $J$  & 11.993 & 0.054 & NOT \\ 
2459951.56267 & $H$  & 10.842 & 0.035 & NOT \\
2459951.56473 & $K_S$& 10.005 & 0.052 & NOT \\ \hline
2460282.038   & $J$  & 11.892 & 0.053 & IRTF \\
2460282.036   & $H$  & 10.693 & 0.020 & IRTF \\
2460282.034   & $K_S$&  9.920 & 0.117 & IRTF \\
2460283.122   & $H$  & 10.665 & 0.020 & IRTF \\
2460283.119   & $K_S$&  9.887 & 0.064 & IRTF \\
2460362.856   & $J$  & 11.824 & 0.037 & IRTF \\
2460362.854   & $K_S$&  9.780 & 0.123 & IRTF \\ \hline\hline
2459211.46620 & $V$  & 17.285 & 0.109 & Lesedi  \\
2459211.46813 & $R$  & 15.971 & 0.124 & Lesedi  \\
2459211.46931 & $I$  & 14.349 & 0.066 & Lesedi \\ 
2459213.55886 & $B$  & 18.853 & 0.151 & Lesedi  \\
2459213.56120 & $V$  & 17.271 & 0.109 & Lesedi  \\
2459213.56272 & $R$  & 15.981 & 0.111 & Lesedi  \\
2459213.56372 & $I$  & 14.331 & 0.067 & Lesedi  \\
2459240.42596 & $B$  & 19.017 & 0.145 & Lesedi  \\
2459240.42951 & $V$  & 17.257 & 0.152 & Lesedi  \\
2459240.43147 & $R$  & 16.057 & 0.106 & Lesedi  \\
2459240.43234 & $I$  & 14.408 & 0.084 & Lesedi  \\ \hline
2459344.52138 & $B$  & 19.313 & 0.098 & NTT  \\
2459344.52464 & $V$  & 17.291 & 0.025 & NTT  \\
2459344.51866 & $R$  & 15.879 & 0.066 & NTT  \\ \hline
2459912.61068 & $B$  & 19.197 & 0.069 & Suhora  \\
2459912.61156 & $V$  & 17.271 & 0.053 & Suhora  \\
2459912.61230 & $R$  & 15.839 & 0.089 & Suhora  \\
2459912.61304 & $I$  & 14.195 & 0.012 & Suhora  \\ 
2459961.41927 & $B$  & 19.215 & 0.129 & Suhora  \\
2459961.42201 & $V$  & 17.248 & 0.032 & Suhora  \\
2459961.42109 & $R$  & 15.775 & 0.079 & Suhora  \\
2459961.42165 & $I$  & 14.130 & 0.017 & Suhora  \\ 
2460235.64325 & $B$  & 19.208 & 0.152 & Suhora  \\
2460235.64434 & $V$  & 17.376 & 0.070 & Suhora  \\
2460235.64519 & $R$  & 15.898 & 0.087 & Suhora  \\
2460235.64564 & $I$  & 14.287 & 0.020 & Suhora  \\ 
2460255.61355 & $B$  & 19.339 & 0.157 & Suhora  \\
2460255.61425 & $V$  & 17.288 & 0.105 & Suhora  \\
2459255.61774 & $R$  & 15.839 & 0.088 & Suhora  \\
2460255.61795 & $I$  & 14.191 & 0.020 & Suhora  \\
2460257.63370 & $B$  & 19.285 & 0.088 & Suhora  \\
2460257.63672 & $V$  & 17.249 & 0.049 & Suhora  \\
2459257.63820 & $R$  & 15.788 & 0.091 & Suhora  \\
2459257.63893 & $I$  & 14.216 & 0.012 & Suhora  \\ 
2460356.34938 & $B$  & 19.172 & 0.148 & Suhora  \\
2460356.35503 & $V$  & 17.198 & 0.047 & Suhora  \\
2459356.35229 & $R$  & 15.763 & 0.085 & Suhora  \\
2459356.35414 & $I$  & 14.158 & 0.014 & Suhora  \\
2460357.33312 & $B$  & 19.200 & 0.079 & Suhora  \\
2460357.33888 & $V$  & 17.156 & 0.045 & Suhora  \\
2459357.34860 & $R$  & 15.778 & 0.082 & Suhora  \\
2459357.34897 & $I$  & 14.146 & 0.016 & Suhora  \\ 
2460378.29557 & $B$  & 19.264 & 0.115 & Suhora  \\
2460378.29793 & $V$  & 17.218 & 0.068 & Suhora  \\
2460378.29848 & $R$  & 15.772 & 0.088 & Suhora  \\
2460378.29970 & $I$  & 14.088 & 0.016 & Suhora  \\ \hline
2459261.29861 & $V$  & 17.547 & 0.042 & RC80  \\
2459261.30076 & $r$  & 16.659 & 0.069 & RC80  \\
2459261.30290 & $i$  & 15.566 & 0.053 & RC80 \\
2459268.30932 & $V$  & 17.578 & 0.042 & RC80  \\
2459268.31147 & $r$  & 16.691 & 0.067 & RC80  \\
2459268.31361 & $i$  & 15.584 & 0.081 & RC80  \\
2459271.27873 & $r$  & 16.731 & 0.082 & RC80  \\
2459271.28089 & $i$  & 15.508 & 0.065 & RC80  \\
2459272.27519 & $r$  & 16.661 & 0.093 & RC80  \\
2459272.27735 & $i$  & 15.488 & 0.061 & RC80  \\
2459274.28487 & $V$  & 17.488 & 0.051 & RC80  \\
2459274.28703 & $r$  & 16.659 & 0.064 & RC80  \\
2459274.28919 & $i$  & 15.468 & 0.061 & RC80  \\
2459511.60812 & $V$  & 17.609 & 0.027 & RC80  \\
2459511.60922 & $r$  & 16.543 & 0.071 & RC80  \\
2459511.60998 & $i$  & 15.491 & 0.074 & RC80  \\
2459542.56258 & $V$  & 17.607 & 0.024 & RC80  \\
2459542.56368 & $r$  & 16.499 & 0.078 & RC80  \\
2459542.56444 & $i$  & 15.400 & 0.071 & RC80  \\
2459586.42710 & $V$  & 17.402 & 0.025 & RC80  \\
2459586.42821 & $r$  & 16.463 & 0.078 & RC80  \\
2459586.42897 & $i$  & 15.348 & 0.083 & RC80  \\
2459616.33608 & $V$  & 17.466 & 0.035 & RC80  \\
2459616.33719 & $r$  & 16.503 & 0.070 & RC80  \\
2459616.33794 & $i$  & 15.393 & 0.074 & RC80  \\
2459870.63185 & $V$  & 17.232 & 0.033 & RC80  \\
2459870.63297 & $r$  & 16.444 & 0.098 & RC80  \\
2459870.63374 & $i$  & 15.298 & 0.080 & RC80  \\
2459904.55552 & $V$  & 17.321 & 0.027 & RC80  \\
2459904.55662 & $r$  & 16.333 & 0.074 & RC80  \\
2459904.55737 & $i$  & 15.270 & 0.066 & RC80  \\
2459938.44617 & $V$  & 17.183 & 0.029 & RC80  \\
2459938.44728 & $r$  & 16.400 & 0.072 & RC80  \\
2459938.44804 & $i$  & 15.256 & 0.071 & RC80  \\
2460231.63856 & $V$  & 17.248 & 0.028 & RC80  \\
2460231.63967 & $r$  & 16.374 & 0.081 & RC80  \\
2460231.64043 & $i$  & 15.342 & 0.078 & RC80  \\
2460261.54642 & $V$  & 17.351 & 0.027 & RC80  \\
2460261.55081 & $r$  & 16.340 & 0.078 & RC80  \\
2460261.55157 & $i$  & 15.259 & 0.069 & RC80  \\
2460282.51737 & $V$  & 17.309 & 0.019 & RC80  \\
2460282.51848 & $r$  & 16.315 & 0.080 & RC80  \\
2460282.51925 & $i$  & 15.277 & 0.074 & RC80  \\
\hline\hline
\end{longtable}
%\end{appendix}
%}
%\begin{appendix} %First appendix
%\section{Determination of MSO colour equations}
%\label{ap1}

%\begin{table*}
%\caption{Colour equation transforming the Mount Suhora $BVRI$ system to the Johnson-Cousins $BVR_CI_C$ system. The coefficients were determined during three photometric nights of 2021 October 29, 30 and 31. Note that two equation options are provided for $VR$ filters.}  
%\centering
%\begin{tabular}{c c c c c c}
%\hline
%$ft$     & $k_{ft}$ & $\beta_{ft}$ & $\mu_{ft}$    & $C_{ft}^{a}$ & $CI$      \\ \hline
%$B$      & 0.2492(50)  & -0.0170(3)    & +0.1579(27) & 2.4487(68) & $B-V$   \\ 
%$V$      & 0.1430(41)  & -0.0125(1)    & -0.1124(20) & 2.5968(55) & $B-V$   \\ 
%$V$      & 0.1395(39)  & -0.0168(1)    & -0.1923(33) & 2.5993(53) & $V-R_C$   \\ 
%$R$      & 0.0738(40)  & -0.0052(1)    & -0.0364(23) & 2.4223(53) & $V-R_C$   \\  
%$R$      & 0.0728(39)  & -0.0035(1)    & -0.0363(23) & 2.4233(53) & $R_C-I_C$   \\  
%$I$      & 0.0304(52)  & -0.0155(1)    & +0.0713(35) & 3.1689(69) & $R_C-I_C$   \\ \hline
%\end{tabular}
%\label{Tab.coeff}
%\end{table*}
%\end{appendix}
\end{document}